\documentclass[acmsmall,screen]{acmart}

\AtBeginDocument{%
  \providecommand\BibTeX{{%
    \normalfont B\kern-0.5em{\scshape i\kern-0.25em b}\kern-0.8em\TeX}}}

\setcopyright{none}

\usepackage[T1]{fontenc}
\usepackage{lmodern}

\usepackage{amsmath}
\usepackage{booktabs}
\usepackage{amsthm}
\usepackage{adjustbox}
\usepackage{multirow} 
\usepackage{algorithmic}
\usepackage{algorithm}
\usepackage{graphicx}  
\usepackage{subfigure}

\usepackage{url}
\setcitestyle{numbers,sort&compress}

\DeclareMathOperator*{\argmin}{arg\,min}

\DeclareUnicodeCharacter{2212}{-}

\begin{document}

\title{COVID-19 Modeling: A Review}

\author{Longbing Cao}
\authornote{Corresponding to: Longbing Cao (Longbing.Cao@uts.edu.au).}
\affiliation{%
  \institution{University of Technology Sydney}
  \city{}
  \state{}
  \postcode{}
}

\author{Qing Liu}
\affiliation{%
  \institution{University of Technology Sydney}
  \city{}
  \state{}
  \postcode{}
}

\renewcommand{\shortauthors}{Cao, et al.}



\begin{abstract}
The SARS-CoV-2 virus and COVID-19 disease have posed unprecedented and overwhelming demand, challenges and opportunities to domain, model and data driven modeling. This paper provides a comprehensive review of the challenges, tasks, methods, progress, gaps and opportunities in relation to modeling COVID-19 problems, data and objectives. It constructs a research landscape of COVID-19 modeling tasks and methods, and further categorizes, summarizes, compares and discusses the related methods and progress of modeling COVID-19 epidemic transmission processes and dynamics, case identification and tracing, infection diagnosis and medical treatments, non-pharmaceutical interventions and their effects, drug and vaccine development, psychological, economic and social influence and impact, and misinformation, etc. The modeling methods involve mathematical and statistical models, domain-driven modeling by epidemiological compartmental models, medical and biomedical analysis, AI and data science in particular shallow and deep machine learning, simulation modeling, social science methods, and hybrid modeling.
\end{abstract}

\maketitle


\section{Introduction}
\label{sec:intro}

Here, we give a brief overview of the COVID-19 pandemic, the global effort on modeling COVID-19, and the scope, motivation and contributions of this comprehensive review.

\subsection{COVID-19 Pandemic}
\label{subsec:pandemic}

The coronavirus disease 2019, designated as COVID-19, is a new epidemic caused by the severe acute respiratory syndrome coronavirus 2 (SARS-CoV-2) virus. SARS-CoV-2 and COVID-19 have overwhelmingly shocked and shaken the entire world. After the first outbreak in Wuhan China in Dec 2019, the disease spread rapidly across the world in only two months due to its strong human-to-human transmission ability. The World Health Organization (WHO) declared COVID-19 a pandemic on 11 March 2020. To date, COVID-19 has infected more than 194M people with 4M having lost their lives\footnote{https://www.worldometers.info/coronavirus/\#countries}. The continuous iterative mutative infections are even more seriously troubling 215 countries and territories with increasingly unexpected resurgences and virus mutations continuously challenging  pandemic containment, vaccinations and treatments. With these widespread and continuous infections, the increasingly contagious mutations and resurgences, and the slow rollout of global vaccination to achieve herd immunity, COVID-19 is and will continuously and fundamentally transform global public health, the economy and society with an increasingly unprecedented impact on every aspect of life. COVID-19 has not only exerted unprecedented pressure on global healthcare systems it has also fundamentally challenged the relevant scientific research on understanding, modeling, diagnosing and controlling the virus and disease. Many questions and challenges arise about COVID-19 in relation to the nature of the coronavirus; the virus' epidemic characteristics, transmission and influence processes; the disease's medical and genomic characteristics, dynamics and evolution, and the pros and cons of existing containment, diagnosis, treatment and precaution strategies. 

Compared with the epidemics that have been defeated in recent decades, such as the severe acute respiratory syndrome (SARS) and the Middle East respiratory syndrome (MERS), SARS-CoV-2 and COVID-19 are more complicated and more transmissible from human to human \cite{world2020transmission}. Its strong uncertainty, transmission and mutation are some of the key factors that make COVID-19 a continuous, unprecedented and evolving pandemic. Another key challenge of COVID-19 is its long incubation period ranging from 1 to 14 days or even longer and high asymptomatic proportion. In the incubation period,  infectious individuals are contagious but show no symptoms. Consequently, susceptible individuals may not be aware that they are infected, hindering their timely identification and contact tracing. Unfortunately, the number of asymptomatic  and mild symptomatic infections is significant. Asymptomatic infectives are a concerning hidden source of the widespread infection and resurgence of COVID-19 \cite{kronbichler2020asymptomatic}, making it extremely difficult to be absolutely clear about the source, diagnosis and mitigation of COVID-19. This brief comparison poses challenging questions to be answered scientifically, including how does COVID-19 differ from other epidemic or endemic diseases, what is the nature of its asymptomatic phase and infection, what are more effective methods to contain and treat the virus and disease, and how does the virus mutate and react to vaccines.

Specifically, to slow the pandemic and bring infections under control, most  governments have implemented many non-pharmaceutical interventions (NPI) such as social distancing, school and university closure, infective isolation or quarantine, banning public events and travel, and business lockdowns, etc. These interventions also incur significant socioeconomic costs and have a wide impact on businesses, triggering various debates on a trade-off between epidemic control and introduced negative impact, relaxed mitigation, or even herd immunity. However, it is quantitatively unclear what makes a better trade-off or timing and the appropriate extent of mitigation, what are the positive and negative impacts of NPI on pandemic control and socioeconomic wellness, and how has COVID-19 influenced other aspects of public life, work, mental and medical health, and the economy on both individual and population (e.g., a city, country to the globe) levels.

The SARS-CoV-2 virus and the COVID-19 disease present significant challenges for health care, the government, society and the economy, and also present both challenges and opportunities for the scientific and research communities. Coronavirus and COVID-19 have reshaped the focus of global scientific attention and efforts, including the exploration of the aforementioned challenges. The scientific research is comprehensive, spreading across almost every discipline from epidemiology to psychology and fostering new research areas and topics such as coronavirus epidemiology and genomics. Of the scientific efforts in 2020, \textit{COVID-19 studies} emerged as the most important and active research area, where the growing volume of COVID-19 data is a valuable intangible asset for evidence-based virus and disease understanding, fostering a critical research agenda and global interest in COVID-19 modeling. \textit{COVID-19 modeling} aims to quantitatively understand and characterize the virus and disease characteristics,  estimate and predict COVID-19 transmission, and identify cases and trends, intervention measures and their effects, and their impacts on social, economic, psychological and political aspects, etc. COVID-19 modeling plays an irreplaceable role in almost every aspect of the fight against the COVID-19 pandemic, in particular, in characterizing the intricate nature of COVID-19 and discovering insights for virus containment, disease treatment, drug and vaccine development, and mitigating its broad socioeconomic impact. These motivate this review of the global reaction to modeling COVID-19.

\subsection{Global Effort on Modeling COVID-19}
\label{subsec:globmodeling}

Hundreds of thousands of studies on coronavirus and COVID-19 and the related issues have been published in the literature in less than two years, as indicated by the WHO-collected global literature on the coronavirus disease COVID-19\footnote{\url{https://search.bvsalud.org/global-literature-on-novel-coronavirus-2019-ncov/}}. There are also an increasing number of publications, which review the relevant progress of specific aspects of coronavirus and COVID-19 studies. 
As of 22 February 2021, approximately 200k references have been published on issues relating to COVID-19 in various fields such as  medical/biological science, computer science, economics, environment, policy, engineering, etc. to understand and study SARS-CoV-2 and COVID-19 and the associated problems. The computer science communities alone have contributed to approximately 7k publications, including work on modeling COVID-19 using so-called AI and data science techniques, in particular classic and deep analytical and machine learning methods. Data-driven discovery \cite{dst_Cao15,eisenstein2018infection} and \textit{COVID-19 data science} play a major role in COVID-19 modeling, aiming to discover valuable knowledge and insights from various kinds of publicly available data including daily cases, texts, biomedical images, mobility, and so on. As further illustrated in this paper, every modeling technique has been applied to COVID-19 in some way. For example, classic epidemic models are tailored for COVID-19 to model its macroscopic transmission and predict the trends of the spread of the virus; generative models with Bayesian hierarchical structures are applied to capture the effects of NPI; deep natural language processing approaches are adopted to understand the growth, nature and spread of COVID-19 and people’s reactions based on the textual data from social media such as Twitter. Modeling also helps to understand and characterize every aspect of coronavirus and COVID-19, from its epidemiological characteristics to the underlying genomic reactions and mutations and drug and vaccine development.

This review also shows that there are many challenges, gaps and opportunities in modeling COVID-19. First, publicly available COVID-19 data is limited with partial, inconsistent, even erroneous, biased, noisy and uncertain observations and statistics due to the limited, imbalanced and non-universal test ability, and non-unified reporting standards and statistical errors, especially at the beginning of the epidemic and in those undeveloped regions and countries. Second, the aforementioned long incubation period from infection to the onset of symptoms and the large number of asymptomatic to mild infections make correct and instant reporting difficult and leads to a significant number of undetected and unreported cases, degrading data quality and trust. Third, coronavirus and COVID-19 exhibit unique complexities, which differ from existing epidemics, including its transmission, infectivity in ethnic populations, external NPIs, people's NPI reactions and behavioral changes as a result of COVID-19 mitigation policies, and the rapid and mysterious mutation and spread of coronavirus. Lastly, the modeled problems and areas are fragmented, and although the modeling techniques and results are highly comprehensive, they are divided and evolving. 

These brief observations indicate the critical need to model COVID-19 and the urgency of forming a comprehensive understanding of the progress being made in COVID-19 modeling, the research gaps and the open issues. This overview is crucial for not only furthering COVID-19 modeling research but also for informing insights on scientific and public strategies and actions to better battle this pandemic and future pandemics.

\subsection{Our Findings}
\label{subsec:findings}

In this review, we seek answers and indications to the following major questions:
\begin{itemize}
    \item What is the research landscape of COVID-19 modeling, i.e., what COVID-19 problems can be modeled and what modeling techniques can address these COVID-19 issues?
    \item How well do AI and data science, specifically machine learning and deep learning, deepen and broaden the understanding and management of the COVID-19 pandemic?
    \item How do varied techniques perform differently in modeling COVID-19?
    \item What are the gaps in modeling COVID-19?
    \item Where can AI and data science make new, more or better difference in containing COVID-19?
\end{itemize}

This comprehensive review obtains a relatively full spectrum of the virus challenges, data issues, techniques, gaps and opportunities in relation to modeling COVID-19. In addition to many specific observations obtained through this review, as discussed in the following sections, here we highlight the following high-to-low level observations and quantitative indications of results in modeling different COVID-19 problems and data. 
\begin{itemize}
    \item \textit{COVID-19 problems and complexities}: as summarized in Section \ref{subsec:objectives}, the spectrum of problems covers typical aspects of epidemic dynamics and transmission, virus and disease diagnosis, infection identification, contact tracing, virus mutation and resurgence, medical diagnosis and treatment, pharmaceutical interventions, pathological and biomedical analysis, drug and vaccine development, non-pharmaceutical interventions, and socioeconomic influence and impact. We further summarize the COVID-19 characteristics and complexities in Section \ref{subsec:disease}, including complex hidden epidemic attributes, high contagion, high mutation, high proportion of asymptomatic to mild symptomatic infections, varied and long incubation periods, ethnic sensitivity, and other high uncertainties, which shows significant differences between SARS-CoV-2 and other existing viruses and epidemics.  
    \item \textit{COVID-19 data and challenges}: the core data is related to the daily reported number of asymptomatic infections, the number of confirmed, recovered and deceased cases, patients' demographics, pathological, clinical and genomic results of the virus and disease tests, and patients' activities and hospitalized information, etc.; external data comprises NPI policies and events, the resident's responses and behaviors, public activities, texts from online and health services, weather, and environment, etc. Since the data spectrum is indeed comprehensive, almost all data complexities widely explored in general modeling have also been involved in modeling COVID, including data uncertainty, dynamics and nonstationarity, various data quality issues such as incompleteness, inconsistency, inequality and incomparability, lack of ground truth information, and limited size of daily reports. For further discussion, see Section \ref{subsec:data}.
    \item \textit{Modeling challenges}: the COVID-19 complexities and data challenges bring about various modeling issues and challenges, as summarized in Section \ref{subsec:modelcomp}, including modeling low-to-poor quality data, modeling small and limited data, learning with weak-to-no prior knowledge and ground truth, modeling hierarchical and diverse forms of heterogeneities and interactions between multi-source core and external data, and disclosing hidden and unknown attributes and dynamics of the virus and disease.
    \item \textit{Modeling techniques}: as summarized in Section \ref{subsec:roadmap} and detailed in Sections \ref{sec:mathmethods} to \ref{sec:hybridmodeling}, the spectrum of techniques is wide enough to cover conventional mathematical and statistical modeling, simulation methods and epidemiological modeling, modern data-driven discovery and machine learning, and the recent advances in deep learning, in addition to social science methods including psychological, economic and behavior modeling methods. Among the 200k publications on COVID-19 with 22k on modeling COVID-19 \cite{who-analysis}, about $4\%$, $6\%$, and $60\%$ of the entire literature involves the research and application on broad computer science, social science, and medical science problems and methods, respectively. Of the 22k publications on modeling, about 50\% applies machine learning, deep learning, mathematical modeling, epidemic modeling to address medical problems. In addition, mathematical models, machine learning, deep learning, and epidemiological models are mostly explored in modeling COVID-19, contributing some $59\%$, $18\%$, $9\%$ and $15\%$ to the 22k publications, respectively. In the top-10 modeling methods, regression methods contribute to 34\%, CCNs to 3\%, Bayesian models to 3\%, SIR models to 4\%, and simulation to 9\% of all modeling publications. 
    \item \textit{Modeling tasks}: on one hand, modeling tasks address almost all of the aforementioned COVID-19 problems and complexities, amounting to $15\%$ on epidemic modeling, $9\%$ on diagnosis and identification, $15\%$ on influence and impact, $7\%$ on simulation, and $1\%$ on resurgence and mutation; on the other hand, the literature covers overwhelming analytical and learning tasks including roughly $2\%$ on unsupervised learning and clustering, $6\%$ on classification, $2\%$ on multi-source and multi-modal data modeling and multi-task learning, and $6\%$ on forecasting and prediction. 
    \item \textit{Epidemic attributes}: as summarized in Section \ref{subsec:disease}, it is estimated that the reproduction number (probably larger than 3 in the original waves and over 2 in the resurgence after receiving vaccination) is much higher than SARS and MERS, the incubation may last for an average of 5 to even beyond 14 days, the asymptomatic infections may be much higher than 20\% with even up to 80\% undocumented infections in some countries, some of virus mutants may have increase the transmission rate by more than 50\% over the original strain.  
    \item \textit{Machine learning performance}: as shown in Section \ref{subsec:machlearning}, shallow machine learning methods report an accuracy of over 90\% in predicting COVID-19 outbreaks, over 96\% in disease diagnosis on clinical reports, over 98\% in diagnosis on medical images, and close to 99\% in diagnosis further involving latent features with specific settings and data; in addition, as shown in Sections \ref{subsec:deeplearning} and \ref{subsec:medianal}, DNN variants achieve significant prediction performance on COVID-19 images and signals, e.g., an accuracy of over 92\% on cough sound using LSTM, over 99\% on chest X-ray images using CNN and imagenet variants, and less than 5\% MAE on real unlabelled lung CT images by attention and gated U-net. 
    \item Non-pharmaceutical intervention effect: as discussed in Section \ref{subsec:NPIeffect}, NPIs such as business lockdowns, school closures, limiting gatherings, and social distancing are crucial to contain the virus outbreaks and reducing COVID-19 case numbers; e.g., reducing the reproduction number by 13\%-42\% individually or even 77\% jointly by these control measures; and resulting in over 40\% transmission reduction by restricting human mobility and interactions.
    \item \textit{Emotional, social and economic impact}: the COVID-19 pandemic has generated an overwhelming negative impact on the public mental health (e.g., significantly increasing anxiety, stress, depression and suicide), economic growth and workforce (e.g., over 20\% estimated annual GDP loss in 2020), public health systems, global supply chain, sociopolitical systems, and information disorder, as discussed in Section \ref{subsec:NPIeffect}.
    \item \textit{Modeling gaps}: as commented in Section \ref{subsec:gaps}, the review also finds various issues and limitations of existing research, e.g., an insufficient, biased and partial understanding of COVID-19 complexities and data challenges; a simple and direct application of modeling techniques on often simple data; lack of robust, generalizable and tailored designs and insights into the virus and disease nature and complexities.
    \item \textit{Future opportunities}: the discussion in Section \ref{subsec:opport} indicates significant new opportunities, e.g., studying rarely to poorly addressed problems such as epidemiologically modeling mutated virus attributes, complex interactions between core and external factors, and the influence of external factors on epidemic dynamics and NPI effect; new directions and methods such as hybridizing multiple sources of data or methods to characterize the complex COVID systems; and novel AI, data science and machine learning research on large-scale simulation of the intricate evolutionary mechanisms in COVID, discovering robust and actionable evidence to dynamically personalize the control of potential resurgence and balance the economic and mental recovery and the virus containment.
\end{itemize}

Note, the above-quoted numerical results are illustrative, which do not represent the state-of-the-art performance. Interested readers may refer to \cite{who-analysis} and specific references for more comprehensive information about how the global scientists have responded to model COVID-19 and \cite{covid-findings} to understand what quantitative results COVID-19 modeling has identified in both the above questioned areas and other areas. 

\subsection{Contributions and Limitations}
\label{subsec:contribution}

Several surveys have been conducted on COVID-19 modeling, which review the progress from specific perspectives, e.g., COVID-19 characteristics \cite{esakandari2020comprehensive}, epidemiology \cite{park2020systematic}, general applications of AI and machine learning \cite{nazrul2020survey,nguyen2020artificial} such as for epidemic and transmission forecasting and prediction \cite{chen2020survey,rahimi2021review,Byambasuren-cov20}, virus detection, spread prevention, and medical assistance \cite{shahid2020machine}, policy effectiveness and contact tracing \cite{mao2020data}, infection detection and disease diagnosis  \cite{kronbichler2020asymptomatic,buitrago2020occurrence}, virology and pathogenesis \cite{Bastosm2516}, drug and vaccine development \cite{Arshadik20}, and mental health \cite{xiong2020impact}. The methods which have been reviewed include epidemiological modeling \cite{park2020systematic}, general AI and machine learning methods \cite{Arshadik20,nguyen2020artificial,nazrul2020survey,mohamadou2020review,chen2020survey,Rasheedj21,Kamalovf21}, data science \cite{latif2020leveraging}, computational intelligence \cite{tseng2020computational}, computer vision and image processing \cite{ulhaq2020covid,Shif21}, statistical models \cite{mohamadou2020review}, and deep learning \cite{Zerouala20,IslamKAZ21}. These reviews paint a partial picture of what happened in their selected areas based on several references and specific techniques. However, there are currently no comprehensive surveys or critical analyses of the intricate challenges posed by the virus, the disease, the data and the modeling. 

This review is the first attempt to provide a comprehensive picture of the problems  by modeling  coronavirus and COVID-19 data. We start by categorizing the characteristics and challenges of the COVID-19 disease, the data and the modeling in Section \ref{sec:char-compl}. A transdisciplinary landscape is formed to categorize and match both COVID-19 modeling tasks and objectives and categorize the corresponding methods and general frameworks in Section \ref{sec:landscape}. The review then focuses on structuring, analyzing and comparing the work on mathematical, data-driven (shallow and deep machine learning), domain-driven (epidemic, medical and biomedical analyses) modeling in Sections \ref{sec:mathmethods}, \ref{sec:datalearning} and \ref{sec:domainmodeling}, respectively. Section \ref{sec:impactmodeling} further discusses the modeling on the influence and impact of COVID-19, Section \ref{sec:simulation} reviews the work on COVID-19 simulations, and the related work on COVID-19 hybrid modeling is reviewed in Section \ref{sec:hybridmodeling}. Lastly, Section \ref{sec:dis-opps} further discusses the significant gaps and opportunities in modeling COVID-19.

This review aims to be \textit{specific} to COVID-19 modeling so pure domain-specific research on its medicine, vaccine, biology and pathology is excluded; more \textit{comprehensive} than the other references to cover problems and techniques from classic to present AI, data science and beyond; \textit{unique} in summarizing the challenges of the COVID-19 disease, data and modeling; \textit{structural} and \textit{critical} by categorizing, comparing, criticizing and generalizing typical modeling methods tailored for COVID-19 modeling from various disciplines and areas; and \textit{insightful} by extracting conclusive and contrastive (to other epidemics) findings about the virus and disease from the references. The review incorporates much discussion on the topics, opportunities and directions to tackle those issues which are rarely or poorly addressed or areas which remain open in the broad research landscape of modeling COVID-19.

However, this review also presents the various limitations and opportunities for further work. (1) As the scope and capacity of the review is limited, we do not cover the domain-specific literature on pure medical, biomedical and social science-oriented topics and methods without involving modeling methods. (2) There are over 10k references closely relevant to modeling COVID-19 and numerous specific modeling techniques from various disciplines and areas \cite{who-analysis}, which could not be fully covered or highlighted in detail in this review. (3) As discussed in the above, different from the narrowly-focused review papers in the area which highlight specific techniques and their relevant references, we only present those mostly used (useful) modeling techniques by summarizing their generalizable formulations. (4) This review does not answer many important questions concerning modelers, governments, policy-makers and domain experts, e.g., what has the modeling told us about the nature of COVID-19, which could be further highlighted in more purposeful reviews and analyses. (5) There are many challenging problems yet to be informed or addressed by the modeling progress, as discussed in Section \ref{sec:dis-opps}. (6) There are increasingly more and newer references including preprints emerging online every day, which poses significant challenges for us to cover all up-to-date important references on modeling COVID-19.

\section{COVID-19 Characteristics and Complexities}
\label{sec:char-compl}

In this section, we summarize the main characteristics and challenges of the COVID-19 disease, the data and the modeling, which are connected to the various modeling tasks and methods reviewed in this paper.

\subsection{COVID-19 Disease Characteristics}
\label{subsec:disease}

Modeling COVID-19 is highly challenging because its sophisticated epidemiological, clinical and pathological characteristics are poorly understood~\cite{holmdahl2020wrong,Hu-cov21,park2020systematic}. Despite common epidemic clinical symptoms like fever and cough~\cite{guan2020clinical}, SARS-CoV-2 and COVID-19 have many other sophisticated characteristics \cite{world2020transmission} that make them more mysterious, contagious and challenging for quantification, modeling and containment. We highlight a few of these below.

\textit{High contagiousness and rapid spread.} 
The high contagiousness of SARS-CoV-2 is one of the most important factors driving the COVID-19 pandemic. In epidemiology, the \textit{reproduction number} $R_0$ denotes the transmission ability of an epidemic or endemic. It is the expected number of cases directly generated by one case in a population where all individuals are susceptible to infection~\cite{fraser2009pandemic}. If $R_0>1$, the epidemic will begin to transmit rapidly in the population, while $R_0 < 1$ indicates that the epidemic will gradually vanish and will not lead to a large-scale outbreak. Different computational methods have resulted in varying reproduction values of COVID-19 in different regions. For example, Sanche et al. \shortcite{sanche2020high} report a median $R_0$ value of 5.7 with a 95\% confidence interval (CI) [3.8, 8.9] during the early stages of the epidemic in Wuhan China. Gatto et al.~\shortcite{gatto2020spread} estimate a generalized reproduction value of 3.60 (95\% CI: 3.49 to 3.84) using the susceptible-exposed-infected-recovered (SEIR)-like transmission model in Italy. de Souza et al.~\shortcite{de2020epidemiological} report a value of 3.1 (95\% CI: 2.4 to 5.5) in Brazil. The review finds that the $R_0$ of COVID-19 may be larger than 3.0 in the initial stage, higher than that of SARS (1.7-1.9) and MERS ($<1$)~\cite{petrosillo2020covid}. It is generally agreed that SARS-CoV-2 is more transmissible than severe acute respiratory syndrome conronavirus (SARS-CoV) and Middle East respiratory syndrome coronavirus (MERS-CoV) although SARS-CoV-2 shares 79\% of the genomic sequence identity with SARS-CoV and 50\% with MERS-CoV~\cite{lu2020genomic,esakandari2020comprehensive,petersen2020comparing,Hu-cov21}. 

\textit{A varying incubation period.} 
The \textit{incubation period} of COVID-19, also known as the pre-symptomatic period, refers to the time from becoming infected by exposure to the virus and symptom onset. A median incubation period of approximately 5 days was reported in~\cite{lauer2020incubation} for COVID-19, which is similar to SARS. In~\cite{park2020systematic}, the mean incubation period was found to range from 4 to 6 days, which is comparable to SARS (4.4 days) and MERS (5.5 days). Although an average incubation period of 5-6 days is reported in the literature, the actual incubation period may be as long as 14 days~\cite{world2020transmission,yu2020familial,lauer2020incubation}. The widely varying COVID-19 incubation period and its uncertain value in a specific hotspot make case identification and infection control very difficult. Unlike SARS and MERS, COVID-19 infected individuals are already contagious during their incubation periods. As it is likely that they are unaware that they are infected and have no-to-mild symptoms during this period, they may easily be the unknown sources of widespread transmission. This has informed screening and control policies, e.g., mandatory 14-day quarantine and isolation, corresponding to the longest predicted incubation time. 

\textit{A large number of asymptomatic and undocumented infections.} 
It is clear that COVID-19 has a broad clinical spectrum which includes asymptomatic and mild illness~\cite{chan2020familial,li2020asymptomatic,petersen2020comparing}. However, the accurate number of asymptomatic and mild-symptomatic infections of both original and new-generations of viruses remains unknown. Asymptomatic infections may not be screened and diagnosed before symptom onset, leading to a large number of undocumented infections and the potential risk of contact with infected individuals~\cite{kronbichler2020asymptomatic}. The review in \cite{Byambasuren-cov20} reports that of those who tested positive in studies which were conducted in seven countries, the proportion who were asymptomatic ranged from 6\% to 41\%, while the study in \cite{Wangyb20} reports that 23\% of those infected by COVID were asymptomatic. Buitrago-Garcia et al.~\shortcite{buitrago2020occurrence} found that most people who are infected with COVID-19 do not remain asymptomatic throughout the course of infection, and only 20\% of infections remain asymptomatic during follow-up, however this estimate requires further verification and study. Ravindra et al.~\shortcite{ravindra2020consideration} analyzed the possibility of different levels of asymptomatic transmission in the community and concluded that asymptomatic human transmission is relevant to the varying incubation periods between people and about 31\% of all populations are asymptomatic, including familial clusters, adults, children, health care workers, and travelers. The study in \cite{Li-sci20} shows that a large percentage (86\%) of infections are undocumented, about 80\% of documented cases are due to transmission from undocumented cases, and the transmission rate of undocumented infections is about 55\% of that of documented cases. 

\textit{High mutation with mysterious strains and high contagion.} 
The four major SARS-CoV-2 variants of concern such as B.1.1.7 (labeled Alpha by WHO) and B.1.351 (Beta) variants have higher transmissibility (B.1.1.7 has approximately 50\% increased transmission)~\cite{priesemann2021action} and reproduction rate (B.1.1.7 has an increase reproduction rate of 1-1.4)~\cite{volz2021transmission}, challenging existing vaccines, containment and mitigation methods. The recently identified variant B.1.617.2 (Delta) in India has even more sophisticated transmissibility and infectious characteristics. The identified variants of concern generally have increased transmissibility (20-50\%), increased detrimental change of epidemiology, more severe virulence and disease presentation (e.g., increased hospitalizations or deaths), and result in the decreased effectiveness of public health and social measures, reduced effectiveness of available diagnostics, vaccines and therapeutics, increased diagnostic detection failures, and reduced neutralization by antibodies generated during previous infection or vaccination \cite{WHO-variants,US-CDC}.

\textbf{Discussion.} While the above summarizes the most recent understanding of  SARS-CoV-2 and COVID-19 complexities, it is also noted that  knowledge on the nature of the virus and its mutation is limited. Without knowing its origins, there is much misinformation about the virus, its contagion and the  interventions required~\cite{roozenbeek2020susceptibility}. 
There is weak to no ground truth about the reality of its infection, symptoms shown in medical imaging, and mitigation and treatment measures. There have been no joint global pathological, epidemiological, biomedical and socioeconomic studies which provide a deep and systematic understanding of the COVID-19 virus and disease complexities, common knowledge, and ground truth.

\subsection{COVID-19 Data Challenges}
\label{subsec:data}

COVID-19 involves multisource, small, sparse and  quality-inconsistent data \cite{latif2020leveraging}\footnote{COVID-19 modeling: https://datasciences.org/covid19-modeling/}. Typical data sources and factors include 
(1) epidemiological factors (e.g., origin, incubation period, transmission rate, mortality, morbidity, and high to least vulnerable population, etc.);
(2) daily new-infected-recovered-death case numbers, their reporting time and region of occurrence; 
(3) quarantine and mitigation measures and policies (e.g., social distancing and border control) relating to communities and individuals; 
(4) clinical, pathological and genomic data (e.g., symptoms, medical facilities, hospitalization records, medical history, medical imaging, pharmaceutical treatments, gene and protein sequences); 
(5) infective demographics (e.g., age, gender, race, cultural background, and habit); 
(6) social activities and mobility; 
(7) domain knowledge and precautionary guidance from authorities on the virus and disease; 
(8) seasonal and environmental factors (e.g., season, geographical location, temperature, humidity, and wind speed); 
(9) news, reports and social media discussions on coronavirus and COVID-19; and
(10) fake news, rumor and misinformation. 
Such COVID-19 data are heterogeneously coded in character, text, number or image; in unordered, temporal/sequential or spatial modes; in static and dynamic forms; and with the characteristics as follows. 

Despite the large volume of existing research, modeling COVID-19 is still in an early stage with many open issues, partially because of the significant complexities of COVID-19 data. The main characteristics of COVID-19 data are summarized below and impose a computational burden on the modeling of COVID-19.

\textit{Acyclic and short-range case numbers which are small in size.} 
The publicly available data for COVID-19 modeling is limited. Except in rare scenarios such as in the US, most countries and regions report a short-range (2-3 months or even shorter such as local hotspot-based outbreaks), low-granularity (typically daily), and small-size (daily case numbers for a short period of time and a small cluster of the population) record of  COVID-19 data. Such data is typically acyclic without obvious seasonal or periodical patterns as in influenza~\cite{coletti2018shifting} and recurrent dengue epidemics in tropical countries~\cite{ulrich2020dengue}.

\textit{Inaccurate statistics on COVID-19 cases.} 
The reported new-infected-recovered-death case numbers are estimated much lower than their real number in most countries and regions. This may be due to many reasons, such as pre-symptomatic and asymptomatic infections, limited testing capability, nonstandard manual recording, different confirmation standards, an evolving understanding  of the disease nature, and other subjective factors. The method of calculating  case statistics may vary significantly from country to country; the actual figures in some countries and regions may even be unknown; and no clear differentiation is made between hotspot and country/region-based case reporting. The gaps between the infection reality and what has been documented may be more apparent in the first wave, in the early stage of outbreaks, and in some countries and regions~\cite{maier2020effective}. As result, the actual infected number and infected regions of COVID-19 pandemic may be much bigger than those publicly reported. 

\textit{Lack of reliable data particularly in an initial outbreak.} The spread of an epidemic in its initial phase can be regarded as transmission under perfect conditions. In its initial phase, the intrinsic epidemiological characteristics of COVID-19, such as reproduction rate, transmission rate, recovery rate and mortality are closer to their true values. For example, the modeling results in~\cite{roda2020difficult} show a wide range of variations due to the lack of reliable data, especially at the beginning of an outbreak.

\textit{Lack of high-quality microlevel data.} Data on COVID-19 cases, including daily infected cases, daily new cases, daily recovery cases, and daily death cases, is collected on a daily basis, while daily susceptible case numbers were also reported in Wuhan. However, macrolevel and low-dimensional data is far from comprehensive for inferring the complex transmission processes accurately and more fine-grained data with various aspects of features and high dimensions are needed. For example, during the initial phase of the Wuhan outbreak, the dissemination of SARS-CoV-2 was primarily determined by human mobility in Wuhan, however no empirical evidence on the effect of key geographic factors on local epidemic transmission was available~\cite{rader2020crowding}. The risk of COVID-19 death varies across various sociodemographic characteristics~\cite{drefahl2020population}, including age, sex, civil status, individual disposable income, region of residence, and country of birth. More specific data is required to address the sociodemographic inequalities related to contracting the COVID-19 virus. To contain the spread of COVID-19, governments propose and initiate a series of similar to different NPIs. No quantitative evidence or systematic evaluation analyzes how these measures affect  epidemic transmission, leading to challenges in inferring NPI-based COVID-19 transmission and mitigation.

\textit{Data incompleteness, inconsistencies, inequality and incomparability.}
Typically, it is difficult to find all-round information about a COVID-19 patient's infection source, demographics, behaviors, social activities (including mobility and in social media), clinical history, diagnoses and treatments, and resurgence if any. COVID-19 public data also presents strong inconsistencies and inequalities across reporting hotspots, countries and regions, updating frequencies and timelinesses, case confirmation standards, collection methods, and stages~\cite{roda2020difficult}. Data from different countries and areas may be unequal and incomparable due to their non-unified statistical criteria, confirmation standards, sampling and coverage methods, and health and medical conditions and protocols. These are also related to or affected by a person's race, living habits, and their applied mitigation policies, etc. 

\textit{Other issues.} 
Comparing public data available from different sources also reveals other issues like potential noise, bias and manipulation in some of the reported case numbers (e.g., due to their nonuniform statistic standards or manual statistical mistakes), missing values (e.g., unreported on weekends and in the early stage of outbreaks), different categorization of cases and stages (e.g., some with susceptible and asymptomatic case numbers), misinformation, and lack of information and knowledge about resurgence and mutation. 

\textbf{Discussion.} While increasing amounts of COVID-19 data are publicly available, they are in fact poor and limited in terms of quality, quantity, capability and capacity to discover deep insights about the nature of COVID-19, its interaction with external factors, and its effects. It is fundamental and urgent to acquire substantially larger and better-quality multisource COVID-19 data. This is crucial so that meaningful modeling can be robustly conducted and evaluated to reveal intrinsic knowledge and insights about the disease and to assist effective pandemic control.

\subsection{COVID-19 Modeling Complexities}
\label{subsec:modelcomp}

The COVID-19 pandemic is essentially an open complex system with significant system complexities \cite{aikp_Cao15,wang2021complex}. Examples are the hidden nature and strong uncertainty, self-organization, dynamics and evolution of the virus, disease and their developments and transmissions; their sophisticated interactions and relations to environments and context; the differentiated virus infections of individuals and communities; and the significant emergence of consequences and impacts on society in almost every part of the world. However, the publicly available small and limited COVID-19 data does not explicitly display a complete picture and a sufficient indication of the above complexities and intrinsic epidemiological attributes, transmission process, and cause-effect relations. It is thus challenging to undertake sound, robust, benchmarkable and generally useful modeling on such potential-limited data. 

\textit{Achieving ambitious modeling objectives on low-quality small COVID-19 data.}
As discussed in Section \ref{subsec:objectives}, many business problems and objectives are expected to be addressed by modeling COVID-19. However, the strong constraints in COVID-19 public data discussed in Section \ref{subsec:data} significantly limit this potential. Modelers have to carefully define learnable objectives, i.e., what can be learned from the data, acquire the essential and feasible data, or leverage data poverty by more powerful modeling approaches. For example, when a model is trained on a country's case numbers, its application to other countries may produce unfair results owing to their data inequalities. Another example is how to combine multisource but weakly connected data for meaningful high-potential analysis and results. 

\textit{Undertaking complex modeling with limited to no domain knowledge and ground truth.}
The weak to no-firm knowledge and ground truth about COVID-19 and its medical confirmation and annotations and poor-quality data limit the capacity and richness of the hypotheses to be tested and modeled on the data. It is not surprising that rather simple and classic analytical and learning models are predominantly applied by medical and biological scientists to verify specific hypotheses, e.g., various SIR models, time-series regression, and traditional machine learning methods~\cite{giordano2020modelling,chen2020time,shahid2020machine}, which occupy the top-80 keywords-based methods in the 200k WHO-collected references. In contrast, statisticians and computer scientists tend to enforce overparameterized models, over-complicated hypotheses, or over-manipulated data, resulting in highly specific results and over- or under-fitting issues.

\textit{Challenges in addressing the COVID-19 disease and data complexities.}
The unique characteristics and complexities of the COVID-19 disease and data discussed in Sections \ref{subsec:disease} and \ref{subsec:data} challenge the existing modeling methods including deep neural learning. Examples are generalized modeling of quality and quantity-limited COVID-19 data from different countries and regions and over evolving time periods~\cite{vespignani2020modelling}, robust modeling of short-range, small-size and incomplete-cycle data, and high-capacity modeling of mixture distributions with exponential growth~\cite{dehning2020inferring}, sub-exponential growth~\cite{maier2020effective}, discontinuous phase transition~\cite{vespignani2020modelling} and instant changes in case developments.

\textit{Disclosing complicated relations and interactions in weakly-coupled multisource data.}
COVID-19 is affiliated with many personal, social, health/medical, political and other factors, dispersedly reflected in explicitly or implicitly related multisource systems. The COVID-19 pandemic is formed and evolves as a dynamic social-technical process and the co-effects of multi-factor interplay. These multi-aspect factors are coupled strongly or weakly, locally or globally, explicitly or implicitly, subjectively or objectively, statically or dynamically, and essentially or accidentally in the virus and disease formation, development, influence, and evolution. Disclosing such sophisticated factor couplings and interactions is significantly challenging as they are not obvious or easily verifiable in observations. Therefore, modeling COVID-19 requires in-depth transdisciplinary cooperation between computer science, bioinformatics, virology, sociology and many other disciplines. A single factor alone cannot disclose the intrinsic and intricate nature of COVID-19 or explain the variability or shape the dynamics of this epidemic. 

\textbf{Discussion.}
The COVID-19 complexities result in significant modeling challenges, resulting from the data, the unclear epidemic transmission mechanisms and processes, and the entanglement between epidemic factors/observations and external objective (e.g. countermeasures) and subjective (e.g. people behavior changes) factors.
COVID-19 modeling goes beyond the transformation and applications of powerful models such as overparameterized deep neural networks, SIR variants and hierarchical Bayesian networks on the highly limited and poorly coupled small COVID-19 data. Careful designs are necessary to address the specific COVID-19 characteristics and complexities of its data and disease, avoid under-/over-fitting, and focus on modeling the complexities in relation to their underlying nature and insight. Complicating models does not necessarily contribute to better or more actionable knowledge and intelligence about the COVID-19 disease and data~\cite{widm_Cao13,booth2020development,tovstiga2020covid}.

\section{COVID-19 Modeling Landscape}
\label{sec:landscape}



To address the aforementioned COVID-19 disease, problems, data and modeling challenges, we present a high-level landscape to categorize and connect the comprehensive objectives and techniques for modeling COVID-19.

\subsection{Objectives of COVID-19 Modeling}
\label{subsec:objectives}

Here, we summarize the main business problems and objectives  in modeling COVID-19. The analysis of the WHO-collected literature \cite{who-analysis} gives us a clear indication of the top business terms in over 200k references and 22k modeling-focused references. The top-ranked keywords include  COVID-19 and coronavirus pandemic outbreak, spread, infection, transmission, factors, symptoms, characteristics, treatment, diagnosis, mortality, their risk and effects, as well as major data analysis and domain-specific research areas and methods. Below, we consolidate the main concerns and objectives of modeling COVID-19. Table \ref{tab:modelgoals} summarizes the associated modeling factors, modeling methods, and references.

\begin{table*}[!ht]
\centering
\caption{Business Problems and Objectives of COVID-19 Modeling.}
\resizebox{1.0\textwidth}{!}{
\begin{tabular}{|  p{0.15\textwidth} | p{0.47\textwidth} | p{0.25\textwidth} | p{0.12\textwidth} |}
    \hline
    \textbf{Objectives} & \textbf{Modeling factors} & \textbf{Approaches} & \textbf{References}  \\  \hline
    Epidemic dynamics and transmission & Epidemiological factors (e.g., origins, incubation period, transmission rate, morbidity, mortality, and highly to least vulnerable population, etc.), daily new-infected-recovered-death case numbers and reporting time, side information about population, etc. & Regression, compartmental models, time/age-dependent compartmental models, probabilistic compartmental models, etc. & \cite{giordano2020modelling,weitz2020modeling,aguiar2020modelling,chen2020time,singh2020age,zhou2020semiparametric,crokidakis2020modeling,peng2020epidemic}  \\ \hline
    Non-pharmaceutical intervention and policies & Intervention measures, quarantine and mitigation measures and policies on communities and individuals, epidemiological factors, daily new-infected-recovered-death case numbers and reporting time, social activities, communications & Regression, customized compartmental models, Bayesian hierarchical models, stochastic compartmental models, etc. & \cite{brauner2020inferring,prem2020effect,aguiar2020modelling,giordano2020modelling,tian2020investigation,maier2020effective,Flaxmans20,dehning2020inferring,lai2020effect} \\ \hline
    Diagnosis, identification and tracing & Clinical, pathological and genomic data (symptoms, medical facilities, hospitalization records, medical history, respiratory signals, medical imaging, physical and chemical measures, gene sequences, proteins), mobility and contacts, etc.  & Regression, statistical learning, shallow (e.g. decision trees, random forest) and deep models, image and signal processing methods, etc. & \cite{wang2020covid,wu2020rapid,zhang2020clinically,chikina2020modeling,basu2020deep,cohen2020predicting,minaee2020deep,das2020truncated,khalifa2020detection,apostolopoulos2020covid,mukherjee2020shallow,mao2020data}  \\ \hline
    Treatment and pharmaceutical interventions  & Clinical measures, hospitalization records, medical test records, drug selection, pharmaceutical treatments, ICU records, ventilator use, healthcare records, etc. & Classifiers, time-series methods, DNNs, etc. & \cite{zheng2020recommendations,wu2020estimating,yan2020prediction,beck2020predicting,litjens2017survey} \\ \hline
    Pathological and biomedical analysis  & Genomic data, protein structures, pathogenic data, drug and vaccine info, immunization response, etc. & Classifiers, outlier detectors, genome analysis, protein analysis, DNNs, etc. &  \cite{Arshadik20,Alakust21,beck2020predicting,Magarr21,senior2020improved,zhavoronkov2020potential,randhawa2020machine,ou2020characterization,walls2020structure,xia2020inhibition} \\ \hline
    Resurgence and mutation & Daily new-infected-recovered-death case numbers and reporting time, quarantine and mitigation measures and policies on communities and individuals, social activities and mobility, seasonal and environmental factors (e.g., season, geographical location, temperature, humidity, and wind speed)  & Compartmental models, simulation models, compartmental models combined with regression, epidemic renormalisation group, etc. & \cite{leung2020first,aravindakshan2020preparing,cacciapaglia2020second,lopez2020end,pedro2020conditions} \\ \hline
    Influence and impact & Quarantine and mitigation measures and policies on communities and individuals, domain knowledge and precaution guidance from authorities, social activities and mobility, demographics (e.g., age, gender, racist, cultural background, habit), related news, reports, social media discussions, and misinformation. & Statistical analysis, questionnaire methods, age-structured SIR/SEIR models, deep neural networks (e.g., BERT and LSTM), simulation models, etc. & \cite{chakraborty2020covid,kreps2020model,li2020analyzing,sharma2020covid,walker2020impact}  \\ \hline
    \end{tabular}
    }
\label{tab:modelgoals}
\end{table*}

\textit{Characterizing and predicting the COVID-19 epidemic dynamics and transmission.} 
An imperative challenge is to understand the COVID-19 epidemic mechanisms, transmission process and dynamics, infer its epidemiological attributes, and understand how the virus spreads spatially and socially~\cite{adiga2020data}. The majority of COVID-19 modeling tasks focus on exploring the source and spectrum of the COVID-19 infection, clinical and epidemiological characteristics, tracking its transmission routes, and forecasting case development trends and the peak number of infected cases and disease transmission~\cite{chan2020familial,wu2020estimating,wu2020nowcasting}. They aim for findings to understand the nature of the virus and disease and inform disease precaution, virus containment, mitigation campaigns, and medical resource planning, etc.

\textit{Modeling the resurgence and mutation.} 
As the SARS-CoV-2 mutation and COVID-19 resurgence are highly uncertain and much more transmissible and infectious, we highlight their relevant research here. However, as our current understanding of the resurgence and mutation is very limited, COVID-19 may  become another epidemic disease which stays with humans for a long time. The WHO-identified four variants of concern which have higher transmissibility, contagion and complexities \cite{grubaugh2020making,grubaugh2021public,WHO-variants,US-CDC}. Imperative research is expected to quantify the resurgence conditions, control potential resurgences after lifting certain restrictions and reactivating businesses and activities~\cite{lopez2020end,pedro2020conditions}, distinguish the characteristics and containment measures between waves~\cite{fan2020decreased,grech2020covid,aleta2020age}, and prepare for and predict resurgence, mutation and their responsive countermeasures~\cite{aravindakshan2020preparing}. 

\textit{Disease diagnosis, infection identification and contact tracing.} 
Given the strong transmission and reproduction rates, high contagion, and sophisticated transmission routes and the unexpected resurgence of COVID-19 and its virus mutation, it is crucial to immediately identify and confirm exposed cases and trace their origins and contacts to proactively implement quarantine measures and contain their potential spread and outbreak \cite{ng2020evaluation}. This is particularly important during the varying incubation periods (usually 2 - 14 days) which are asymptomatic to mildly symptomatic yet highly contagious. In addition to chemical and clinical approaches, identifying COVID-19 by analyzing biomedical images, genomic sequences, symptoms, social activities, mobility and media communications is also essential \cite{udugama2020diagnosing}. 

\textit{Modeling the efficacy of medical treatment and pharmaceutical interventions.} 
The general practices of timely and proper COVID-19 medical treatments, drug selection and pharmaceutical measures, and ICU and ventilation etc. play fundamental roles in fast recovery, mitigating severe symptoms and reducing the mortality rate of both the original and increasingly-mutated virus strains. However, the lack of best practices and standardized protocols and specifications of medical and pharmaceutical treatments on the respective virus variants in terms of patient's demographic and ethnic context and the wide dispersal of online misinformation of drug use may also contribute to  global imbalance in containing COVID-19. Research is required to select and discover suitable drugs, which best match the patient's diagnosis and ethnic contexts with suitable medical treatments to mitigate critical conditions and mortality in a timely manner, etc. \cite{zheng2020recommendations,wu2020estimating,yan2020prediction,beck2020predicting}.

\textit{Modeling the efficacy of non-pharmaceutical intervention and policies.} 
Various NPIs, such as travel bans, border control, business and school shutdowns, public and private gathering restrictions, mask-wearing, and social distancing are often implemented to control the outbreak of COVID-19. Different governments tend to enforce them in varied combinations and levels and ease them within different timeframes and following different procedures, resulting in different outcomes. Limited research results have been reported to verify the effects of these measures and their combinations on containing the virus spread and case number development, the balance between  enforcement levels and containment results, and the response sensitivity of the restrictions in relation to the population's ethnic context~\cite{tian2020investigation,dehning2020inferring,brauner2020inferring}. Limited results are available on the threshold and effects of COVID-19 vaccinations and herd immunity. More robust results will inform medical and public health policy-making on medication, business and society.

\textit{Understanding pathology and biomedical attributes for drug and vaccine development.} 
By involving  domain knowledge and techniques such as virology, pathogenesis, genomics and proteomics, pathological and biomedical analyses can be conducted on pathological test results, gene sequences, protein sequences, physical and chemical properties of SARS-CoV-2, drug and vaccine information and their effects. Accordingly, it is necessary to conduct domain-driven analysis and model correlated drugs and vaccines with genomic and protein structures to select and develop COVID-19 drug and vaccines, to understand the drug-target interactions, and to diagnose and identify infection, understand virus mutation, etc. \cite{beck2020predicting,randhawa2020machine,hu2020prediction}. More research is needed on COVID-19 immunity responses, drug and vaccine development, and mutation intervention.

\textit{Modeling COVID-19 influence and impact.}
While the COVID-19 pandemic has changed the world and has had a significant and overwhelming influence on almost all aspects of  life, society and the economy, quantifying its influence and impact has rarely been studied. COVID-19 negative impact modeling may include (1) economic impact on growth and restructuring~\cite{world2020global}; (2) social impact on people's stress, psychology, emotions, behavior and mobility~\cite{pedrosa2020emotional,xiong2020impact}; and (3) transforming business processes and organizations, manufacturing, transport, logistics, and globalization~\cite{vo7impact,seetharaman2020business}. In contrast, it would also be interesting to model its `opportunity' and influence on (1) enhancing the wellbeing and resilience of individuals, families, society and work-life balance~\cite{prime2020risk}; (2) digitizing and transforming work, study, entertainment and shopping~\cite{soto2020covid}; (3) restructuring supply-demand relations and supply chains for better immediate availability and to satisfy demand~\cite{del2020supply}; (4) promoting research and innovation on intervening in global black-swan disasters like COVID-19 and its impacts~\cite{zhang2020topic}; and (5) enhancing trust and development in science, medicine, vaccination and hygiene~\cite{plohl2021modeling}. Other impact modeling tasks include analyzing the relations between the COVID-19 containment effect and  socioeconomic level (e.g., income level particularly in relation to lower-income and disadvantaged groups), healthcare capacity and quality, government crisis management capabilities, citizen-government-cooperation, and public health and hygiene habits.

\subsection{Categorization of COVID-19 Modeling}
\label{subsec:roadmap}

The flow of COVID-19 modeling has strong features such as: (1) multi-disciplinary techniques of mathematics and statistics, epidemiology, broad AI and data science including shallow and deep learning, and social science; (2) epidemiological methods to explore business problems and research areas; (3) domain, model and data-driven approaches consisting of various families of domain knowledge, models and methods which are widely applied in all business problems; (4) case studies and hypothesis tests to highlight the results of particular methods or modeling specific settings, scenarios or data. 

\begin{figure*}[!ht]
	\centering
	\includegraphics[width=0.99\linewidth]{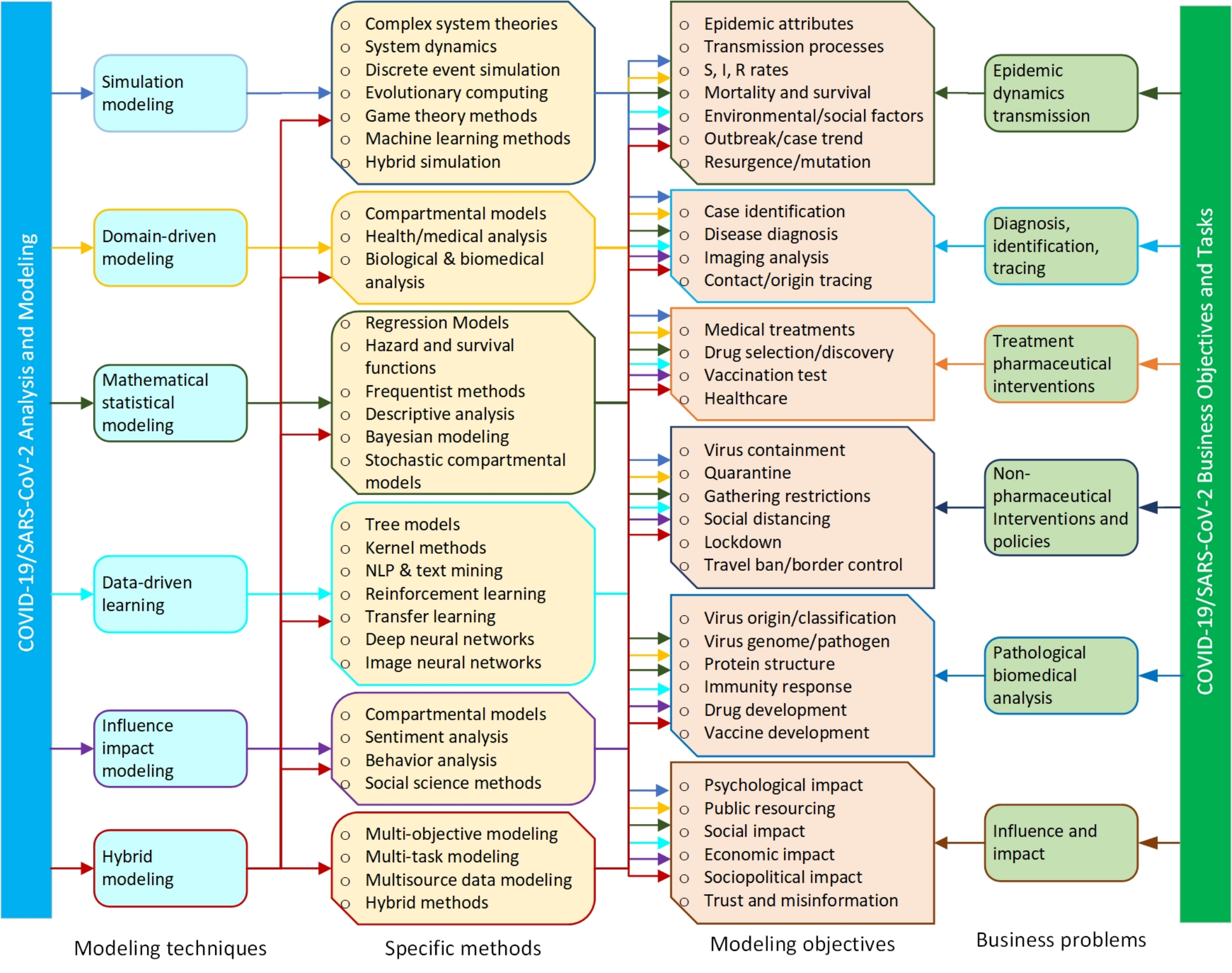}
	\caption{The transdisciplinary research landscape of COVID-19 modeling: From right: business problems and modeling objectives to the left: modeling techniques and methods.}
	\label{fig:covid19-modeling}
\end{figure*}

The keyword-based analysis of the 22k WHO-collected modeling references in \cite{who-analysis} shows that epidemiological modeling, mathematical and statistical modeling, artificial intelligence and data science, and simulation modeling play predominant roles in understanding, characterizing, simulating, analyzing and predicting COVID-19 issues. We thus categorize the research landscape of COVID-19 modeling into six: domain-driven modeling, mathematical/statistical modeling, data-driven learning, influence/impact modeling, simulation modeling, and hybrid methods in this review. Fig. \ref{fig:covid19-modeling} summarizes the transdisciplinary research landscape connecting the aforementioned six categories of modeling techniques and their respective modeling methods to major COVID-19 business problems and their modeling objectives in Section \ref{subsec:objectives}. 

\begin{itemize}
    \item \textit{COVID-19 mathematical/statistical modeling}: developing and applying mathematical and statistical models such as time-series analysis (e.g., regression models and hazard and survival functions) and statistical models (e.g., descriptive analytics, statistical processes, latent factor models, temporal hierarchical Bayesian models, and stochastic compartmental models) to estimate COVID-19 transmission processes, symptom identification, disease diagnosis and treatment, sentiment analysis, misinformation analysis, and resurgence and mutation.
    \item \textit{COVID-19 data-driven learning}: developing and applying data-driven classic (e.g., tree models such as random forests and decision trees, kernel methods such as support vector machines (SVMs), NLP and text analysis, and classic reinforcement learning) and deep (e.g., deep neural networks, transfer learning, deep reinforcement learning, and variational deep neural models) analytics and learning methods on COVID-19 data to characterize, represent, classify, and predict COVID-19 problems, such as case development, mortality and survival forecasting, medical imaging analysis, NPI effect estimation, and genomic analysis.  
    \item \textit{COVID-19 domain-driven modeling}: developing and applying domain knowledge and domain-specific models for COVID-19; examples are epidemiological compartmental models to characterize the COVID-19 epidemic transmission processes, dynamics, transmission and risk, and the influence of external factors on COVID-19 epidemics, resurgence and mutation; and medical, pathological and biomedical analysis for infection diagnosis, case identification, patient risk and prognosis analysis, medical imaging-based diagnosis, pathological and treatment analysis, and drug development. 
    \item \textit{COVID-19 influence/impact modeling}: developing and applying methods to estimate and forecast the influence and impact of SARS-COV-2 variations and COVID-19 diseases and their interventions, treatments and vaccination on epidemic transmission dynamics, virus containment, disease treatment, public resources including healthcare systems, social systems, economy, and human psychological health and behaviors. 
    \item \textit{COVID-19 simulation modeling}: developing and applying simulation models such as theories of complex systems, agent-based simulation, discrete event analysis, evolutionary learning, game theories, and Monte-Carlo simulation to simulate the COVID-19 epidemic evolution and the effect of interventions and policies on the COVID-19 epidemic.
    \item \textit{COVID-19 hybrid modeling}: hybridizing and ensembling multiple models to tackle multiple business problems and objectives, multiple tasks, and multisource data and those individual objectives, tasks and data sources that cannot be better understood by single approaches. 
\end{itemize}

It is worth mentioning that each of the above modeling techniques and their specific methods may be applicable to address different business problems and modeling objectives, as shown in Fig. \ref{fig:covid19-modeling}. Below, we review the progress of the above six categories of COVID-19 modeling by (1) summarizing the typical modeling techniques and (2) categorizing their typical applications in modeling diverse COVID-19 issues.

\section{COVID-19 Mathematical Modeling}
\label{sec:mathmethods}

Mathematical and statistical models are overwhelmingly used to estimate and predict the transmission dynamics and reveal the truth of epidemic in a formalized and quantitative way. Accurate COVID-19 mathematical models are indispensable for the COVID-19 epidemic forecasting and decision making, amounting to 13k of 22k references on modeling COVID-19. Here, we review two sets of main mathematical methods: time-series analysis and statistical modeling, and their applications in COVID-19 modeling.

\subsection{Time-series Analysis}
\label{subsec:timeseries}

\subsubsection{Time-series modeling} We here focus on two typical methods that are predominantly customized for modeling COVID-19: regression models and hazard and survival functions.

\textbf{Regression models.} Typical regression models such as logistic regression and auto-regressive integrated moving average (ARIMA) variants are widely used in epidemic and COVID-19 modeling. \textit{Logistic growth models} can model the number of COVID-19 infected cases. For a population of $S$ with the infection rate $\beta$, the growth scale of infection number $I$ can be modeled by 
\begin{equation}
    \partial_t I = \beta I (1 - I/S)
\end{equation}
over time \textit{t}. Accordingly, with historical COVID-19 cases of a place at a time period, an S-shaped curve can be derived to describe and forecast the growth distributions of the COVID-19 infections and the peak infected number by adjusting the constant rate $\beta$. Logistic models are weak to incapable of modeling other states of COVID-19 cases and there are many challenges, such as nonstationary characteristics discussed in Section \ref{subsec:data}, while some may be better modeled by more sophisticated regression models.

ARIMA and its variants also model the temporal movement of COVID-19 case numbers with more flexibility than logistic ones. For example, the number $I$ of infected cases can be modeled by $ARIMA(p,d,q)$ which factorizes the number into consecutive past numbers with errors:
\begin{equation}
    I_t = \alpha_1 I_{t-1} + \alpha_2 I_{t-2} + \dots + \alpha_p I_{t-p} + \varepsilon_t + \theta_{t-1} \varepsilon_{t-1} + \dots + \theta_{t-q} \varepsilon_{t-q} + c
\end{equation}
over the number of time lags (order) of autoregression $p$, the order of moving average $q$, the degree of differencing $d$, and time $t$ with a constant $c$. $I_{t-p}$ refers to the infected cases at time $t-p$ with weight $\alpha_p$, $\varepsilon_{t-q}$ refers to the error between $I_{t-q}$ and $I_{t-q-1}$ with weight $\theta_{t-q}$. Adjusting parameters like $p,d$ and $q$ can simulate/capture some of the time series characteristics (e.g., the process and trend of the infection series by $p$ and $q$, seasonality by $d$, and volatile movement by the distribution of error terms). Similarly, ARIMA models can be used to simulate and forecast the number of recoveries and deaths in relation to COVID-19. 

In addition, ARIMA and its variants can be integrated with other modeling methods to characterize other aspects of COVID-19 time series. For example, the wavelet decomposition of frequency-based nonstationary factors can model the oscillatory error terms of ARIMA-based modeling of COVID-19 infected cases  \cite{chakraborty2020covid}). Another example is to combine the decision tree method with regression to form a regression tree and identify mortality-sensitive COVID-19 factors \cite{chakraborty2020real}.

\textbf{Hazard and survival functions}. Hazard functions and survival functions are often used to model the mortality and survival (recovery) rates of patients using  time-to-event analysis. A hazard function models the mortality probability $h(t|\mathbf{x})$ of a COVID-19 patient with the factor vector $\mathbf{x}$ ($\in \mathcal{R}^d$) of dying at discrete time \textit{t}: 
\begin{equation}
    h(t|\mathbf{x}) = p(T=t|T \geq t; \mathbf{x})
\end{equation}
On the contrary, the survival function models the probability $S(t|\mathbf{x})$ of surviving until time \textit{t}:
\begin{equation}
    S(t|\mathbf{x}) = P(T>t| \mathbf{x}) 
\end{equation}
In discrete time, $S(t) = \prod^{t}_{t_i=1}(1-h_{t_i})$ where $h_{t_i}$ is the mortality probability at time $t_i$. In continuous time, $S(t|\mathbf{x}) = 1 - F(t|\mathbf{x})$ where $F(\cdot)$ is the cumulative distribution function until time \textit{t}. For covariates $\mathbf{x}$ with their relations represented by an either linear or nonlinear function $f(\cdot;\mathbf{\delta})$ with parameters $\mathbf{\delta}$, the mortality rate of COVID-19 can be modeled by a Cox proportional hazard model \cite{Cox1972}:
\begin{equation}
    h(t|\mathbf{x}) = h_0(t)exp(f(\cdot;\mathbf{\delta}))
\end{equation}
where $h_0(t)$ is the baseline hazard function, the function $f(\cdot;\mathbf{\delta})$ can be implemented by a linear function such as a linear transform or a nonlinear function such as a deep convolutional network. For example, in \cite{schwab2021real}, $f(\cdot;\mathbf{\delta})$ is implemented by a shallow neural network with a leaky rectified linear unit-based activation of the input and then another tangent transformation. In the case of time-varying covariates $\mathbf{x}_t$, the above hazard, survival and transform functions should be time sensitive as well. 

Typically, the performance of mathematical modeling is measured by metrics such as mean absolute error (MAE), root mean square error (RMSE), the improvement percentage index (IP), and symmetric mean absolute percentage error (sMAPE) in terms of certain levels of confidence intervals.

\subsubsection{COVID-19 time-series modeling}

Time-series analysis contributes the most (about 3k of the 22k WHO-listed references) to COVID-19 modeling. As shown in \cite{who-analysis}, regression models, linear regression, and logistic regression are mostly applied in COVID-19 modeling. Many linear and nonlinear, univariate, bivariate and multivariate analysis methods have been intensively applied for the regression and trend forecasting of new, susceptible, infectious, recovered and death case numbers. Popular methods include linear regression models such as ARIMA and GARCH ~\cite{singh2020prediction,tandon2020coronavirus}, logistic growth regression~\cite{wang2020prediction}, COX regression~\cite{schwab2021real}, multivariate and polynomial regression~\cite{ahmad2020number,chen2020roles,guptaa2020seir}, generalized linear model and visual analysis~\cite{muthusami2020statistical}, support vector regression (SVR)~\cite{gupta2020prediction}~\cite{ribeiro2020short}, regression trees~\cite{chakraborty2020real}, hazard and survival functions \cite{schwab2021real}, and more modern LSTM networks. In addition, temporal interpolation methods such as best fit cubic, exponential decay and Lagrange interpolation, spatial interpolation methods such as inverse distance weighting, smoothing methods such as moving average, and spatio-temporal interpolation \cite{CaiR20} are applicable to fit and forecast COVID-19 case time series.
We illustrate a few tasks below: COVID-19 epidemic distributions, case number and trend forecasting, and COVID-19 factor and risk analysis. 

\textit{COVID-19 case number and trend forecasting and epidemic distributions}. Regression-centered time series analysis has been widely applied to forecast case number developments and trends. For COVID-19 prediction, Singh et al.~\shortcite{singh2020prediction} apply ARIMA to predict the COVID-19 spread trajectories for the top 15 countries with confirmed cases and conclude that ARIMA with a weight to adjust the past case numbers and the errors has the ability to correct model prediction and is better than regression and exponential models for prediction. However, ARIMA lacks flexible support for volatility and in-between changes during the prediction periods~\cite{singh2020prediction}. Gupal et al.~\shortcite{guptaa2020seir} adopt polynomial regression to predict the number of confirmed cases in India. Almeshal et al.~\shortcite{almeshal2020forecasting} utilize logistic growth regression to fit the actual infected cases and the growth of infections per day. Wang et al.~\shortcite{wang2020prediction} model the cap value of the epidemic trend of  COVID-19 case data using a logistic model. With the cap value, they derive the epidemic curve by adapting time series prediction. To find the best regression model for case forecasting, Ribeiro et al.~\shortcite{ribeiro2020short} explore and compare the predictive capacity of the most widely-used regression models including ARIMA, cubist regression (CUBIST), random forest, ridge regression, SVR, and stacking-ensemble learning models. They conclude that SVR and stacking ensemble are the most suitable for the short-term COVID-19 case forecasting in Brazil. In addition, linear regression with Shannon diversity index and Lloyd’s index are applied to analyze the relations between the meta-population crowdedness in city and rural areas and the epidemic length and attack rate \cite{rader2020crowding}.

\textit{COVID-19-specific factor and risk analysis}. Time-series analysis may be used to (1) analyze the influence of specific and contextual factors on COVID-19 infections and COVID-19 epidemic developments including infection, transmission, outbreak, hospitalization, and recovery, e.g., on  COVID-19 survival, mortality and recovery; and (2) analyze the influence and impact of external and contextual factors of COVID-19 outbreak on the population, health, society and the economy, case developments and containment. For example, to investigate the potential risk factors associated with fatal outcomes from COVID-19, Schwab et al.~\shortcite{schwab2021real} present an early warning system assessing COVID-19 related mortality risk with a variation of the Cox proportional hazard regression model. Chen et al.~\shortcite{chen2020risk} adapt the Cox regression model to analyze the clinical features and laboratory findings of hospitalized patients. Charkraborty et al.~\shortcite{chakraborty2020real} design the wavelet transform optimal regression tree (RT) model, which combines various factors including case estimates, epidemiological characteristics and healthcare facilities to assess the risk of COVID-19. The advantage of RT is that it has a built-in variable selection mechanism from high dimensional variable space and can model arbitrary decision boundaries. 

\textit{Correlation analysis between COVID-19 epidemic dynamics and external factors}. Much research has been conducted on analyzing the relationships between COVID-19 transmission and dynamics and external and contextual factors. For example, Cox proportional hazard regression models are used to analyze high risk sociodemographic factors such as gender, individual income, education level and marital status that may be associated with a patient's death \cite{drefahl2020population}, and logistic regression models are applied to analyze the relations between COVID-19 (or SARI with unknown aetiology) and socioeconomic status (per-capita income) \cite{de2020epidemiological}. To reveal the impact of meteorological factors, Chen et al.~\shortcite{chen2020roles} examine the relationships between meteorological variables (i.e., temperature, humidity, wind speed and visibility) and the severity of the outbreak indicated by the confirmed case numbers using the polynomial regression method; while Liu et al.~\shortcite{liu2020impact} fit the generalized linear models (GLM) with negative binomial distribution to estimate the city-specific effects of meteorological factors on confirmed case counts. In \cite{Poirier-cov20}, Loess regression does not show an obvious relation between the COVID-19 reproduction number, weather factors (humidity and temperature) and human mobility. Lastly, linear models including linear regression, Lasso regression, ridge regression, elastic net, least angle regression, Lasso least angle regression, orthogonal matching pursuit, Bayesian ridge, automatic relevance determination, passive aggressive regressor, random sample consensus, TheilSen regressor and Huber regressor are applied to analyze the potential influence of weather conditions on the spread of coronavirus \cite{Malkiz20}.

\textbf{Discussion}.
Time-series methods excel at characterizing sequential transmission processes and temporal case movements and trends. They lack the capability to involve other multisource factors and disclose deep insights into why case numbers evolve in a certain way and how to intervene in the infection, treatment and recovery. 

\subsection{Statistical Modeling}
\label{subsec:statmodel}

Statistical learning, in particular Bayesian models, play a critical role in stochastic epidemic and infectious disease modeling \cite{Brauer-me19}. It takes generative stochastic processes to model epidemic contagion in epidemic modeling~\cite{andersson2012stochastic,o1999bayesian}. In contrast to compartmental models, statistical models involve prior knowledge about an epidemic disease and their results have confidence levels corresponding to distinct assumptions (i.e., possible mitigation strategies), which better interpret and more flexibly model COVID-19 complexities. Below, we summarize typical statistical models and their applications in COVID-19 statistical modeling.

\subsubsection{Statistical models}
\label{subsubsec:statmodels}
Statistical models are widely applied to COVID-19 modeling tasks including (1) simulating and validating the state distributions and transitions of COVID-19 infected individuals over time, (2) modeling latent and random factors affiliated with the COVID-19 epidemic processes, movements and interactions, (3) forecasting short-to-long-term transmission dynamics, (4) evaluating the effect of non-pharmaceutical interventions (NPI), and (5) estimating the impact of COVID-19 such as on socioeconomic aspects. Typical methods include descriptive analytics, Bayesian hierarchical models, probabilistic compartmental models, and probabilistic deep learning. Below, we introduce some common statistical settings and corresponding statistical models in both frequentist and Bayesian families that are often applied in  COVID-19 statistical modeling.

Taking a stochastic (vs. deterministic) state transition assumption, various statistical processes can be assumed to simulate and estimate the state-specific counts (case numbers) and the probability of state transitions (e.g., between infections and deaths) during the COVID-19 spread. The stochastic processes and states (e.g., its infection and mortality) of a COVID-19 outbreak are influenced by various explicit and latent factors. Examples of explicit (observable) factors include a person's demographics (e.g., age and race), health conditions (e.g., disease history and hygienic conditions), social activities (e.g., working environment and social contacts), and the containment actions (e.g., quarantined or not) taken by the person. Latent factors may include the person's psychological attitude toward cooperation (or conflict) with containment, health resilience strength to coronavirus and the containment influence on the outcome (e.g., infected or deceased). 

\begin{figure*}[ht]
	\centering
	\subfigure[A general graphical model of statistical COVID-19 modeling]{
    \begin{minipage}[t]{0.35\linewidth}
    \centering
    \includegraphics[width=\linewidth]{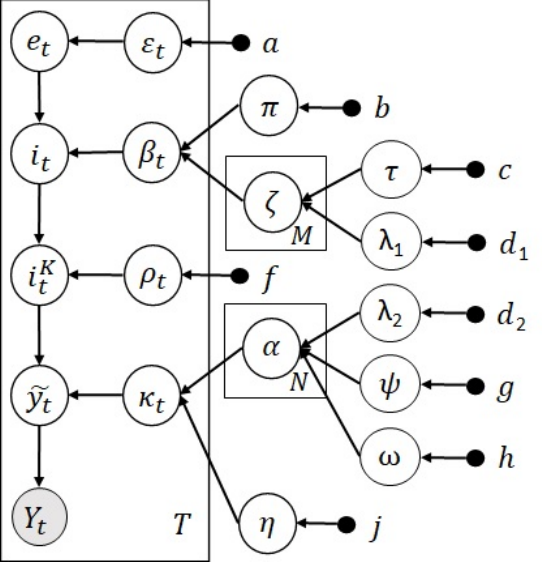}
    \end{minipage}%
    }%
	\subfigure[The graphic model for the method proposed in \cite{Flaxmans20}]{
    \begin{minipage}[t]{0.5\linewidth}
    \centering
    \includegraphics[width=\linewidth]{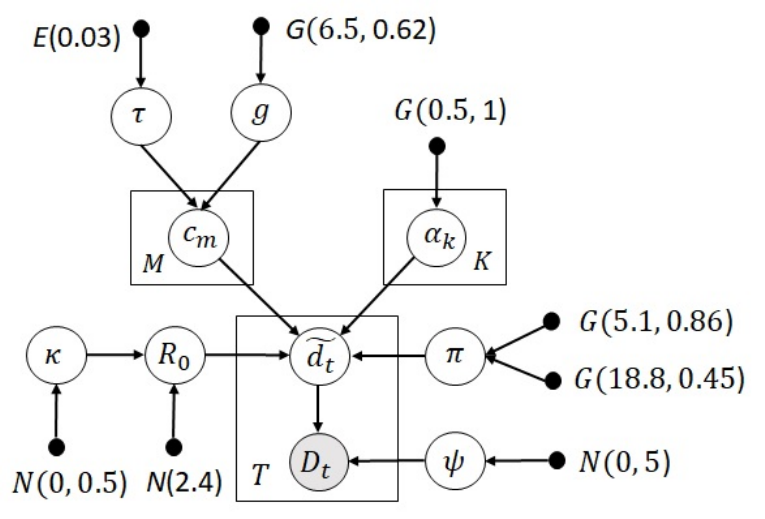}
    \end{minipage}%
    }%
	\caption{A general and a specific hierarchical Bayesian model for COVID-19.}
	\label{fig:statmodel}
\end{figure*}

Fig. \ref{fig:statmodel}(a) illustrates a general graphical model of the temporal hierarchical Bayesian modeling of COVID-19 case numbers for estimation and forecasting. The reported case number $Y_t$ (e.g., death toll or infected cases) at time $t$ can be estimated by $\tilde{y}_t$, which is inferred from the documented (declared) infections $i^K_t$ and removed (e.g., recovered and deceased) rate $\kappa_t$. The documented infection number $i^K_t$ is inferred from the infected population $i_t$ and the test rate $\rho_t$. $i_t$ is inferred from the exposed population $e_t$ and the infection rate $\beta_t$, $e_t$ is determined by its exposed rate $\epsilon_t$. Further, we assume the removed rate $\kappa_t$ is influenced by various medical treatments $\alpha$, determined by auxiliary variables including socioeconomic condition $\lambda_2$, the treatment effectiveness $\psi$, and the public health quality $\omega$. The infection rate $\beta_t$ is determined by NPIs $\zeta$, which are further influenced by the NPI execution rate $\tau$ and the socioeconomic factor $\lambda_1$. The priors of the corresponding parameters are $a$, $b$, $c$, $d_1$, $d_2$, $f$, $g$, $h$ and $j$, which may follow specific assumptions.

For the statistical settings and hypotheses, typical statistical distributions of COVID-19 state-specific counts are applied to (1) \textit{infection modeling}, e.g., by assuming a Bernoulli process ($B(n, \varepsilon)$ with the probability $\varepsilon$ of exposure to infections over $n$ contacts) and then a Poisson process at points of infections with exponentially-distributed infectious periods ($Pois(\beta)$ with the rate $\beta$ referring to the infection rate within the infectious period); (2) \textit{mortality modeling}, e.g., by assuming a negative binomial distribution ($NB(\mu, \sigma)$) or a Poisson distribution ($Pois(\gamma))$ with the rate $\gamma$ parameterized on the mortality rate $\gamma_D$ and population $N$). Further, the basic reproduction number $R_0$ may be estimated by $N \varepsilon \beta/(\beta + \gamma)$, the infections will be under control if $R_0$ is less than a given threshold (e.g., 1). 

\begin{subequations}\label{eq:statmodel}
    \begin{alignat}{2}
        &\text{The standard deviation:} 
            &&\kappa \sim \text{Normal} (0, 0.5) \\
        &\text{The initial reproduction number:} 
            &&R_{0,m} \sim \text{Normal} (2.4, |\kappa|) \\
        &\text{The intervention impact:} 
            &&\alpha_k \sim \text{Gamma} (0.5, 1) \\
        &\text{The time-varying reproduction number:} 
            &&R_{t,m} = R_{0,m} \text{exp} (-\sum_{k=1}^{6} \alpha_k I_{k,t,m}) \\
        &\text{The distribution rate:} 
            &&\tau \sim \text{Exponential} (0.03) \\
        &\text{The 6 sequential days of infections:} 
            &&c_{1,m},...,c_{6,m} \sim \text{Exponential} (\tau) \\
        &\text{The daily serial interval:} 
            &&g \sim \text{Gamma} (6.5, 0.62) \\
        &\text{The number of infections:} 
            &&c_{t,m} = R_{t,m} \sum_{\tau=0}^{t-1} c_{\tau,m} g_{t-\tau} \\
        &\text{The time from infection to death:} 
            &&\pi_m \sim \text{ifr}_m (\text{Gamma} (5.1, 0.86) + \text{Gamma} (18.8, 0.45)) \\
        &\text{The variance latent variable:} 
            &&\psi \sim \text{Normal}^{+} (0, 5) \\
        &\text{The expected number of deaths:} 
            &&d_t = \sum_{\tau=0}^{t-1} c_{\tau,m} \pi_{t-\tau,m} \\
        &\text{The observed daily deaths:} 
            &&D_t \sim \text{Negative Binomial} (d_{t,m}, d_{t,m} + \frac{d_{t,m}^2}{\psi})
    \end{alignat}
\end{subequations}

The above hierarchical statistical model in Fig. \ref{fig:statmodel}(a) can be customized to estimate and forecast COVID-19 case numbers in terms of specific hypotheses, settings and conditions. For example, Fig. \ref{fig:statmodel}(b) shows the graphic model for the hierarchical model proposed in \cite{Flaxmans20} to estimate the death number $D$ from its inferred variable $\tilde{d}_t$ and inferred from auxiliary variable $\psi$. The inferred death number $\tilde{d}_t$ is sampled from the basic reproduction number $R_{0}$ with a normally-distributed prior $Normal(2.4, |\kappa|)$ parameterized by its variance variable $\kappa$ and the probability of infected death $\pi$ determined by two Gamma priors. In addition, $\tilde{d}_t$ is also influenced by the number of new infections $c$ with two latent variables, the distribution rate of the Exponential distribution $\tau$ and the daily serial interval $g$ and a variable $\alpha$ as a parameter of the reproduction rate. Fig. \ref{fig:statmodel}(b) also shows the prior distributions of the auxiliary variables, for example, assuming the variable describing the time from infection to death $\tau$ following an exponential prior $Exponential(0.03)$. The hierarchical statistical model in Fig. \ref{fig:statmodel}(b) to estimate the death number can be described by the following equations.

Another major set of COVID-19 statistical modeling incorporates statistical hypotheses and settings into other epidemic models such as compartmental models to approximate some state distributions or estimate some parameters. A typical application reformulates SIR-based models as a system of stochastic differential equations, e.g., by assuming Gamma-distributed probability density of the exposed $E$ and infected $I$ states in Section \ref{subsec:compartment}. Lastly, modeling the influence of mitigation strategies on the COVID-19 case numbers is also a typical statistical modeling problem.

\subsubsection{COVID-19 Statistical Modeling}
\label{subsubsec:covidstatmodel}
The contagion of an epidemic like COVID-19 is complex and uncertain. Statistical or probabilistic modeling naturally captures this uncertainty around epidemics better than other models. In COVID-19, for example, hierarchical Bayesian distributions with hidden states and parameters are used to model the causal relationships in their transmission~\cite{niehus2020using,Flaxmans20}, and probabilistic compartmental models~\cite{zhou2020semiparametric,osthus2019dynamic,hebert2020macroscopic} integrate the transmission mechanisms of epidemics with the statistics of observed case data. Below, we summarize the relevant applications of descriptive analytics, Bayesian statistical modeling and stochastic compartmental modeling of the COVID-19 epidemic statistics, epidemic processes, and the influence of external factors such as NPIs on the epidemic. Table \ref{tab:statlearning} further summarizes various applications of COVID-19 statistical modeling.

\textit{COVID-19 descriptive analytics.} 
Descriptive analytics are the starting point of COVID-19 statistical analysis, which are typically seen in non-modeling-focused references and communities. Typically, simple statistics such as the mean, deviation, trend and change of COVID-19 case numbers are calculated and compared. For example, the statistics of asymptomatic infectives are reported in \cite{kronbichler2020asymptomatic}. In \cite{Bherwanih20}, change point analysis detects a change in the exponential rise of infected cases and Pearson’s correlation between the change and lockdown implemented across risky zones. In addition, case statistics may be calculated in terms of specific scenarios, e.g., a population's mobility \cite{HuangWFZSL20} or workplace \cite{Baolf21}.

\textit{Bayesian statistical modeling of COVID-19 epidemic processes.}
Bayesian statistical modeling can model stochastic COVID-19 epidemic processes, specific factors that may influence the COVID-19 epidemic process, causality, partially-observed data (e.g., under-reported infections or deaths), and other uncertainties. For example, stochastic processes are adopted to model conventional epidemic contagion~\cite{andersson2012stochastic,o1999bayesian}. Niehus et al.~\shortcite{niehus2020using} use a Bayesian statistical model to estimate the relative capacity of detecting imported cases of COVID-19 by assuming the observed case count to follow a Poisson distribution and the expected case count to be linearly proportional to daily air travel volume. To capture the complex relations in the COVID-19 pandemic, the causal relationship in the transmission process can be modeled by hierarchical Bayesian distributions~\cite{niehus2020using,Flaxmans20}. In \cite{Eshragha20}, a special case of the continuous-time Markov population process, i.e., a partially-observable pure birth process, assumes a binomial distribution of partial observations of infected cases and estimates the future actual values of infections and the unreported percentage of infections in the population.

\textit{Bayesian statistical modeling of external factors on COVID-19 epidemic.}
Another important application is to model the influence of external factors on COVID-19 epidemic dynamics. For example, Flaxman et al.~\shortcite{Flaxmans20} infer the impact of NPIs including case isolation, educational institution closure, banning mass gatherings and/or public events and social distancing (including local and national lockdowns) in 11 European countries and estimate the course of COVID-19 by back-calculating infections from observed deaths by fitting a semi-mechanistic Bayesian hierarchical model with an infection-to-onset distribution and an onset-to-death distribution. In addition, case numbers, especially deaths, their model also jointly estimates the effect sizes of interventions.

\textit{Stochastic compartmental modeling of COVID-19 epidemic.}
Stochastic compartmental models can simulate stochastic hypotheses of specific aspects (e.g., probability of a state-based population or of a state transition) of the COVID-19 epidemiological process and the stochastic influence of external interventions on the COVID-19 epidemic process. Such probabilistic compartmental models integrate the transmission mechanisms of epidemics with the characteristics of observed case data~\cite{zhou2020semiparametric,dehning2020inferring,osthus2019dynamic,hebert2020macroscopic}. For example, in \cite{Wangl20}, a COVID-19 transmission tree is sampled from the genomic data with Markov chain Monte Carlo (MCMC)-based Bayesian inference under an epidemiological model, the parameters of the offspring distribution in this transmission tree are then inferred, and the model infers the person-to-person transmission in an early outbreak. Based on probabilistic compartmental modeling, Zhou et al.~\shortcite{zhou2020semiparametric} develop a semiparametric Bayesian probabilistic extension of the classical SIR model, called BaySIR, with time-varying epidemiological parameters to infer the COVID-19 transmission dynamics by considering the undocumented and documented infections and estimates the disease transmission rate by a Gaussian process prior and the removal rate by a gamma prior. To estimate the all-cause mortality effect of the pandemic, Kontis et al.~\shortcite{kontis2020magnitude} apply an ensemble of 16 statistical models (autoregressive with holiday and seasonal terms) on the vital statistics data for a comparable quantification of the weekly mortality effects of the first wave of COVID-19 and an estimation of the expected deaths in the absence of the pandemic. Other similar stochastic SIR models can also be found such as by assuming a Poisson time-dependent process on infection and reproduction \cite{hong2020estimation}, a beta distribution of infected and removed cases \cite{wang2020epidemiological}, and a Poisson distribution of susceptible, exposed, documented infected and  undocumented infected populations in a city~\cite{Li-sci20}.

\textit{Statistical influence modeling of COVID-19 interventions and policies.}
Apart from modeling the transmission dynamics or forecasting case counts, Bayesian statistical models are also applied in some other areas, e.g., to estimate the state transition distributions by applying certain assumptions such as of the susceptible-to-infected (i.e., the infection rate) or infected-to-death (mortality rate) transition. For example, Cheng et al.~\shortcite{cheng2020covid} use a Bayesian dynamic item-response theory model to produce a statistically valid index for tracking the government response to COVID-19 policies. Dehning et al.~\shortcite{dehning2020inferring} combine the established SIR model with Bayesian parameter inference with MCMC sampling to analyze the time dependence of the effective growth rate of new infections and to reveal the effectiveness of interventions. With the inferred central epidemiological parameters, they sample from the parameter distribution to evolve the SIR model equations and thus forecast future disease development. In~\cite{wang2020epidemiological}, a basic SIR model is modified by adding different types of time-varying quarantine strategies such as government-imposed mass isolation policies and micro-inspection measures at the community level to establish a method of calibrating cases of under-reported infections.

\textbf{Discussion}. Statistical modeling and a Bayesian statistical framework allow us to elicit informative priors for parameters that are difficult to estimate due to the lack of data reflecting the clinical characteristics of COVID-19, offer coherent uncertainty quantification of the parameter estimates, and capture nonlinear and non-monotonic relationships without the need for specific parametric assumptions~\cite{zhou2020semiparametric}. Compared with compartmental models, statistical models usually converge at different confidence levels for different assumptions (i.e., possible mitigation strategies), providing better interpretability and flexibility for characterizing the COVID-19 characteristics and complexities discussed in Section \ref{sec:char-compl}. However, the related work on COVID-19 statistical modeling is limited in terms of addressing COVID-19-specific characteristics and complexities, e.g., asymptomatic effect, the couplings between mitigation measures and case numbers, and the time-evolving and nonstationary case movement.

\begin{table*}[htb]
\centering
\caption{Examples of COVID-19 Mathematical and Statistical Learning.}
\small
\resizebox{1.0\textwidth}{!}{
\begin{tabular}{| p{0.15\textwidth}  | p{0.45\textwidth} | p{0.4\textwidth} |}
    \hline
    \textbf{Objectives} & \textbf{Approaches} & \textbf{Data}  \\ \hline 
    Infection diagnosis  & Descriptive analytics, Bayesian models, continuous-time Markov processes \cite{kronbichler2020asymptomatic,niehus2020using,Eshragha20,BrownC20,udugama2020diagnosing} & Case numbers, demographics, biomedical test, medical imaging, sensor data, etc.  \\ \hline
    Transmission processes & Bayesian inference, stochastic compartmental models, state-space model, MCMC \cite{zhou2020semiparametric,osthus2019dynamic,hebert2020macroscopic,Wangl20,hong2020estimation,Li-sci20}  & Case numbers, demographics, genomic data, external factors, etc.   \\ \hline
    Medical treatment  & Descriptive analytics, Bayesian models \cite{kontis2020magnitude} & Health/medical data, case numbers, etc. \\ \hline
    NPI evaluation  & Bayesian models, temporal and hierarchical Bayesian model, stochastic compartmental models, compartmental models with Bayesian inference \cite{Flaxmans20,dehning2020inferring,cheng2020covid,aguiar2020modelling,brauner2020inferring}  & NPI policies, case numbers, external data (e.g., social activities), etc.  \\ \hline
    Sentiment and emotion impact  & Descriptive analytics, latent models for sentiment/topic modeling, time-series analysis like regression variants \cite{VegtK20,xiong2020impact,pedrosa2020emotional,prime2020risk} & Questionnaire data, social media data, external factors like wellbeing, etc. \\ \hline
    Social, economic and workforce influence  & Descriptive analytics, time-series analysis, numerical methods, stochastic compartmental models \cite{HuangWFZSL20,Kraemerm20,Keelingm20,vo7impact,soto2020covid,del2020supply,walker2020impact,Baolf21,Myersk20,Lijj20}  & Case numbers, data related to economy, trade, supply chain, logistics, social activities, workforce, technology, transport, mobility, sustainability and public resources, etc.  \\ \hline    
    Misinformation & Descriptive analytics, time-series models, numerical methods, statistical language models \cite{LengZSWSSZCD21,Agleyj21,roozenbeek2020susceptibility}  & Fact data, online texts, social media, case numbers, etc.  \\ \hline
    \end{tabular}
    }
\label{tab:statlearning}
\end{table*}

\section{COVID-19 Data-driven Learning}
\label{sec:datalearning}

This section reviews the related work on data-driven discovery, i.e., applying classic (shallow) and deep machine learning methods, AI and data science techniques on COVID-19 data, to discover interesting knowledge and insights through characterizing, representing, analyzing, classifying and predicting COVID-19 problems.

\subsection{Shallow and Deep Learning}
\label{subsec:shadeeplearning}

Classic \textit{shallow machine learning} methods have been predominantly applied to COVID-19 classification, prediction and simulation, as shown by the WHO-based literature statistics in \cite{who-analysis}. Typical shallow learning methods include artificial neural networks (ANN), SVM, decision trees, Markov chain models, random forest, reinforcement learning, and transfer learning. These tools are easy to understand and implement and they are more applicable than other sophisticated methods (e.g., deep models and complex compartmental models) for the often small COVID-19 data. They are well explained in the relevant literature (e.g.,  \cite{nazrul2020survey,nguyen2020artificial,chen2020survey}) and interested readers can refer to them and other textbooks for technical details. Though different machine learning methods may be built on their respective learning paradigms \cite{dst_Cao15}, their main learning tasks and processes for COVID-19 modeling are similar, including (1) selecting discriminative features $\mathbf{x}$, (2) designing a model $f$ (e.g., a random forest classifier) to predict the target $y$: $\tilde{y}(\mathbf{x}) = f(\theta, \mathbf{x}, \mathbf{b})$ with parameters $\theta$ and bias term $\mathbf{b}$, and (3) optimizing the model to fit the COVID-19 data by defining and optimizing an objective function $\mathcal{L} = \argmin_{\theta} (y - \tilde{y}(\mathbf{x}))$ for the goodness of fit between expected $\tilde{y}$ and actual $y$ target (e.g., infective or diseased case numbers). 

\textit{Deep learning} as represented by deep neural networks 
is a more advanced COVID-19 modeling typically favored by computing researchers. Typical models applied in COVID-19 modeling include (1) convolutional neural networks (CNN) and their extensions in particular for images such as ImageNet and ResNet; (2) sequential networks such as LSTM, recurrent neural networks (RNN), memory networks and their variants; (3) textual neural networks such as BIRT, Transformer and their variants; (4) unsupervised neural networks such as autoencoders and generative adversarial networks (GAN); and (5) other neural learning mechanisms such as attention networks. 

Typical approaches for COVID-19 deep modeling can be represented by a general deep interaction and prediction framework as follows. It models (1) temporal dependencies over sequential \textit{case} ($\mathbf{x}$, which may consist of categories of case numbers $s$, $i$ and $r$ or their rates) evolution, (2) interactions and influence between external containment \textit{actions} ($\mathbf{a}$, which may consist of various control measures such as masking and social distancing) and case developments, and (3) the influence of personal \textit{context} ($\mathbf{c}$, which may consist of demographic and health circumstances and symptomatic features on COVID-19 infections) over time $t$. As COVID-19 case developments are sequential and stochastic to be influenced by many external factors, the framework combines autoencoders for the influence of unknown and stochastic asymptomatic and unreported case dynamics on reported numbers $\mathbf{x}$, RNN for sequential evolution of case numbers, control measures and personal context, and contextual attention for exterior containment strategies applied on case control to model complex interactions between various sources of underlying and control factors in COVID-19 sequential developments.

\begin{figure}[htb!]
	\centering
	\includegraphics[width=0.7\linewidth]{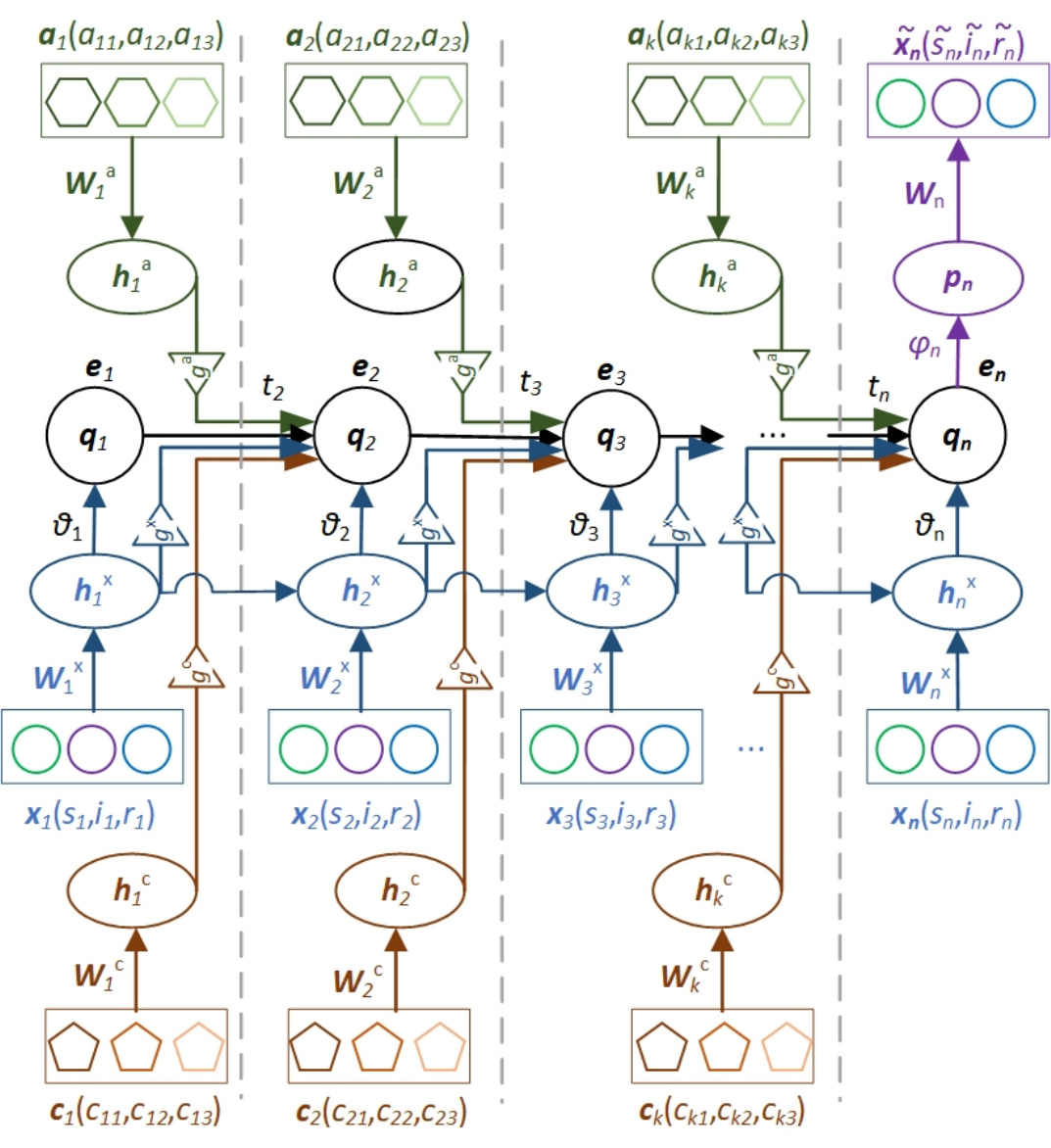}
	\caption{A time-varying case-action-context neural interaction network for COVID-19 deep sequential modeling.}
	\label{fig:deepmodel}
\end{figure}

Fig. \ref{fig:deepmodel} illustrates a deep sequential case-action-context interaction network for this purpose. The formulation of the key variables and their interactions are shown below. In practice, networks $\mathbf{h}$ can be based on an LSTM, RNN, Transformer or other deep networks, the gating function unit $g$ can be implemented by a gated recurrent unit (GRU) or other gating functions to determine the influence of control actions/context on case number movement. The transformation from input case vector $\mathbf{x}$ to action and context ($\mathbf{a}$ and $\mathbf{c}$) adjusted case representation $\mathbf{e}$ through network $q_{\theta}$ can be treated as an encoder, while the estimation (reconstruction) of $\mathbf{x}_t$ from $\mathbf{e}$ by network $p_{\varphi}$ is a decoding or prediction process. 

\begin{subequations}\label{eq:deepmodel}
    \begin{align}
        \mathbf{h}^{x}_t \sim h^{x}(\mathbf{x}_t, \mathbf{W}^{x}_t, \mathbf{b}^{x}_t | \mathbf{h}^{x}_{t-1}) \\ 
        \mathbf{h}^{c}_t \sim h^{c}(\mathbf{c}_t, \mathbf{W}^{c}_t, \mathbf{b}^{c}_t) \\ 
        \mathbf{h}^{a}_t \sim h^{a}(\mathbf{a}_t, \mathbf{W}^{a}_t, \mathbf{b}^{a}_t) \\
        \mathbf{g}^x_t \sim g(\mathbf{x}_t, \mathbf{h}_t) \\
        \mathbf{e}_t \sim q_{\theta} (\mathbf{e}_t | \mathbf{e}_{t-1}, \mathbf{h}^{x}_t, \mathbf{h}^{x}_{t-1}, \mathbf{g}^x_t, \mathbf{g}^c_t, \mathbf{g}^a_t) \\
        \mathbf{\tilde{x}}_t \sim p_{\varphi} (\mathbf{\tilde{x}}_t | \mathbf{e}_t, \mathbf{h}^{x}_t)
    \end{align}
\end{subequations}

The interaction and prediction network in Fig. \ref{fig:deepmodel} can be implemented in terms of an autoencoder (where $q$ and $p$ refer to encoding and decoding networks, e.g., \cite{Ibrahimm21}) or LSTM/RNN-based prediction (with $q$ for representation and $p$ for estimating the next input, e.g., \cite{Shahid20}) framework. Accordingly, the objective function can be defined in terms of the discrepancy $\mathcal{J}$ between $\mathbf{x}_t$ and $\mathbf{\tilde{x}}_t$ (i.e., $\argmin_{\theta, \varphi} \mathcal{J}(\mathbf{x}_t, \mathbf{\tilde{x}}_t)$) or the KL-divergence ($\mathcal{D}$) with loss $\mathcal{L}$ (where $\mathbf{h}_t$ and $\mathbf{h}_{t-1}$ refer to the representations of input $\mathbf{x}$ interacting with actions $\mathbf{a}$ under the context $\mathbf{c}$ through gating $g$ integration).
\begin{equation}
    \mathcal{L}(\theta, \varphi; \mathbf{x}_t) \sim - \mathcal{D}(q_\theta (\mathbf{e}_t|\mathbf{h}_t, \mathbf{h}_{t-1}) || p_{\varphi}(\mathbf{e}_t)) + \mathcal{E}_{q_{\theta}}(\mathbf{e}_t|\mathbf{h}_t, \mathbf{h}_{t-1}) [log p_{\varphi}(\mathbf{h}_t, \mathbf{h}_{t-1} | \mathbf{e}_t)]
\end{equation}

\subsection{COVID-19 Shallow Learning}
\label{subsec:machlearning}

Here, \textit{COVID-19 shallow learning} refers to the application of general shallow or classic machine learning methods to the analytics and modeling of COVID-19 problems and data. It forms the second popular set of modeling methods (about 4k of 22k WHO-listed references) that model COVID-19 outbreak, risk, transmission, uncertainty, anomalies, complexities, classification, variation, and prediction and more specifically case forecasting, medical diagnostics, contact tracing, and drug development \cite{Kamalovf21}. General machine learning methods including ANN, tree models such as decision trees and random forest, kernel methods like SVM, transfer learning, NLP and text mining methods, evolutionary computing like genetic algorithms and fuzzy set, and reinforcement learning are mostly applied in addressing the above COVID-19 tasks by medical, biomedical, computing and social scientists \cite{shahid2020machine,mohamadou2020review,Rasheedj21}, as discussed below. 

\textit{Machine learning for COVID-19 outbreak prediction and risk assessment.} Typical classifiers like ANN, SVM, decision trees, random forest, regression trees, least absolute shrinkage and selection operator (LASSO), and self-organizing maps are applied to forecast COVID-19 spread and outbreak and their coverage, patterns, growth and trends; estimate and forecast the confirmed, recovered and death case numbers or the transmission and mortality rates; and cluster infected cases and groups, etc. For example, in \cite{Kasilingamd20}, logistic regression, decision trees, random forest and SVM are applied to estimate the growth trend and containment sign on the data consisting of factors about health infrastructure, environment, intervention policies and infection cases with accuracy between 76.2\% and 92.9\%. Evolutionary computing such as genetic algorithm, particle swarm optimization, and gray wolf optimizer forecast COVID-19 infections \cite{niazkar2020covid,salgotra2020time,tseng2020computational}.

\textit{Machine learning for COVID-19 diagnosis on clinical attributes.} The machine learning of COVID-19 clinical reports such as blood test results can assist in diagnosis. For example, in \cite{Khandayam20}, clinic attributes and patient demographic data are extracted by term frequency/inverse document frequency (TF/IDF), bag of words (BOW) and report length from textual clinic reports. The extracted features are then classified in terms of COVID, acute respiratory distress syndrome (ARDS), SARS and both COVID and ARDS by SVM, multinomial naıve Bayes, logistic regression, decision tree, random forest, bagging, Adaboost, and stochastic gradient boosting, reporting an accuracy of 96.2\% using multinomial naıve Bayes and logistic regression. In \cite{Brinatid20}, hematochemical values are extracted from routine blood exam-based clinic attributes, which are then classified into positive or negative COVID-19 infections by decision trees, extremely randomized trees, KNN, logistic regression, naive Bayes, random forest, and SVM. It reports an accuracy of 82\% to 86\%. The work in \cite{wu2020rapid} applies random forest to identify COVID-19 infections.

\textit{Machine learning for COVID-19 diagnosis on respiratory data.} Machine learning can be conducted on COVID-19 patient's respiratory data such as lung ultrasound waves and breathing and coughing signals to extract respiratory behavioral patterns and anomalies. For example, logistic regression, gradient boosting trees and SVMs distinguish COVID-19 infections from asthmatic or healthy people on the Android app-based collection of coughs and breathing sounds and symptoms with AUC at 80\% \cite{BrownC20}. 

\textit{Machine learning for COVID-19 diagnosis on medical imaging.} A very intensive application of classic machine learning methods is to screen COVID-19 infections on CT, chest X-ray (CXR)  or PET images. For example, in \cite{ChandraVSJN21}, the majority voting-based ensemble of SVM, decision tree, KNN, naive Bayes and ANN is applied to classify normal, pneumonia and COVID-19-infected patients on CXR images with an accuracy of 98\% and AUC of 97.7\%. In \cite{Faridaa20}, the simple applications of SVM, naive Bayes, random forest and JRip on CT images screen COVID-19 diseases with a reported accuracy of 96.07\% by naive Bayes combined with random forest and JRip, in comparison with 94.11\% by CNN. 

\textit{Machine learning for COVID-19 diagnosis on latent features.}
Further, shallow learners are applied to detect and diagnose COVID-19 infections on latent features learned by shallow to deep representation models on COVID-19 medical images. For example, in \cite{Kanghy20}, ANN-based latent representation learning captures latent features from gray, texture, histogram, number, intensity, surface and volume features in CT images, then classifiers including SVM, logistic regression, Gaussian naive Bayes, KNN and ANN are applied to differentiate COVID-19 infections from community-acquired pneumonia with 95.5\% accuracy reported. In \cite{OzturkOB21}, latent features are extracted from CXR and CT images to form a gray level co-occurrence matrix (GLCM), local binary gray level co-occurrence matrix (LBGLCM), gray level-run length matrix (GLRLM) and segmentation-based fractal texture analysis (SFTA)-based features, which are then oversampled by the synthetic minority over-sampling technique (SMOTE) and further selected by a stacked autoencoder (sAE) and principal component analysis (PCA), before SVM is applied to achieve 94.23\% accuracy. In \cite{TOGACAR20}, MobileNetV2 and SqueezeNet extract features from CXR images, which are then processed by social mimic optimization to classify coronavirus, pneumonia, and normal images with 99.27\% accuracy by SVM. Lastly, in \cite{Tuncert20}, a residual exemplar local binary pattern (ResExLBP)-based method extracts features from CXR images, which are then selected by an iterative relief-based method before decision trees, linear discriminant, SVM, KNN and subspace discriminant are applied on the selected features to detect COVID-19 infection with an accuracy of 99.69\% to 100.0\%.  

\textit{Modeling the influence of external factors on COVID-19.}
Various machine learning tasks are undertaken to analyze the relation and influence of external and contextual factors on COVID-19 epidemic attributes. For example, ensemble methods including random forest, extra trees regressor, AdaBoost, gradient boosting regressor, extreme gradient boosting (XGBoost), light gradient boosting machine (LightGBM), CatBoost regressor, kernel ridge, SVM, KNN, MLP and decision trees indicate potential association between COVID-19 mortality and weather data \cite{Malkiz20}.

\textit{Machine learning-driven drug and vaccine development for COVID-19.}
Machine learning methods are applied to analyze the drug-target interactions, drug selection, and the effectiveness of drugs and vaccines on containing COVID-19. 
For example, machine learning methods including XGBoost, random forest, MLP, SVM and logistic regression are used to screen thousands of hypothetical antibody sequences and select nine stable antibodies that potentially inhibit SARS-CoV-2 \cite{Magarr21,Arshadik20}. 


\subsection{COVID-19 Deep Learning}
\label{subsec:deeplearning}

Deep learning has been intensively applied to modeling COVID-19 as discussed in Section \ref{subsec:shadeeplearning}, with about 2k of 22k references on modeling COVID-19. Typical applications involve COVID-19 data on daily infection case numbers, health and clinic records, hospital transactions, medical imaging, respiratory signals, genomic and protein sequences, and exterior data such as infective demographics, social media communications, news and textual information, etc. Below, we first discuss a common application of deep learning for COVID-19 epidemic description and forecasting, and then briefly review other applications.

\textit{Deep learning of the COVID-19 epidemic.}
Deep neural networks are intensively applied to characterize and forecast COVID-19 epidemic outbreak, dynamics and transmission. Examples are predicting the peak confirmed numbers and peak occurrence dates, forecasting daily confirmed, diseased and recovered case numbers, and forecasting $N$-day (e.g., $N=\{7, 14, 10, 30, 60~days\}$) infected/confirmed, recovered and death case numbers (or their transmission/mortality rates) through modeling short-range temporal dependencies in case numbers by applying LSTM, stacked LSTM, Bi-LSTM, convolutional LSTM-like RNNs, and GRU \cite{Devaraj21,Shahid20}. Other work models the transmission dynamics and predicts daily infections of COVID-19 using a variational autoencoder (VAE), encoder-decoder LSTM or LSTM with encoder and Transformer \cite{KimKKSMNP020} and modified auto-encoder \cite{Pereira-cov20}, GAN and their variants, tracks its outbreak~\cite{hu2020artificial}, predicts the outbreak size by encoding quarantine policies as the strength function in a deep neural network~\cite{dandekar2020neural}, estimates global transmission dynamics using a modified autoencoder~\cite{hu2020forecasting}, predicts epidemic size and lasting time, and combines medical information with local weather data to predict the risk level of a country by a shallow LSTM model~\cite{pal2020neural}. In \cite{Zerouala20}, a comparative analysis shows that VAE outperforms simple RNN, LSTM, BiLSTM and GRU in forecasting COVID-19 new and recovered cases.

\textit{Broad deep COVID-19 learning.}
In addition, we highlight several other typical application areas of COVID-19 deep learning: 
\begin{enumerate}
    \item Characterizing symptoms of coronavirus infections, e.g., by pretrained neural networks (e.g., \cite{schuller2020covid}), with more discussion in Section \ref{subsubsec:diaganal}; 
    \item Analyzing health and medical records, blood sample-based test reports, and respiratory sounds and signals for diagnosis and treatment e.g. by CNN, LSTM and GRU \cite{Rasheedj21}, with more discussion in Section \ref{subsubsec:prognosisanal};
    \item Analyzing medical imaging for diagnosis, quarantine and treatment by convolutional neural networks (CNN, e.g., ImageNet and ResNet), GAN and their mutations \cite{IslamKAZ21,Rasheedj21}, with more discussion in Section \ref{subsubsec:medanal};    
    \item Analyzing COVID-19 genomic and protein sequence and interaction analysis by RNN, CNN and their variants for drug and vaccine development, tracing infection sources, and analyzing virus structures and evolution, with more details in Section \ref{subsubsec:pathanal};
    \item Repurposing and developing drugs and vaccines by generative autoencoders, generative tensorial reinforcement learning and generative adversarial networks \cite{zhavoronkov2020potential} for generative chemistry discovery;
    \item Analyzing COVID-19 impact on sentiment and emotion by RNN, Transformer-based NLP neural models and their derivatives \cite{li2020analyzing,duong2020ivory}; 
    \item Characterizing the COVID-19 infodemic by NLP and text mining including misinformation identification~\cite{sharma2020covid}, enhancing epidemic modeling using social media data~\cite{kim2019incorporating}, and analyzing the COVID-19 research progress and topic evolution~\cite{zhang2020topic}; 
    \item Other topics such as analyzing the influence and effect of countermeasures, e.g., the effect of quarantine policies on outbreak using DNNs~\cite{dandekar2020neural}, with more discussion in Section \ref{sec:impactmodeling}.
\end{enumerate}

Table \ref{tab:machlearning} illustrates some typical applications of shallow and deep learning methods for modeling COVID-19. More discussion on COVID-19 deep learning can be found in Section \ref{subsec:medianal}. 

\textbf{Discussions.} Most of the existing studies on shallow and deep COVID-19 modeling directly apply the existing shallow machine learning methods and deep neural networks on COVID-19 data, as shown in reviews like \cite{litjens2017survey,senior2020improved,wang2020covid}. Our literature review also shows that deep neural models are widely applicable to COVID-19 modeling tasks, which are unnecessarily overwhelmingly applied to all possibilities and significantly outperform time-series forecasters and shallows machine learners. In fact, sometimes, deep models may even lose their advantage over traditional modelers such as ensembles, as shown in Table \ref{tab:machlearning}, 
and Table \ref{tab:imaging}. 

\begin{table*}[htb]
\centering
\caption{Examples of COVID-19 Shallow and Deep Learning$^a$.}
\small
\resizebox{1.0\textwidth}{!}{
\begin{tabular}{| p{0.15\textwidth}  | p{0.52\textwidth} | p{0.3\textwidth} |}
    \hline
    \textbf{Objectives} & \textbf{Approaches} & \textbf{Data}   \\ \hline
    Transmission \& external factor impact & Shallow learners like SVM, ANN, decision tree, random forest and ensemble methods \cite{Kasilingamd20}, evolutionary computing methods such as particle swarm optimization \cite{,niazkar2020covid,salgotra2020time,tseng2020computational}, DNN variants such as LSTM, GRU, VAE, GAN and BiLSTM, etc. \cite{Devaraj21,Shahid20,KimKKSMNP020,Pereira-cov20,hu2020artificial,dandekar2020neural,Zerouala20}  & Epidemic case numbers and external data such as meteorological data, environmental data (e.g., humidity), social activity and mobility data, etc.  \\ \hline
    Infection diagnosis  & Shallow learners \cite{Brinatid20,wu2020rapid,BrownC20,ChandraVSJN21}, CNN and RNN variants like LSTM and GRU, and pretrained CNN-based image nets like ResNet, MobileNetV2 and SqueezeNet, etc. \cite{Faridaa20,TOGACAR20,Tuncert20,Jiangz20}, text analysis models \cite{Khandayam20}   &  Pathological and clinical records, respiratory signals (e.g. coughing and breathing signals and patterns in ultrasound or thermal video), computed tomography (CT) and CXR images, etc. \\ \hline
    Mortality and survival analysis  & Shallow learners like SVM, ANN, decision tree, regression tree, random forest and ensemble methods like XGBoost \cite{schwab2021real,chakraborty2020real,schwab2021real}, CNNs, pretrained CNN-based image nets, RNN variants like LSTM and GRU \cite{Devaraj21,Shahid20,Zhanggp21} & Medical imaging including CT and CXR images, clinical records, patient demographics, case numbers, external data, etc.  \\ \hline    
    Medical treatment  & Shallow machine learning methods and DNNs, etc. \cite{zheng2020recommendations,wu2020estimating,yan2020prediction,beck2020predicting} & Health/medical records, pharmaceutical treatments, ICU data, etc.  \\ \hline
    Genomic and protein analysis, drug/vaccine development  & Shallow classifiers like SVM and ensembles, frequent pattern and sequence analysis methods, CNN variants, RNN variants, attention networks, GAN, autoencoders, reinforcement learning, NLP models like Transformer, etc. \cite{Arshadik20,metskyc20,beck2020predicting,Zhavoronkova20,zhavoronkov2020potential,hu2020prediction,metsky2020crispr,Nawazm21,Alakust21}  & Genomic data, proteomic data, drug-target interactions, molecular reactions, etc.  \\ \hline
    Resurgence and mutation  & Shallow learners like linear discriminant, SVM, KNN and subspace discriminant, combining classifiers with compartmental models, DNN variants, sequence analysis$^b$, etc. \cite{randhawa2020machine,aravindakshan2020preparing} & Resurgence case numbers, virus strain genome and protein sequences, NPI data, external data, etc.  \\ \hline
    NPI evaluation  & Various Bayesian models$^c$, combining compartmental models with classifiers or estimators, DNNs$^b$, etc. \cite{Fangyq20,gatto2020spread}  & Case numbers, NPI policies, external data, etc.  \\ \hline
    Sentiment and emotion impact  & NLP models like LDA and topic models and DNN variants like BERT and Transformer variants, etc. \cite{li2020analyzing,li2020analyzing,duong2020ivory,WangLCZ20,Nemesl21,Chenaghlu20,miner2020chatbots}  & Social media data, news feeds, Q/A data, external factors, etc.  \\ \hline
    Socioeconomic influence  & Relation (e.g., correlation and causality) analysis$^b$, topic modeling by NLP models \cite{zhang2020topic} & Social, economic and workforce activities, case numbers, etc.  \\ \hline
    Misinformation analysis  & Classic NLP models, correlation analysis, shallow learners, outlier detectors, DNN variants like BERT and Transformer mutations \cite{MicallefHKAM20,LengZSWSSZCD21,sharma2020covid} & Social media, online texts, Q/A data, news feeds, etc.  \\ \hline
    \end{tabular}
    }
\label{tab:machlearning}
\footnotesize{$^a$ See Table \ref{tab:imaging} for deep COVID-19 medical imaging analysis; $^b$ such methods are applicable but not much work is reported in the literature; $^c$ See Table \ref{tab:statlearning} for NPI effect modeling.}
\end{table*}

\section{COVID-19 Domain-driven Modeling}
\label{sec:domainmodeling}

As a complex social-technical issue, COVID-19 modeling brings many specific challenges and research questions from the relevant domains and for domain-specific research communities. In this section, we focus on two major and mostly relevant domains of COVID-19: epidemic modeling, and medical and biomedical analysis. 

\subsection{COVID-19 Epidemic Modeling}
\label{subsec:compartment}

\subsubsection{Epidemiological compartmental models}
\label{subsubsec:comparmentmodels}

Epidemiological modeling portrays the state-space, interaction processes and dynamics of an epidemic in terms of its macroscopic population, states and behaviors. Compartmental models are widely used in characterizing COVID-19 epidemiology by incorporating epidemic knowledge and compartmental hypotheses into imitating the multi-state COVID-19 population transitions. An individual in the COVID-19 epidemic sits at one state (compartment) at a time-point and may transit this state to another at a state transmission rate. The individuals of the closed population are respectively labeled per their compartments and migrate across compartments during the COVID-19 epidemic process, which are modeled by (ordinary) differential equations. 

\begin{figure}[htb!]
	\centering
	\includegraphics[width=0.75\linewidth]{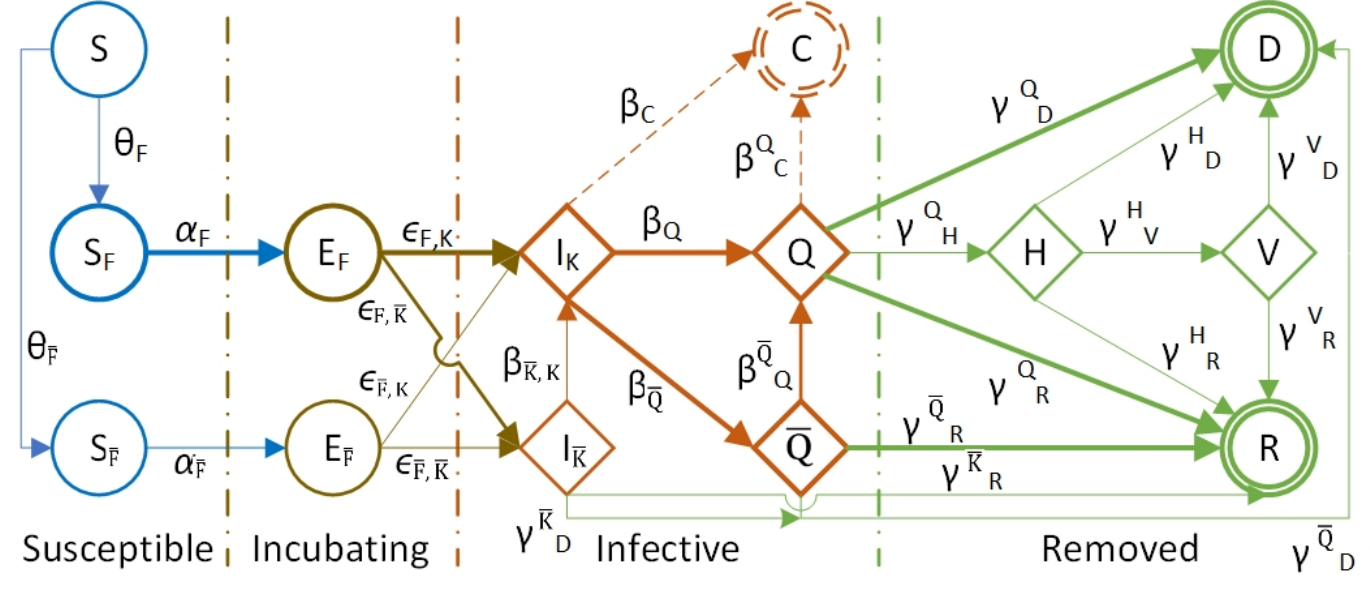}
	\caption{COVID-19 epidemiological compartmental modeling. The COVID-19 epidemiological process can be categorized into four major phases: susceptible (S), exposed/incubating (E), infective (I) and removed (R), and other optional states. A \textit{circle} denotes the initial states, a \textit{diamond} denotes the optional states, and a \textit{double circle} denotes the final states. \textit{Blue} indicates the noninfectious states, \textit{red} the infectious states, and \textit{green} the states with patients removed or to be removed from the infection. The \textit{thick} lines denote the main states and state transition paths while the \textit{thin} lines are minor ones which may be ignored in modeling.}
	\label{fig:compartmentmodeling}
\end{figure}

By consolidating various COVID-19 epidemiological characteristics, hypotheses and compartmental models, Fig. \ref{fig:compartmentmodeling} illustrates a typical COVID-19 state-space and evolution system with major (thick) and minor (thin) states and state transition paths that can be sequentially categorized into four phases: 
\begin{itemize}
    \item Susceptible (S): Individuals ($S$) are susceptible to infection under free (uncontained or unrestrained, $S_F$ at uncontained rate $\theta_F$) or contained (restrained, $S_{\bar{F}}$ at containment rate $\theta_{\bar{F}}$) conditions at a respective transmission rate $\alpha_{F}$ or $\alpha_{\bar{F}}$; 
    \item Exposed (E): Free or contained susceptibles are exposed to infection from those who are infected but in the incubation period (which could be as long as 14 days), and may be noninfectious and free ($E_F$) or infectious and contained ($E_{\bar{F}}$) at a respective exposure rate $\varepsilon_{F}$ or $\varepsilon_{\bar{F}}$; 
    \item Infective (I): Those exposed become infectious and may be detected (registered/documented and known to medical authorities, $I_K$) or undetected (unreported/undocumented and unknown to medical management, $I_{\bar{K}}$); also, some initially undetected infectives may be further detected and converted to detected infectives at rate $\beta_K^{\bar{K}}$; some documented infectives may be symptomatic and quarantined ($Q$) at quarantine rate $\beta_Q$ while others may be asymptomatic and unquarantined ($\bar{Q}$) at unquarantined rate $\beta_{\bar{Q}}$; there may be some rare cases (at rate $\beta_C$) who carry the virus and infection for a long time with or without symptoms, called lasting \textit{carriers} ($C$); in addition, some initially asymptomatic cases may transfer to symptomatic and quarantined at rate $\beta^{\bar{Q}}_{Q}$; 
    \item Removed (R): Unquarantined infectives may recover at recovery rate $\gamma^{\bar{Q}}_R$ or die at mortality rate $\gamma^{\bar{Q}}_D$ respectively, the same for quarantined infectives at rate $\gamma^Q_R$ or $\gamma^Q_D$ and unknown/undetected infectives at rate $\gamma^{\bar{K}}_R$ or $\gamma^{\bar{K}}_D$; some quarantined infectives may present acute symptoms even with life threat, who are then hospitalized (H) at rate $\gamma^Q_H$ or even further ventilated (V) at rate $\gamma^H_V$; hospitalized infectives may recover or die at rate $\gamma^H_R$ or $\gamma^H_D$, the same for ventilated at rate $\gamma^V_R$ or $\gamma^V_D$. 
\end{itemize}
 
In practice, the above COVID-19 state-space may be too complicated to model and not all states are characterizable by the available data. Accordingly, a focus is on those main states and their transitions when the corresponding data is available, for example, susceptible, exposed, infectious (which consists of both detected and undetected), recovered, and diseased. Below, we illustrate the differential equations of the states \textit{S}, \textit{E}, \textit{I}, \textit{Q}, $\bar{Q}$, \textit{R} and \textit{D} which are the main states of a closed COVID-19 population. Here, (1) $S$, $E$, $I$, $Q$, $\bar{Q}$, $R$ and $D$ represent the fraction of the population at each state; (2) $S$ and $E$ are impacted by containment measures at the containment rate $\theta$ and $I$ at $\theta_I$ who are contained and recovered; and (3) the state transitions take place at the rates shown in Fig. \ref{fig:compartmentmodeling}. 


\begin{subequations}\label{eq:compartmodel}
    \begin{align}
        \partial_t S = -\alpha SI - \theta S \label{s} \\
        \partial_t E = \alpha SI - \epsilon E - \theta E \label{E} \\
        \partial_t I = \epsilon E - \beta_Q I - \beta_{\bar{Q}} I - \theta_I I \label{I} \\
        \partial_t Q = \beta_Q I + \beta_Q^{\bar{Q}} \bar{Q} - \gamma_D^Q D - \gamma_R^Q R \\
        \partial_t \bar{Q} = \beta_{\bar{Q}} I - \beta_Q^{\bar{Q}} \bar{Q} - \gamma_R^{\bar{Q}} \bar{Q} - \gamma_D^{\bar{Q}} \bar{Q} \label{bar_Q$} \\
        \partial_t R = \gamma_R^Q R + \gamma_R^{\bar{Q}} \bar{Q} + \theta S + \theta E + \theta_I I \label{R} \\
        \partial_t D = \gamma_D^Q Q + \gamma_D^{\bar{Q}} \bar{Q} \label{D} \\
        S + E + I + Q + \bar{Q} + R + D = 1
 \end{align}
\end{subequations}


\subsubsection{COVID-19 epidemiological modeling}
\label{subsubsec:covidcompartmodels}

We summarize and discuss the related work on the major tasks of COVID-19 epidemiological modeling, and highlight the work on modeling COVID-19 epidemic transmission processes, dynamics, external factor influence, and resurgence and mutation. 

\textit{COVID-19 epidemiological modeling tasks.}
Epidemiological models dominate COVID-19 modeling (about 3.5k publications of the 22k reported in the WHO literature) by epidemic researchers and computing scientists through the expansion or hybridization with other models such as statistical models and machine learning methods. COVID-19 compartmental modeling aims to answer several epidemiological problems: (1) the growth (spread) of COVID-19 and its case number movements at different epidemiological states to forecast case numbers in the next days or periods; (2) the basic reproduction rate $R_0$ that informs the contagion and transmission level and control strategies; (3) the sensitivity and effect of control measures on infection containment and case movements; and (4) the sensitivity and effect of strategies for herd immunity and mass vaccination. Accordingly, various compartmental models are customized to cater for specific assumptions, settings and conditions of modeling COVID-19, as discussed in Section \ref{subsubsec:covidcompartmodels}.
For (1), with historical case numbers of a country or region and the initial settings of hyperparameters, we can estimate the parameters and further predict the number over time, e.g., the number of infections and deaths in a country or city. 
Regarding (2), with the state-space shown in Fig. \ref{fig:compartmentmodeling}, to resolve these differential equations, we first obtain the population projection matrix $A$ corresponding to all states and their transition probabilities. The projection matrix $A$ can be converted to a state transition matrix $T$ (where each element $T_{ij}$ is the probability of an individual transferring from state $i$ at time $t$ to state $j$ at time $j+1$) and a fertility (reproductive) matrix $F$ (where an element $F_{ij}$ refers to the reproduced number of $i$-state offsprings of an individual at state $j$), i.e., $A = T + F$. Further, we can calculate the fundamental matrix $N$: $N = (I - T)^{-1}$ with identity matrix $I$ to represent the expected time spent in each state and that to death. Then, we can obtain another matrix $R$: $R = FN$ with each entry referring to the expected lifetime production number of $i$-state offspring by an individual at stage $j$ \cite{Caswell01,singh2018mathematical}. Its dominant eigenvalue is the net reproduction rate $R_0$.
With regard to (3), since control measures such as social distancing and lockdown may influence the growth of case numbers and reproduction and transmission rates, we can analyze the sensitivity of adjusting related parameters on the case numbers and rates. To explore the opportunities for herd immunity and mass vaccination in (4), the herd immunity rate and vaccination rate are expected to be greater than $1 - \frac{1}{R_0}$ to eradicate the disease. 

In addition to these major problems, below, we further discuss two applications of epidemiological modeling in COVID-19: modeling its transmission (which is also the most explored area) and resurgence and mutation (which is a recent challenge). More discussion on modeling the NPI effect on COVID-19 transmission and epidemic is in Section \ref{subsec:NPIeffect}.

\textit{Modeling COVID-19 epidemic transmission process.} 
Studies on modeling COVID-19 epidemic transmission mainly focus on evaluating the epidemiological attributes (e.g., infection rate, recovery rate, mortality, reproduction number, etc.), predicting the infection and death counts, and revealing the transmission, spread and outbreak trends under experimental or real-world scenarios. As illustrated in Table \ref{tab:epidmodels}, various compartmental models are available to characterize COVID-19. For example, the SIDARTHE compartmental model considers eight stages of infection: susceptible ($S$), infected ($I$), diagnosed ($D$), ailing ($A$), recognized ($R$), threatened ($T$), healed ($H$) and extinct ($E$) to predict the course of the epidemic and to plan an effective control strategy~\cite{giordano2020modelling}. A new compartment is introduced to the classic SIR model to quantify those who are symptomatic, quarantined infecteds~\cite{maier2020effective}. Further, a stochastic SHARUCD model framework contains seven compartments: susceptible ($S$), severe cases prone to hospitalization ($H$), mild, sub-clinical or asymptomatic ($A$), recovered ($R$), patients admitted to the intensive care units ($U$), and the recorded cumulative positive cases ($C$), which include all new positive cases for each class of $H$, $A$, $U$, $R$, and deceased ($D$)~\cite{aguiar2020modelling}. In addition, several models involve new compartments to represent asymptomatic features to mild symptoms \cite{aguiar2020modelling,weitz2020modeling} and undocumented cases \cite{Li-sci20}.

\textit{Modeling COVID-19 epidemic dynamics, transmission and risk.}
Classic compartmental models assume constant transmission and recovery rates between state transitions. This assumption is taken in many SIR variants tailored for COVID-19, which cannot capture the disease characteristics in Section \ref{subsec:disease}. 
To cater for COVID-19-specific characteristics especially when mitigation measures are involved, the classic susceptible-infectious-recovered (SIR)~\cite{kermack1927contribution} and susceptible-exposed-infectious-recovered (SEIR) models~\cite{aron1984seasonality}, which were applied to modeling other epidemics like measles and Ebola, are tailored for COVID-19. Since COVID-19 transmission contains more states, especially with interventions, SIR/SEIR models are extended by adding customized compartments like quarantine, protected, asymptomatic and immune~\cite{giordano2020modelling,weitz2020modeling,aguiar2020modelling,crokidakis2020modeling,maier2020effective}. 
Accordingly, to capture the evolving COVID-19 epidemiological attributes including time-variant infection, mortality and recovery rates, time-dependent compartmental models are proposed. For example, a time-dependent SIR model adapts the change of infectious disease control and prevention laws as city lockdowns are imposed and traffic halt with the control parameters infection rate $\beta$ and recovery rate $\gamma$ modeled as time-variant variables~\cite{chen2020time}. Dynamical modeling is also considered in temporal SIR models with temporal susceptible, insusceptible, exposed, infectious, quarantined, recovered and closed (or death) cases in \cite{peng2020epidemic}. An early-stage study of a dynamic SEIR model estimates the epidemic peak and size, and an LSTM further forecasts its trend after taking into account public monitoring and  detection policies \cite{Yangzf20}.

\textit{Modeling the influence of external factors on the COVID-19 epidemics.}
COVID-19 epidemic dynamics reflect the time-varying states, state transition rates, and their vulnerability to contextual and external factors such as a person's ethnicity and public health conditions and social contacts and networking \cite{Liu12680}. 
To depict the influence of external factors, more complex compartmental models involve the relevant side information (e.g., NPIs, demographic features such as age stratification and heterogeneity, and social activities such as population mobility) into their state transitions. Examples include an age-sensitive SIR model~\cite{chikina2020modeling} which integrates known age-interaction contact patterns into the examination of potential effects of age-heterogeneous mitigations on an epidemic in a COVID-19-like parameter regime, an age-structured SIR model with social contact matrices and Bayesian imputation~\cite{singh2020age}, and an age-structured susceptible-exposed-infectious-recovered-dead (SEIRD) model that identifies no significant susceptibility difference between age groups \cite{Omori-cov20}. More about the NPI influence on the COVID-19 epidemic is in Section \ref{subsec:NPIeffect}. 
In addition, environmental factors, especially humidity and temperature, may affect COVID-19 virus survival and the epidemic's transmission~\cite{xie2020association,wang2020temperature,bu2020analysis,oliveiros2020role} despite inconsistent conclusions. In \cite{daSilvar20}, variational mode decomposition decomposes COVID-19 case time series into multiple components and then a Bayesian regression neural network, cubist regression, KNN, quantile random forest and support vector regression (SVR) are combined to forecast six-day-ahead case movements by involving climatic exogenous variables.

\textit{Modeling COVID-19 resurgence and mutation.} 
Our current understanding of COVID-19 resurgence and mutation is very limited while the British, South African, Indian and other newly-emergent mutations show higher contagion and complexities~\cite{grubaugh2020making,grubaugh2021public}. COVID-19 may indeed become another epidemic disease which remains with humans for a long time. Imperative research is expected to quantify the virus mutation and disease resurgence conditions, forecast and control potential resurgences and future waves after lifting certain mitigation restrictions and reactivating businesses and social activities~\cite{lopez2020end,pedro2020conditions}, distinguish the epidemiological characteristics, age sensitivity, and intervention and containment measures between waves~\cite{grech2020covid,aleta2020age}, compare the epidemiological wave patterns between countries experiencing mutations and resurgences and compare COVID-19 wave patterns with influenza wave patterns \cite{fan2020decreased}, predict resurgences and mutations (e.g., by estimating the daily confirmed case growth when relaxing interstate movement, mobility and contact restrictions and social distancing by SEIR-expanded modeling) and prepare for  countermeasures on future waves~\cite{aravindakshan2020preparing}.  
Limited research results are available in the literature on the above broad issues. For example, a comparative analysis in \cite{Bontempie21} shows the differences in the second COVID-19 wave in Europe in Italy and indicates the different causing strategies taken by them in implementing facemasks, social distancing, business closures and reopenings. In \cite{Cacciapagliag}, building on fitting the first wave data, an epidemic renormalisation group approach further simulates the dynamics of disease transmission and spreading across European countries over weeks by modeling the European border control effects and social distancing in each country.
In \cite{leung2020first}, an SIR model estimates the scenarios of incurring a potential second wave in China and the potential case fatality rate if containment measures such as travel ban and viral reintroduction from overseas importation are relaxed for certain durations in a population with a certain epidemic effect size and cumulative count after the first wave. 
In \cite{pedro2020conditions}, an SEIR model incorporates social distancing to model the mechanism (closure releasing) of forming the second wave, the epidemiological conditions (ranges of transmission rate and the inverse of the average infectious duration) for triggering the second and third waves, and the socioeconomic (economic loss due to lockdown) and intervention (novel social behavior spread) factors on case numbers.  
In \cite{lopez2020end}, a revised stochastic SEIR model estimates different resurgence scenarios reflected on infections when applying time-decaying immunity, lockdown release, or increasing implementation of social distancing and other individual NPIs.


\textbf{Discussion}. COVID-19 compartmental models excel at modeling epidemiological hypotheses, processes and factors with domain knowledge and interpretation. Such models often assume constant state-space transitions, capture average behaviors and the contagion of a closed population, and are sensitive to initial states and parameters. Challenges and opportunities exist in expanding its traditional frameworks to address the specific COVID-19 complexities and challenges in Section \ref{sec:char-compl}. Examples are time-varying, non-IID dynamics and complex couplings between interior and exterior factors related to COVID-19 populations, management groups and contexts. Other important issues include understanding how vaccination and specific vaccines affect coronavirus mutation and discovering the relationships between interior and exterior factors and the resurgence and mutations.

\begin{table*}[htb]
\centering
\caption{Examples of COVID-19 Epidemiological Modeling.}
\small
\resizebox{1.0\textwidth}{!}{
\begin{tabular}{|  p{0.13\textwidth}  | p{0.63\textwidth} | p{0.25\textwidth} |}
    \hline
    \textbf{Objectives} & \textbf{Factors and Settings} & \textbf{Data}  \\ \hline
    Epidemic transition and spread & SIR variants like SIDARTHE with eight phases: susceptible, infected, diagnosed, ailing, recognized, threatened, healed and extinct as well as severe symptoms ~\cite{giordano2020modelling}, variational mode decomposition with shallow regressors \cite{daSilvar20}, etc.  & Case numbers, external data, etc. \\ \hline
    Epidemic dynamics & Time-dependent compartment transmissions, time-varying state transition rates~\cite{chen2020time,peng2020epidemic}  & Case numbers, reporting time information, etc. \\ \hline
    Asymptomatic transmission & Asymptomatic to mild symptoms \cite{weitz2020modeling}, undocumented \cite{Li-sci20}, SHARUCD differing mild and asymptomatic from severe infections \cite{aguiar2020modelling}, undocumented cases \cite{Li-sci20}, epidemiological interventions with serological tests, age-dependent and asymptomatic settings \cite{weitz2020modeling}, etc. & Case numbers, reporting information, symptoms, demographics, etc. \\ \hline
    External factor's epidemic influence  & Age-sensitive SIR model~\cite{chikina2020modeling}, age-structured SIR model with social contacts~\cite{singh2020age}, age-structured SEIRD \cite{Omori-cov20}, public monitoring and detection policies \cite{Yangzf20}, ethnicity, public health conditions and social contacts \cite{Liu12680}, environmental factors~\cite{xie2020association,wang2020temperature,bu2020analysis,oliveiros2020role} & Case numbers, demographics, health conditions, social activities, environmental factors, etc.\\ \hline
    NPI influence  & Lockdown and social distancing \cite{chen2020time}, lockdown \cite{aguiar2020modelling}, quarantine \cite{crokidakis2020modeling}, symptomatic and quarantined infecteds~\cite{maier2020effective}, self-protection and quarantine \cite{peng2020epidemic}, etc.  & Case numbers, NPIs, health conditions, test results, demographics, etc. \\ \hline
    Resurgence  & Second waves~\cite{aravindakshan2020preparing}, wave difference~\cite{fan2020decreased,grech2020covid}, reopening business and social activities~\cite{pedro2020conditions}, time-decaying immunity and easing lockdown and social distancing \cite{lopez2020end}, age sensitivity~\cite{aleta2020age}, NPI influence on future waves~\cite{aravindakshan2020preparing}, travel ban and virus importation \cite{leung2020first} & Case numbers, multi-wave data, NPI and external data, etc. \\ \hline
    Herd immunity  & Compartmental model for simulating `shield immunity' in a population \cite{weitz2020modeling}  & Case numbers, serological tests, etc. \\ \hline
    \end{tabular}
    }
\label{tab:epidmodels}
\end{table*}

\subsection{COVID-19 Medical and Biomedical Analyses}
\label{subsec:medianal}

COVID-19 medical and biomedical analyses reveal the intrinsic and intricate characteristics, patterns and outliernesses of SARS-CoV-2 virus and COVID-19 disease. A wide range of research issues may benefit from such analyses, including but not limited to: COVID-19 infection diagnosis, prognosis and treatment, virology and pathogenesis analysis, potential therapeutics development (e.g., drug repurposing and vaccine development), genomic similarity analysis and sourcing, and contact tracing. In this section, we summarize the medical and biomedical modeling of COVID-19 infection diagnosis and case identification, risk and prognosis analysis, medical imaging analysis, pathological and treatment analysis and drug development. 

\subsubsection{COVID-19 infection diagnosis, test and case identification} 
\label{subsubsec:diaganal}

Given the high transmission and reproduction rates, high contagion, and sophisticated and unclear transmission routes of COVID-19 and its virus strains such as the Delta strain, it is crucial to immediately identify and confirm exposed cases, test positive or negative infections, identify the infected virus variant types, and trace their origins and contacts so as to timely and proactively implement appropriate quarantine measures and contain their potential spread and outbreak~\cite{ng2020evaluation}. This is particularly important during the varying incubation periods which are often asymptomatic to mildly symptomatic yet highly contagious particularly for the virus variants. The SARS-CoV-2 diagnosis and test methods include (1) chemical and clinical methods, typically nucleic acid-based molecular diagnosis and antibody-based serological detection; (2) medical imaging-driven analysis, such as symptom inspection from CXR and CT images; (3) clinical diagnoses and tests like respiratory signal analysis, such as on the abnormal patterns of the lung's ultrasound waves and coughing and breathing signals; and (4) other noninvasive methods such as by involving SARS-CoV-2 and its disease data and external data~\cite{ChauC20,chen2020survey}. Data-driven discovery also plays an increasingly important role in improving COVID-19 diagnosis. Due to the virus and disease complexities, alternative and complementary to the chemical and clinical diagnosis approaches, COVID-19 identification~\cite{udugama2020diagnosing} can benefit from analyzing biomedical images, genomic analysis, symptom identification and discrimination, and external data including social contacts, social activities, mobility and media communications, etc. by data-driven discovery~\cite{dst_Cao15}. 
\begin{itemize}
    \item \textit{Nucleic acid-based diagnosis test} (NAT) \cite{FengW20,AfzalA20,YuCY21} refers to various molecular diagnosis test methods, including non-isothermal amplification (e.g., the real-time reverse transcription polymerase chain reaction (RT-PCR) test, which is the golden standard of COVID-19 diagnosis), isothermal amplification (e.g., CRISPR-based), and sequencing-based tests. Such methods may benefit from modeling techniques including gene and protein sequence analysis and drug-target and virus-host interaction analysis. It is highly sensitive and usable for large-scale operations, but it is expensive as typically it is done using specific test materials and in labs, and is less accurate as it is subject to the varied quality and quantity of specimen collections. The challenges are to reduce its false-negative and false-positive rates supplemented by other diagnosis tools and develop scalable fast test tools. 
    \item \textit{Antibody-based serological diagnosis} \cite{Bastosm2516,LeeC20,PeelingR20} is to detect anti-SARS-CoV-2 immunoglobulins i.e. the antibodies produced in response to COVID-19 infections by validating the specificity and sensitivity of chemiluminescent immunoassays, enzyme-linked immunosorbent assays and lateral flow immunoassays against SARS-CoV-2. It is an alternative or complement to NATs for acute infection diagnosis with easier and cheaper operations at any time. It, however, may produce poor-performing results which are unreliable for decision-making, it may take time to get the results, and it might be difficult for early large-scale diagnosis. There is an urgent need to develop more accurate serological test methods and tools. Machine learning methods such as CNNs could improve test performance e.g. by analyzing the test results, involving external data on patient demographics and clinical results, and integrating various test results \cite{Mendelse2019893118}.
    \item \textit{Clinical diagnosis and analysis} involves clinical reports, domain knowledge and clinicians in identifying  COVID-19-specific symptoms, indications and infections, differentiating them from other similar diseases such as influenza, and confirming positive, negative, severe or fatal conditions. Such diagnoses are conducted by blood tests, cough sound judgment, breathing pattern detection, and external factors by involving external data, etc. AI, machine learning and analytics methods are increasingly being used to classify COVID-19 from other diseases, predict infections, recovery and mortality rates, numbers or timing, etc. \cite{Brinatid20,Khandayam20,BrownC20}. For further discussion, see Section \ref{subsec:machlearning}. 
    \item \textit{Clinical medical imaging analysis} for COVID-19 inspection on COVID-19-sensitive medical images, typically by DNN-based image analysis, can complement the aforementioned chemical and medical methods by detecting abnormal and discriminative symptoms and patterns sensitive to COVID-19 in patient's CXR and CT images. Both typical deep and shallow learning methods are widely applied, which also present inconsistencies and biases in their applications, experiments, results and actionability \cite{RobertsM21}. For further discussion, see Section \ref{subsubsec:medanal}.
    \item \textit{Data-driven prediction} on COVID-19 related data such as blood test results, respiratory signals, and external data that may indicate symptoms, patterns or anomalies of COVID-19 infections. Shallow and deep learning and mathematical modeling methods are applied to classify the symptom types, differentiate COVID-19 infections from other diseases, or detect outliers that may indicate COVID-19 infections. For example, in~\cite{wu2020rapid}, a random forest algorithm-driven assistant discrimination tool extracts 11 top-ranking clinically available blood indices from 49 blood test samples to identify COVID-19 infectives from suspected patients. In~\cite{schuller2020covid}, computer audition is used to recognize COVID-19 patients under different semantics such as breathing, dry/wet coughing or sneezing, and speech during colds, etc. AI4COVID-19~\cite{imran2020ai4covid} combines the deep domain knowledge of medical experts with smart phones to record cough/sound signals as the input data to identify suspect COVID-19 infections with 92.8\% accuracy reported. In~\cite{mukherjee2020shallow}, a shallow LSTM model combines medical information and local weather data to predict the risk level of the country.
\end{itemize}

\textbf{Discussion.} A comparison of the diagnosis methods is shown in Table~\ref{tab:diagmeth}. As commented in various reviews \cite{VandenbergO21,BinnickerM20,TangYW20,Bastosm2516,MinaM20}, COVID-19 diagnosis and tests still suffer from various limitations and challenges. The issues include concerns about result quality, implementation scalability, actionability for determining isolation and quarantine strategies, and trustfulness of accepting medical findings as general clinical specifications. An increasing number of studies appear promising by incorporating advanced data science and AI techniques to complement  medical and chemical test approaches and tools, to integratively enhance preanalytical and postanalytical test results, and strengthen the interpretability and actionability of the results for clinicians, microbiological staff and public health authorities.

\begin{table*}[htb]
\centering
\caption{Examples of Deep COVID-19 Medical Imaging Analysis.}
\small
\resizebox{1.0\textwidth}{!}{
\begin{tabular}{| p{0.22\textwidth}  | p{0.1\textwidth} | p{0.28\textwidth} | p{0.4\textwidth} |}
    \hline
    \textbf{Method} & \textbf{Task} & \textbf{Data} & \textbf{Performance} \\ \hline
    Domain extension transfer learning with pretrained CNN~\cite{basu2020deep}  & Diagnosis & Italian Society of Medical Radiology and Interventional (25 cases)$^a$; Radiopaedia.org (20 cases)$^b$; COVID-19 images (180 cases)~\cite{cohen2020covid}; A Spanish hospital (80 cases)$^c$ & Overall accuracy $90.13\%\pm0.14$ \\ \hline
    Shallow CNN~\cite{mukherjee2020shallow}  & Diagnosis & COVID-19 images~\cite{cohen2020covid} (321 cases); Kaggle non-COVID-19 CXR images (5856 cases)$^d$ & Highest accuracy $99.69\%$, sensitivity 1.0, AUC 0.9995 \\ \hline
    CNN-based truncated InceptionNet~\cite{das2020truncated}  & Diagnosis & COVID-19 images~\cite{cohen2020covid} (162 cases); Kaggle CXR images (5863 cases); Tuberculosis CXR images$^e$ & Accuracy $99.96\%$ (AUC of 1.0) in classifying COVID-19 cases from combined pneumonia and healthy cases; accuracy $99.92\%$ (AUC of 0.99) in classifying COVID-19 cases from combined pneumonia, tuberculosis and healthy CXRs \\ \hline
    DarkCovidNet~\cite{ozturk2020automated} built on Darknet-19 & Diagnosis & COVID-19 images~\cite{cohen2020covid} (127 cases); CXR images~\cite{wang2017chestx} & Accuracy  $98.08\%$ for binary classes and $87.02\%$ for multi-class cases  \\ \hline
    CoroNet built on Xception pre-trained on ImageNet~\cite{khan2020coronet}  & Diagnosis & COVID-19 images~\cite{cohen2020covid}; Kaggle CXR images & Accuracy $89.6\%$, precision $93\%$ and recall $98.2\%$ for COVID vs pneumonia bacterial, pneumonia viral and normal \\ \hline
    COVID-CAPS based on capsule network~\cite{afshar2020covid}  & Diagnosis & COVID-19 images~\cite{cohen2020covid}, Kaggle CXR images & Accuracy $95.7\%$, sensitivity $90\%$, specificity $95.8\%$, and AUC 0.97 \\ \hline
    VGG16 and transfer learning~\cite{nishio2020automatic}  & Diagnosis & COVID-19 images~\cite{cohen2020covid}, RSNA Pneumonia Detection Challenge data$^f$ & Accuracy $83.6\%$ for COVID-19 pneumonia vs non-COVID-19 pneumonia and healthy; sensitivity $90\%$ for COVID-19 pneumonia \\ \hline
    COVID-Net~\cite{wang2020covid}  & Diagnosis & COVID-19 images~\cite{cohen2020covid}, COVID-19 CXR Dataset Initiative$^g$, ActualMed COVID-19 CXR Dataset Initiative$^h$, RSNA data, COVID-19 radiography data$^i$ & Accuracy $93.3\%$, sensitivity $91.0\%$, positive predictive value $98.9\%$ \\ \hline
    Inf-Net with decoder and attention~\cite{cohen2020covid}  & Segment & COVID-19 CT Segmentation$^j$, COVID-19 CT Collection & Dice similarity coefficient 0.739, Sensitivity 0.725, Specificity 0.960  \\ \hline
    DeepPneumonia built on ResNet-50~\cite{song2020deep}  & Diagnosis & Private data & Sensitivity 0.93, AUC 0.99; AUC 0.95 and sensitivity 0.96 for COVID-19 vs. bacteria pneumonia-infections  \\ \hline
    \end{tabular}
    }
\label{tab:imaging}
\footnotesize{$^a$\url{https://www.sirm.org/category/senza-categoria/covid-19/}}; \footnotesize{$^b$\url{https://radiopaedia.org/search?utf8=\%E2\%9C\%93&q=covid&scope=all&lang=us}}; \footnotesize{$^c$\url{https://twitter.com/ChestImaging/status/1243928581983670272}}; \footnotesize{$^d$\url{https://www.kaggle.com/paultimothymooney/chest-xray-pneumonia}}; \footnotesize{$^e$\url{https://ceb.nlm.nih.gov/tuberculosis-chest-X-ray-image-data-sets/}}; \footnotesize{$^f$\url{https://www.kaggle.com/c/rsna-pneumonia-detection-challenge}}; \footnotesize{$^g$\url{https://github.com/agchung/Figure1-COVID-chest-xray-dataset}}; \footnotesize{$^h$\url{https://github.com/agchung/Actualmed-COVID-chest-xray-dataset}}; \footnotesize{$^i$\url{https://www.kaggle.com/tawsifurrahman/covid19-radiography-database}}; \footnotesize{$^j$\url{https://medicalsegmentation.com/covid19/}}.
\end{table*}

\begin{table*}[htb]
\centering
\caption{Comparison of COVID-19 Infection Diagnosis Methods.}
\small
\resizebox{1.0\textwidth}{!}{
\begin{tabular}{| p{0.15\textwidth} | p{0.3\textwidth} | p{0.35\textwidth} | p{0.2\textwidth} |}
    \hline
     \textbf{Methods} & \textbf{Pros} & \textbf{Cons} & \textbf{Data}  \\ \hline
     Nucleic acid-based diagnosis test~\cite{YuCY21} & High sensitivity, suitable for large-scale operation & Preliminary assessment by technicians, professional data analysis, expensive, less accurate, false-negative or false-positive results & Nasal, nasopharyngeal or oropharyngeal swab, aspiration, saliva or wash specimens \\ \hline
     Serological diagnosis~\cite{Bastosm2516} & Easy and cheap to implement, no requirement of experts & Unstable performance, time-inefficient, unscalable for early diagnosis & Serum or plasma samples \\ \hline
     Clinical diagnosis analysis~\cite{Brinatid20} & Diagnosis from mixed clinical reports and tests, on-demand, verifiable by domain experts & Require professional tools and domain knowledge & Blood and respiratory test samples, etc. \\ \hline
     Medical imaging inspection~\cite{RobertsM21} & Fast and automated detection, data-driven analysis & Need trained experts, costly in labeling and early detection, train data scarcity & CT and CXR images    \\ \hline
     Data-driven prediction \cite{chen2020survey} & Algorithmic prediction by data-driven analytics and learning on data relevant to the COVID-19 diagnosis & Biases from data and predictors & Any relevant data including clinical test results and genomic/protein sequences  \\ \hline
    \end{tabular}
 }
\label{tab:diagmeth}
\end{table*}

\subsubsection{COVID-19 patient risk and prognosis analysis} 
\label{subsubsec:prognosisanal}

COVID-19 patient risk assessment identifies the risk factors and parameters associated with patient infections, disease severity, and recovery or fatality to support accurate and efficient prognosis, resource planning, treatment planning, and intensive care prediction. This is crucial for early interventions before patients progress to more severe illness stages. Moreover, risk and prognosis prediction for patients can help with effective health and medical resource allocation when intense monitoring, such as that involving ICU and ventilation and more urgent medical interventions are needed and prioritized. Machine learning models and data-driven discovery can also play a vital role in such risk factor analysis and scoring, prediction, prioritization and planning of prognostic and hospitalization resources and facilities, treatment and discharge planning, and the influence and relation analysis between COVID-19 infection and disease conditions and the external environment and context (e.g., weather conditions and socioeconomic statuses). 

Techniques including mathematical models, and shallow and deep learners are applicable on health records, medical images, and external data. For example, LightGBM and Cox proportional-hazard (CoxPH) regression models incorporate quantitative lung-lesion features and clinical parameters (e.g., age, albumin, blood oxygen saturation, CRP) for prognosis prediction~\cite{zhang2020clinically}, their results showing that lesion features are the most significant contributors in clinical prognosis estimation. Supervised classifiers like XGBoost are applied on electronic health records to predict the survival and mortality rates of severe COVID-19 infectious patients~\cite{yan2020machine,schwab2021real} for the detection, early intervention and potential reduction of mortality of high-risk patients. In~\cite{qi2020machine}, logistic regression and random forest are used to model CT radiomics on features extracted from pneumonia lesions to predict feasible and accurate COVID-19 patient hospital stay, which can be treated as one of the prognostic indicators. Further, shallow and deep machine learning methods are applied to screen COVID-19 infections on respiratory data including lung ultrasound waves, coughing and breathing signals. For example, in \cite{Jiangz20}, a bidirectional GRU network with attention differentiates COVID-19 infections from normal on face-based videos captured by RGB-Infrared Sensors with 83.69\% accuracy. Lastly, external data can be involved for risk analysis; e.g., the work in \cite{Malkiz20} analyzes the association between weather conditions and COVID-19 confirmed cases and mortality.

\subsubsection{COVID-19 medical imaging analysis} 
\label{subsubsec:medanal}

A rapidly growing body of research literature on COVID-19 medical image processing is available, which involves both shallow and deep learning methods especially pretrained CNN-based image nets in learning tasks such as feature extraction, region of interest (ROI) segmentation, infection region/object detection, and disease/symptom diagnosis and classification, etc. Typical COVID-19 medical imaging data includes CXR and CT images of lung (lobes or segments), lesion, trachea and bronchus. The most commonly used DNNs are pretrained or customized CNN, GAN, VGG, Inception, Xception, ResNet, DenseNet and their variants \cite{IslamKAZ21,Shif21}. 

Further, CNN-based transfer learning models, deep transfer learning and GAN are applied on CXR images to detect COVID-19 pneumonia and its segmentation and severity~\cite{cohen2020predicting,minaee2020deep,khalifa2020detection}. On chest CT images, CNNs like ResNet, DenseNet and VGG16 and the inception transfer model are applied to classify COVID-19 infected patients and detect and localize COVID-19 pneumonia and infection regions~\cite{zheng2020deep,Polat21,apostolopoulos2020covid,Wang-cov21,song2020deep}. 

The application of DNNs in COVID-19 medical imaging analysis show significant performance advantages. For example, several references report close-to-perfect prediction performance of pretrained DNNs on CXR images (e.g., achieving accuracy and F-score 100 \cite{Mohamedl20}, AUC 100 \cite{Punn_2020} and 99.97 \cite{maguolo2020critic}, and accuracy and F-score 98 \cite{narin2020automatic}), in contrast to the lower performance of customized networks on CT images (e.g., with accuracy 99.68 \cite{Hasana2020}, AUC 99.4 \cite{Faridaa20} and F-score 94 \cite{Faridaa20,song2020deep}). The highly promising medical imaging analysis results provide strong evidence and support to further case confirmation, medical treatment, hospitalization resource planning, and quarantine, etc. 

Table~\ref{tab:imaging} illustrates various DNNs applied on medical imaging for COVID-19 screening and abnormal infection region segmentation, etc. For example, various CNNs such as shallow CNN, truncated InceptionNet, VGG19, MobileNet v2, Xception, ResNet18, ResNet50, SqueezeNet, DenseNet-121, COVIDX-Net with seven different architectures of deep CNN models, GoogleNet, AlexNet and capsule networks~\cite{das2020truncated,mukherjee2020shallow,basu2020deep,hemdan2020covidx,afshar2020covid,wang2020covid} are applied to analyze CXR images for screening COVID-19 patients, assisting in their diagnosis, quarantine and treatments, and differentiating COVID-19 infections from normal, pneumonia-bacterial and pneumonia-viral infections.

\subsubsection{COVID-19 pathological and treatment analysis and drug development}
\label{subsubsec:pathanal}


The modeling of COVID-19 pathology and treatment aims to characterize virus origin and spread, infection sources, pathological findings, immune responses, and drug and vaccine development, etc. The formulation of molecular mechanisms and pathological characteristics underlying viral infection can inform the development of specific anti-coronavirus therapeutics and prophylactics, which disclose the structures, functions and antigenicity of SARS-CoV-2 spike glycoprotein~\cite{walls2020structure}. The pathological findings pave the way to design vaccines against the coronavirus and its mutations. For example, the higher capacity of membrane fusion of SARS-CoV-2 compared with SARS-CoV is shown in~\cite{xia2020inhibition}, suggesting the fusion machinery of SARS-CoV-2 as an important target of developing coronavirus fusion inhibitors. Further, human angiotensin coverting enzyme 2 (hACE2) may be the receptor for SARS-CoV-2~\cite{ou2020characterization} informing drug and vaccine development for SARS-Cov-2. In~\cite{walls2019unexpected}, a structural framework for understanding coronavirus neutralization by human antibodies can help understand the human immune response upon coronavirus infection and activate coronavirus membrane fusion. The kinetics of immune responses to mild-to-moderate COVID-19 discloses clinical and virological features~\cite{thevarajan2020breadth}. 
Data-driven analytics are applied in COVID-19 virology, pathogenesis, genomics and proteomics and collecting pathological testing results, gene sequences, protein sequences, physical and chemical properties of SARS-CoV-2, drug information and its effect, together with their domain knowledge. This plays an important role in discovering and exploring feasible drugs and treatments, drug discovery, drug repurposing, and correlating drugs with protein structures for COVID-19 drug selection and development. For example, a pre-trained MT-DTI (molecule transformer-drug target interaction) deep learning model based on the self-attention mechanism identifies commercially available antiviral drugs by finding useful information in drug-target interaction tasks~\cite{beck2020predicting}. The GAN-based drug discovery pipeline generates novel potential compounds targeting the SARS-CoV-2 main protease in \cite{Zhavoronkova20}. In~\cite{zhavoronkov2020potential}, 28 machine learning methods including generative autoencoders, generative adversarial networks, genetic algorithms, and language models generate molecular structures and representations on top of generative chemistry pipelines and optimize them with reinforcement learning to design novel drug-like inhibitors of SARS-CoV-2. Further, multitask DNN screens candidate biological products ~\cite{hu2020prediction}. In \cite{metsky2020crispr,metskyc20}, CNN-enabled CRISPR-based surveillance supports a rapid design of nucleic acid detection assays. 

For genome and protein analysis, frequent sequential pattern mining identifies frequent patterns of nucleotide bases, predicts nucleotide base(s) from their previous ones, and identifies the genome sequence locations where nucleotide bases are changed \cite{Nawazm21}. In \cite{Alakust21}, a bidirectional RNN classifies and predicts the interactions between COVID-19 non-structural proteins and between the SARS-COV-2 virus proteins and other human proteins with an accuracy of 97.76\%.

Classic and deep machine learning methods such as classifiers SVM and XGBoost, sequence analysis, multi-task learning, deep RNNs, reinforcement learning such as deep Q-learning network, and NLP models are applied to SARS-COV-2 therapy discovery, drug discovery, and vaccine discovery \cite{Arshadik20}. Examples are the rule-based filtering and selection of COVID-19 molecular mechanisms and targets; virtual screening of protein-based repurposed drug combinations; identifying the links between human proteins and SARS-COV-2 proteins; developing new broad-spectrum antivirals, and molecular docking; identifying functional RNA structural elements; discovering vaccines such as predicting potential epitopes for SARS-COV-2 and vaccine peptides by LSTM and RNNs, and analyzing protein interactions, molecular reactions by neural NLP models such as Transformer variants. Table \ref{tab:drugvaccinedev} briefly illustrates the applications of modeling in supporting COVID-19 treatments and drug and vaccine development.

\textbf{Discussion.} Most of the literature on COVID-19 medical and biomedical analytics directly applies existing mathematical models, shallow and pretrained deep models. There are gaps and opportunities in characterizing COVID-19-specific characteristics and domain knowledge into tailored modeling and training deep neural networks on usually small and quality-limited COVID-19 data and involving multimodal COVID-19 data to discover more informative medical and biomedical insights.  

\begin{table*}[htb]
\centering
\caption{Examples of Modeling for COVID-19 Treatment and Drug/Vaccine Development.}
\small
\resizebox{1.0\textwidth}{!}{
\begin{tabular}{| p{0.1\textwidth} | p{0.7\textwidth} | p{0.2\textwidth} |}
    \hline
    \textbf{Objective} & \textbf{Approaches} & \textbf{Data}  \\ \hline
     Treatment & Data-driven diagnosis-informed treatment e.g. pathological analysis, medical imaging analysis, immune reaction, genomic and proteomic analysis \cite{IslamKAZ21,zhang2020clinically,IslamKAZ21,Shif21} & Pathological, clinical, virological, genomic, proteomic data  \\ \hline
     Drug development & Correlating drugs with protein structures and molecule transformer for drug-target interactions \cite{beck2020predicting}, DNNs like GANs and multitask DNNs for drug discovery \cite{hu2020prediction,Zhavoronkova20}, machine learning and language models to generate molecular structures and drug-like inhibitors \cite{zhavoronkov2020potential}  &  Virological, genomic, proteomic data \\ \hline
     Vaccine development & Sequence analysis and sequential modeling like LSTM and RNN variants and NLP models like Transformer variants for functional RNA structures, vaccine epitopes and peptides, protein interactions and molecular reactions \cite{Arshadik20} & Genomic and proteomic data \\ \hline
\end{tabular}
}\label{tab:drugvaccinedev}
\end{table*}

\section{COVID-19 Influence and Impact Modeling}
\label{sec:impactmodeling}

COVID-19 has had an unprecedented and overwhelming influence and impact on all aspects of our life, society and economy, posing significant health, economic, environmental and social challenges to the entire world and human population~\cite{chakraborty2020covid}. Over 3k references of the 22k literature involve the topic of influence and impact modeling. In this section, we review and summarize the modeling and analysis methods and results on many broad areas affected by SARS-CoV-2 and COVID-19. These include the modeling of the effect of COVID-19-sensitive NPIs and the COVID-19 healthcare, psychological, economic and social influence and impact.

\subsection{Modeling COVID-19 Intervention and Policy Effect}
\label{subsec:NPIeffect}

On one hand, pharmaceutical measures, drug and vaccine development play fundamental roles \cite{zheng2020recommendations}. On the other, to control the outbreaks of COVID-19 and its further influence on various aspects of life, governments adopt various NPIs such as travel restrictions, border control, business and school shutdown, public and private gathering restrictions, mask-wearing, and social distancing. 
For example, travel bans and lockdown are issued to decrease cross-boarder population movement; social distancing and shutdowns minimize contacts and community spread; schooling closures and teleworking reduce  indoor gatherings and workplace infections. Although these control measures flatten the curve, they also undoubtedly change the regular mobility and activities of the population, normal business and economic operations, and the usual practices of our daily businesses. 

A critical modeling issue is to characterize, estimate and predict how such NPIs influence COVID-19 epidemic dynamics, infection spread, case development, and population structure including deceased, medical resource and treatment allocation, and human, economic and business activities. Accordingly, various modeling tasks involve epidemiological, statistical and social science modeling methods and their hybridization (typically stochastic compartmental models) to evaluate and estimate the effects, typically by aligning the NPIs with case numbers for correlation and dependency modeling. Below, we summarize a few aspects of NPI influence.

\textit{Modeling the effect of NPIs on COVID-19 epidemic dynamics.}
This typically models the correlations between COVID-19 cases and NPIs, the NPI influence on COVID-19 epidemic factors including transmission rate and case numbers, and the NPI influence on improving recovery rates and lowering death rates. 
Various SIR and statistical modeling variants evaluate the effects of such control measures and their combinations on containing the virus spread and controlling infection transmission (e.g., per transmission rate) and estimate the corresponding scenarios (distributions) of case number development~\cite{tian2020investigation,dehning2020inferring,brauner2020inferring}. For example, in \cite{peng2020epidemic}, a generalized SEIR model includes the self-protection and quarantine measures to interpret the publicly released case numbers and forecast their trend in China. The effect of control measures, including city lockdowns and travel bans implemented in the first 50 days in Wuhan and their effect on controlling its outbreak across China in terms of infection case numbers estimated by an SEIR model before and after the controls is described in \cite{tian2020investigation}. 

Often, various NPIs are jointly implemented to contain a COVID-19 epidemic. It may be reasonable that multiple NPIs cooperatively reduce the epidemic effective reproduction number~\cite{brauner2020inferring,lai2020effect,Flaxmans20,prem2020effect}. In \cite{brauner2020inferring}, a temporal Bayesian hierarchical model incorporates auxiliary variables describing the temporal implementation of NPIs, which infers the effectiveness of individually (estimated 13\% to 42\% reduction of reproduction number) and conjunctionally (77\% reduction of reproduction number) implementing NPIs such as staying-at-home, business closures, shutting down educational institutions and limiting gathering sizes in terms of their influence on the reproduction number. In \cite{Flaxmans20}, a hierarchical Bayesian model infers the impact and effectiveness of NPI (including case isolation, school closure, mass gathering ban, social distancing) on the infections, reproduction number $R_0$, effect sizes of population, and death tolls in 11 European countries and suggests continued interventions to keep the epidemic under control. 

\textit{Modeling NPI influence on public resources including healthcare systems.}
The implementation of NPIs affects the demand, priority and effectiveness of anti-pandemic public health resources and the planning and operations of healthcare systems. For example, in \cite{Fangyq20}, an SEIR model and a polynomial regressor simulates the effect of early detection, isolation, treatment, adequate medical supplies, hospitalization and therapeutic strategy on COVID-19 transmission, in addition to estimating the reproductive number and confirmed case dynamics. The SIDARTHE model \cite{giordano2020modelling} simulates  possible scenarios and the necessity of implementing countermeasures such as lockdowns and social distancing together with population-wide testing and contact tracing to rapidly control the pandemic. The SHARUCD model \cite{aguiar2020modelling} predicts the COVID-19 transmission response (in terms of infection cases, growth rate and reproduction number) to the control measures including partial lockdown, social distancing and home quarantining and differentiates asymptomatic and mild-symptomatic from severe infections, which could inform the prioritization of healthcare supplies and resources.

\textit{Modeling NPI influence on human activities.}
This explores the relations between COVID-19 NPIs and human mobility, travel, and social and online activities. For example, in \cite{Kraemerm20}, the alignment between human mobility and case number development in Wuhan and China presents the effect of travel restrictions on case reduction and COVID-19 spread. In \cite{Keelingm20}, a simple SEIR model analyzes the tracing contacts in UK social network data, estimates the scenarios of COVID-19 infection control and subsequent untraced cases and infections, and shows the efficacy of close contact tracing in identifying secondary infections. 
In \cite{gatto2020spread}, MCMC parameter estimation and a metacommunity Susceptible–Exposed–Infected–Recovered (SEIR)-like disease transmission model shows the need for planning emergency containment measures such as restrictions on human mobility and interactions to control COVID-19 outbreak (by 42\% to 49\% transmission reduction). In \cite{Grantzk20}, mobile phone data is collected and analyzed to inform COVID-19 epidemiologically relevant behaviors and response to interventions. Weitz et al.~\shortcite{weitz2020modeling} develop and analyze an epidemiological intervention model that leverages serological tests to identify and deploy recovered individuals as focal points for sustaining safer interactions by interaction substitution, developing the so-called `shield immunity' at the population scale. 
It is shown that the change of contact patterns could dramatically decrease the probability of infections and reduce the transmission rate of COVID-19~\cite{zhang2020changes,feehan2021quantifying,latsuzbaia2020evolving}. 

\textbf{Discussion.} The many diverse applications of SIR-based modeling of COVID-19 invention and policy effects enable an epidemiological explanation. Such methods assume each NPI independently acts on  case movement. This leaves open issues including characterizing the effectiveness of individual NPIs by assuming they are coupled with each other and cooperatively contribute to flatten the curves; and exploring the interactions between NPIs, case development, and external factors including people's behaviors and environmental factors without disentangling them (opposite to the method of DNNs-based decoupled, homogeneous and independent representations and learning).

\subsection{Modeling COVID-19 Psychological Impact}
\label{subsec:psychimpact}

A common concern is the influence of COVID-19  on individual and public psychological and mental health~\cite{xiong2020impact}. Typical tasks are to characterize, classify and predict  social-media-based individual and public emotion and sentiment and their mental health. These may be sensitive to the COVID-19 outbreak, health and medical mitigation, NPI measures, government governance, public healthcare system performance, vaccine, resurgence and coronavirus mutations, and the `new normal' including working from home and online education, etc. The data involved are from social media and networks such as Twitter, Facebook, Wechat, Weibo, YouTube, Instagram and Reddit; online news feeds, discussion boards, blogs and Q/A; and instant messaging such as mobile messaging and apps. 

Negative sentiments \cite{WangLCZ20,Nemesl21}, opinion and topic trends, online hate speech \cite{VishwamitraH0CC20}, psychological stress, men's and women's worries \cite{VegtK20}, responsive emotions \cite{HouZY2020} and behaviors and events \cite{Chenaghlu20} can be characterized, clustered or classified on short and long texts by simply applying NLP processing techniques. Examples are extracting TF-IDF and part-of-speech features, shallow NLP and text analysis models including BOW and latent Dirichlet allocation (LDA), and neural text modelers including DNN variants such as BioBERT, SciBERT and Transformer variants on the word, sentence or corpus level. For example, in \cite{xiong2020impact}, the preferred reporting items for systematic reviews and meta-analyses guidelines are used to review the COVID-19 impact on public mental health, disclosing the extent of symptoms and risk factors associated with anxiety (6.33\% to 50.9\%), depression (14.6\% to 48.3\%), posttraumatic stress disorder (7\% to 53.8\%), psychological distress (34.43\% to 38\%) and stress (8.1\% to 81.9\%) in the surveyed population of 8 countries. 

\textbf{Discussion.} The existing modeling of COVID-sensitive psychological influence often misses psychological knowledge because it is purely driven by data; the analytical results are based on a cohort of infected people owing to its anonymous nature; no work is reported on fusing various sources of data including online misinformation to infer the predominant drivers of specific mental stress such as vaccination hesitation; and the targeted analysis of specific mental issues in vulnerable groups, such as COVID-driven teenage suicide and racism.


\subsection{Modeling COVID-19 Economic Impact}
\label{subsec:ecoimpact}

The COVID-19 pandemic has incurred overwhelming and devastating impact on regional and global economy and business activities including trade, tourism, education exchange, logistics, supply chain, workforce and employment. It seems that no economy on the interconnected globe is immune from the negative consequences of COVID-19~\cite{chudik2020economic}. A critical modeling task is to quantify how COVID-19 influences various aspects of the economy and businesses, how to manage and balance COVID-19 control measures (including NPIs and vaccination rollouts) and government relief and recovery programs, and how to sustain and recover business and economic activities without seriously suffering from uncontrollable outbreaks and resurgences for better sustainability in the COVID new normal.

\textit{Modeling the COVID-19 impact on economic growth.}
A rapidly growing body of research investigates the heterogeneous, non-linear and uncertain macroeconomic effects of COVID-19 across regions and sectors in individual countries, as well as on a global scale. It is estimated that COVID-19 and SARS-CoV-2 may cause over 2\% monthly GDP loss and a 50\% to 70\% decline in tourism \cite{chakraborty2020covid}. 
In \cite{Pichlera20}, a sectoral macroeconomic model analyzes the short-term effects of intervention measures such as lockdown, social distancing and business reopening on economic outcomes such as production network, supply and demand, inventory dynamics, unemployment and consumption and estimates their influence on the relations between reproduction number and GDP. 
The study in \cite{walker2020impact} illustrates the relations between a country's income levels, public healthcare availability and capacity and the COVID-19 infected patient's demography and social patterns in low- to middle-income countries.

\textit{Modeling the COVID-19 impact on workforce and sustainability.}
COVID-19 drives the new normal of working, including a hybrid work mode, cloud-based enterprise operations, the shift from centralized infrastructures (including IT) to cloud-based ICT and home-based workplaces, and new ways of ensuring sustainability including engaging and supporting clients through online operations and services and AI-enabled cost-effective planning, production, logistics and services. 
In \cite{Baolf21}, the descriptive statistics of the daily activities of Baidu developers show the positive and negative impacts of working from home on developer productivity, particularly on large and collaborative projects. 
The survey conducted in \cite{Myersk20} shows the various impacts of COVID-19-driven work from home on the scientific workforce, including the time spent on work, parenting distraction, and impact on laboratory-based projects.
The analysis in \cite{Lijj20} in Australia shows the impact of government welfare support responses to COVID-19-infected people and businesses on mitigating potential unemployment, poverty and income inequality and the sustainability of such support measures.

\textbf{Discussion.} The existing modeling objectives, tasks and methods are highly preliminary, specific and limited. Expectations include macro-, meso- and micro-level modeling of economic impact by involving their economic-financial variables and activities, contrastive analysis with similar historical events and periods, and data-driven discovery of insights for a sustainable tradeoff between mitigation and economic growth in the new normal, to name a few.

\subsection{Modeling COVID-19 Social Impact}
\label{subsec:socimpact}

The COVID-19 pandemic has had significant impact on public health, welfare, social, political and cultural systems, including restricting human activities, affecting people's well-being, causing an overwhelming burden on public health systems, reshaping sociopolitical systems, and disturbing social regularity such as incurring online information disorder. This section reviews the relevant modeling work on such social impacts.

\textit{Modeling the COVID-19 influence on human behaviors.}
In addition to the COVID-19-sensitive NPI influence on human activities as discussed in Section \ref{subsec:NPIeffect}, SARS-CoV-2 and COVID-19 have fundamentally reshaped people's social activities and habits. For example, Baidu-based daily transportation behaviors and simple statistics were collected which show high-level mobility patterns such as visiting venues, origins, destinations, distances, and transport time during the COVID-19 epidemic in China \cite{HuangWFZSL20}. 
In \cite{Grantzk20}, large-scale mobile phone data such as call detail records, GPS locations, Bluetooth data and contact tracing apps are collected and analyzed by off-the-shelf tools to extract statistic metrics and patterns of behaviors, mobility and interactions. The results may inform population behaviors, individual contacts, movement paths and mobility patterns, and networking, in addition to evaluating the effectiveness of NPIs and informing COVID-19 responses such as contact tracing. 
In \cite{HouZY2020}, social media data from Sina Weibo, the Baidu search engine, and 29 Ali e-commerce marketplaces were collected and analyzed using keyword-based linguistic inquiries and statistics like word frequencies and Spearman’s rank correlation coefficient analysis. Keywords are extracted to show people's behavioral responses to COVID-19 outbreaks, public awareness and attention to COVID-19 protection measures, concerns about misinformation and rumors about ineffective treatments, and the correlation between risk perception and negative emotions. 

\textit{Modeling the COVID-19 influence on public health systems.} 
The sudden COVID-19 endemic or pandemic and its mysterious resurgence has resulted in the imperative, nonscheduled and overwhelming rationing demand on healthcare and medical professionals, public health and medical resources and supplies including oxygen, hospital beds and facilities, ICU facilities and ventilators, medical waste processing equipment, hygiene protection equipment such as medical masks and sanitization chemicals, and intervention materials and devices. How to plan, prioritize, ration and manage these resources, assess their supply/demand and effects to prioritized hotspots and regions and optimize their reorganization per local and global needs and population-wide well-being are some challenging issues to model and optimize. In \cite{Emanuele20}, recommendations are made to allocate medical resources to both COVID-19 and non-COVID-19 patients to maximize benefits, prioritize health workers, avoid a first-come, first-served approach, in a way that is evidence-based and involves science and research.

\textit{Modeling the COVID-19 influence on sociopolitical systems.} 
The COVID-19 influence on social and political systems is unprecedented. This influence extends to the confidence and trust in existing sociopolitical systems such as public and moral values, national interests, social welfare systems, human services, political relations, globalization, scientific exchange and collaborations, science-driven epidemic mitigation policies and strategies, and the impact on social governance and disaster management. 
For example, in \cite{Bullockj20}, an identity fusion theory-based online sampling and a moral foundations theory-based computer simulation show the correlations between nationalism, religiosity, and anti-immigrant sentiment from a socio-cognitive perspective during the COVID-19 pandemic in Europe. 
The surveys undertaken in \cite{kreps2020model} show that the scientific uncertainty of  COVID-19-oriented modeling and findings affect the public and political trust in science-based policy making in the US and suggest more careful science communications. 
The work in \cite{Shresthan20} evaluates the impact of COVID-19 on globalization and global health, in particular, mobility, trade, travel, event management, food and agriculture, and a pandemic vulnerability index quantitatively measures the potential impact on global health and the countries most impacted.

\textit{Modeling misinformation and disorder in the COVID-19 infodemic.} 
The COVID-19 infodemic has been accompanied by a large volume of misinformation (partially or entirely inaccurate or misleading information), biased (polarized), questionable or unverified information, rumor and propaganda. Such information is harmful for correctly understanding, recognizing, intervening, and preventing the COVID-19 pandemic. Its diffusion is usually fast, its spread is often wide, and its impact is typically devastating. Modeling the COVID-19 misinformation and information disorder involves tasks such as detecting and ranking misinformation, classifying them, undertaking fact checks and cross-references, tracing their sources and transmission paths, discovering their diffusion and propagation networks and paths, and estimating their effects on the COVID-19 epidemic spread and control.
For example, in \cite{Cinellim20}, skip-gram is used to represent the words collected from Twitter, Instagram, YouTube, Reddit and Gab; the converted vector representations are then clustered by partitioning them around medoids and cosine distance-based similarity analysis to extract the topics of concern. An SIR model is then applied to estimate the basic reproduction number of the social media-based COVID-19 infodemic. A comparative analysis then estimates and compares the platform-dependent interaction patterns, information spread (w.r.t. reproduction rate), questionable and reliable information sourcing and differentiation, and rumor amplification across the above platforms. In \cite{Wangy21}, SVM classifies credible and misinformation from Twitter texts and a correlation analysis shows the predominant credible information on wearing masks and social distancing can lead their misinformation with a time lag. In \cite{Agleyj21}, bivariate (ANOVA) and multivariate logistic regression identifies similar belief profiles of political orientation, religious commitment, and trust in science in survey-based narratives and compares the profiles of those who are disinformed or conspiratorial with scientific narratives. Further, the statistics on Weibo tweets show the COVID-19 misinformation evolution related to topics and events such as city lockdowns, cures, preventive measures, school reopening and foreign countries, the bias involving cures and preventive methods, and sentiment evolution such as fear of specific topics \cite{LengZSWSSZCD21}. The work in \cite{MicallefHKAM20} applies SVM, logistic regression and BERT to classify COVID-19 misinformation and counter-misinformation tweets, characterizes the type, spread and textual properties of counter-misinformation, and extracts the user characteristics of the citizens involved.

\textbf{Discussion.} Typical research on COVID-19 influence and impact modeling only involves local and regional COVID-19 data and their affected objects, hence the resultant conclusions are limited in the ability to indicate their applicability to general practice and broad pandemic control. More robust results are expected to inform medical and public health policy-making on medication, business and society. No-to-rare outcomes are available on how NPIs influence the threshold and effects of COVID-19 vaccinations and herd immunity and on how to balance NPIs and  economic and social revivification. It is difficult to find actionable evidence and guidelines on what policies should be implemented and what tradeoff is appropriate in balancing a COVID-19 outbreak and containing resurgence with economic and social business recovery.

\begin{table*}[htb]
\centering
\caption{Examples of COVID-19 Influence and Impact Modeling.}
\small
\resizebox{1.0\textwidth}{!}{
\begin{tabular}{|  p{0.12\textwidth}  | p{0.15\textwidth} | p{0.43\textwidth} | p{0.3\textwidth} |}
    \hline
    \textbf{Aspects} & \textbf{Objectives} & \textbf{Approaches}  & \textbf{Data}  \\ \hline
    \multirow{3}{*}{NPI effect} & on epidemic dynamics & SIR variants, statistical models, Bayesian hierarchical models, etc. \cite{tian2020investigation,dehning2020inferring,brauner2020inferring,peng2020epidemic,lai2020effect,Flaxmans20,prem2020effect} & COVID-19 case data, NPIs \\ \cline{2-4}
    & on public resources & SIR variants, statistical models, polynomial regressors, etc. \cite{Fangyq20,giordano2020modelling,aguiar2020modelling} & Case data, NPIs, public resource (incl. healthcare) data \\ \cline{2-4}
    & on human behaviors & SIR variants, statistical models e.g. MCMC, relation modeling, etc. \cite{Kraemerm20,Keelingm20,gatto2020spread,Grantzk20,weitz2020modeling,zhang2020changes,feehan2021quantifying,latsuzbaia2020evolving} & Case data, NPIs, human activities (incl. mobile phone data and mobility), etc. \\ \hline
    \multirow{3}{0.12in}{\\ Psychological influence} & on individual mental health & Psychology, systematic reviews, classic and neural NLP models e.g. BOW, LDA, SciBERT, Transformer variants, etc. \cite{xiong2020impact,WangLCZ20,Nemesl21,VegtK20,Chenaghlu20} & Identity, social media data, news feeds, Q/A, surveys, instant messaging, behaviors, NPIs, etc.  \\ \cline{2-4}
     & on public mental health  & Psychology, systematic reviews, classic and neural NLP models e.g. BOW, LDA, SciBERT, Transformer variants, etc. \cite{xiong2020impact,WangLCZ20,Nemesl21,VishwamitraH0CC20,HouZY2020} & Social media data, news feeds, Q/A, surveys, instant messaging, public emotion, activities and events, NPIs, etc.  \\ \cline{2-4}
    & on mental health  & Psychology, systematic reviews and meta-analyses, classic and neural NLP models, statistics, etc. \cite{xiong2020impact}  & Social media data, questionnaires, instant messaging, behavior and events, NPIs, etc. \\ \hline
    \multirow{2}{0.12in}{\\Economic impact} &  on economic growth & Time series analysis, descriptive analytics, macroeconomic modes, relational models, etc. \cite{chakraborty2020covid,Pichlera20,walker2020impact} & Economic data, case data, NPIs, etc. \\ \cline{2-4}
    & on workforce and sustainability & Descriptive analytics, time-series analysis, relational models, etc.  \cite{Baolf21,Myersk20,Lijj20} & Work and sustainability-related data, performance, employment, surveys, social welfare data, etc. \\  \hline
    \multirow{4}{*}{Social impact} & on human behaviors & Descriptive analytics, pattern analysis, social media/network analysis, NLP models, etc. \cite{HuangWFZSL20,Grantzk20,HouZY2020} & Public, online and household activities, gathering, mobility data,  mobile phone data, social media data, etc. \\ \cline{2-4}
    & on public health systems & Descriptive analytics, relational models, etc. \cite{Emanuele20}  & Public health and medical data, public hygiene data, case data, etc. \\ \cline{2-4}
    & on misinformation & Classifiers, classic and neural NLP models, social media/network analysis, sentiment/topic modeling, time-series analysis, outlier detection, etc. \cite{Cinellim20,Wangy21,Agleyj21,LengZSWSSZCD21,MicallefHKAM20} & Social media data, news feeds, Q/A, cross/fact-check, etc. \\ \cline{2-4}
    & on socio-political systems & Descriptive analytics, sociopolitical methods, survey analysis, etc. \cite{Bullockj20,kreps2020model,Shresthan20} & Social and political data, case data, surveys, questionnaires, sociopolitical events, etc. \\  \hline
    \end{tabular}
    }
\label{tab:influimpactdmodels}
\end{table*}

\section{COVID-19 Simulation Modeling}
\label{sec:simulation}

Despite being a small focus (over 1.5k of the 22k publications on modeling), simulation is an essential means to understand, imitate, replicate and test the working mechanisms, the epidemic transmission processes, the evolution and mutation of COVID-19 and its virus SARS-CoV-2, the interactions and self-organization between factors, the effect of mitigation measures and various interior and contextual factors, and resource planning and optimization such as  healthcare resource allocation. Typical simulation methods include dynamic systems, state-space modeling, discrete event simulation, agent-based modeling, reinforcement learning, Monte-Carlo simulation, and hybrid simulation~\cite{currie2020simulation}. Below, we summarize the relevant work on simulating the COVID-19 epidemic evolution and the effect of interventions and policies on COVID-19 epidemic development.

\textit{Simulating the COVID-19 epidemic evolution.}
One important but unclear question is how does the COVID-19 evolve over time in the community. 
What-if analyses can be applied to estimate infection case numbers and their evolution under various hypotheses tests \cite{zhou2020semiparametric}. Typical methods include SIR variants, statistical and mathematical models, e.g., introducing control measure-sensitive variables into such models to estimate their effects on infections, reproduction number, transmission rate, and outbreak control after implementing or relaxing certain interventions. For example, in \cite{Fongsj20}, composite Monte Carlo simulation conducts the what-if analysis of future COVID-19 epidemic development possibilities on top of the estimation made by a polynomial neural network on COVID-19 cases, then fuzzy rule induction outputs decision rules to inform epidemic growth and control.
In \cite{Gomezj21}, an agent-based simulation system simulates a COVID-19 patient's demographic, mobility and infectious disease state (susceptible, exposed, seriously-infected, critically-infected, recovered, immune and dead) information and their dynamic interactions between each other (agents, i.e., people in epidemiology) in certain environments (home, public transport stations, and other places of interest), and evaluates the effect of adjusting individual and social distancing (separation) on  epidemics (e.g., numbers in each state).

\textit{Simulating the policy effects on the COVID-19 epidemic.}
Another important task is to simulate how interventions, interior and external factors, and other policies and control measures of interest influence the dynamics of the COVID-19 epidemic. 
For example, a discrete-time and stochastic agent-based simulation system (Australian Census-based Epidemic Model)~\cite{chang2020modelling} incorporates 24 million software agents, where each agent mimics an Australian individual in terms of their demographics, occupation, immunity and susceptibility to COVID-19, contact rates in their social contexts, interactions, commuting and mobility patterns, and other aspects, which are informed by census data from the Australian government. The system evaluates various scenarios by adjusting the level of restrictions on case isolation, home quarantine, international air travel, social distancing and school closures and their effects on COVID-19 pandemic consequences in terms of the reproductive number, the generation period, the growth rate of cumulative cases, and the infection rate for children. The simulation provides evidence to help the Government understand how COVID-19 is transmitted and what policies should be implemented to control COVID-19 in Australia. In \cite{maier2020effective}, an SIR model is extended by adding variables reflecting symptomatic infections and the quarantine of susceptibles, which then estimates the case development distribution as subexponential after implementing the quarantine. In~\cite{ye2020alpha}, an attributed heterogeneous information network incorporates the representations of external information about the COVID-19 disease features, the  population's demographic features, mobility and public perception of sentiment into a GAN model, which then assesses the hierarchical community-level risks of COVID-19 to inform interventions and minimize disruptions. 

\textbf{Discussion.} Although we mention many aspects and questions that could be (better) addressed by simulation, very limited research is available in this direction. In addition to the above two aspects closely relevant to COVID-19 epidemic dynamics, other important topics include simulating the mutation and resurgence of the coronavirus and COVID-19 in communities with different social, ethnic and economic conditions; the influence of individual and compound COVID-19-sensitive policies on social, economic and psychological aspects; and the tradeoff between the strength and width of mitigation strategies and their impact on society and the economy.

\section{COVID-19 Hybrid Modeling}
\label{sec:hybridmodeling}

Hybrid COVID-19 modeling can be categorized into the following families: (1) multi-objective modeling: to address multiple problems and multiple business and modeling objectives at the same time, such as jointly understanding COVID-19 epidemic dynamics and the corresponding effective NPI policies; (2) multi-task modeling: to handle multiple modeling tasks, e.g., simultaneously forecasting daily confirmed, death and recovered case numbers; (3) multisource (multimodal etc.) COVID-19 data modeling: to involve multiple sources of internal and external data for modeling, e.g., supplementing environmental and demographic data with case numbers and complementing case numbers with medical imaging and social mobility data; (4) hybrid methods for COVID-19 modeling: typically by sequentializing (i.e., multi-phase) or parallelizing multiple tasks or methods from different disciplines and areas, e.g., integrating statistical methods, shallow or deep learning methods, and evolutionary computing methods into compartmental models; and (5) hybrid modeling with multi-methods from various disciplines on multisource COVID-19 data for multi-objective or multi-task modeling.

\textit{COVID-19 multi-objective modeling} is commonly seen in COVID-19 modeling, as shown in Sections \ref{sec:mathmethods}-\ref{sec:simulation}, where, multiple business problems and learning objectives are involved in one research or case study. Examples are forecasting COVID-19 transmission and its sensitivity to external factors such as the patients' age groups, hygiene habits and environmental factors; modeling the influence of NPIs and people's ethnic conditions on case movements; modeling the influence of NPIs on both case trends and public psychological health; and  survival/mortality rate estimation and the influence analysis of dependent factors such as the patients' health conditions. Typical methods include multivariate analysis, probabilistic compartmental models, simulation systems, multi-objective evolutionary learning methods, and DNN variants. For example, in \cite{Poirier-cov20}, a regression model estimates the relations between reproduction number and environment factors and human movements. In \cite{dehning2020inferring}, Bayesian inference of an SIR model infers the effect of various interventions on new infections. In \cite{pedro2020conditions}, an SEIR models the relations between case trends and epidemic conditions, socioeconomic effect, and interventions. In \cite{xiong2020impact}, systematic reviews and meta-analyses review the work on the relations between COVID-19 symptom severity, risk factors and public emotions. 

\textit{COVID-19 multisource data modeling} serves various purposes such as predicting COVID-19 epidemic spread and transmission, medical diagnosis and treatment, and government and community interventions by combining data from respective modalities, sources or views. 
Examples of multisource data are combining COVID-19 case numbers with NPI data; people's demographics, health conditions, mobility, social and business activities, social networking and media information; health and medical records, diagnosis information, treatments, pharmaceutical interventions, and pathological tests; social and public activities and events, economic data, and sociopolitical data; and online, social media and mobile apps-based messaging, news, Q/A, and discussion groups.
Typical methods include data fusion-based learning, mixed representations-based learning, clustering and classification on mixed data types, DNN variants, etc. \cite{Kanghy20}.
For example, a novel variational-LSTM autoencoder model in \cite{Ibrahimm21} predicts the coronavirus spread in various countries by integrating historical confirmed case numbers with urban factors (about location, urban population, population density, and fertility rate) and governmental measures and responses (school, workplace and public transport closures, public events cancellation, contact tracing, public information campaigns, international travel controls, fiscal measures, and investment in health care and vaccines). In \cite{Malkiz20}, COVID-19 case numbers and weather data are combined to analyze the correlation between COVID-19 confirmed cases and mortality and weather factors.
NLP methods can extract and analyze related news, which are then input to LSTM networks to update the infection rate in a susceptible–infected epidemic model~\cite{zheng2020predicting}, which shows to beat the susceptible–infected epidemic model and its combination with LSTM. In~\cite{soures2020sirnet}, coupling LSTM with an epidemic model forecasts COVID-19 spread on case data, population density and mobility.

\textit{COVID-19 hybrid methods} integrate various methods for single or multiple-objective/task/source learning. In addition to ensemble learning by integrating the results from multiple learners such as ensemble trees and XGBoost, often multiple methods are sequentially involved to learn specific tasks or data over phases; other common tasks are to integrate compartmental models with other methods such as statistical models, classifiers and DNNs for the improved forecasting of COVID-19 epidemic dynamics and attributes. For example, in \cite{Zhanggp21}, a hybrid model predicts the infected and death cases by integrating a genetic algorithm to optimize infection rates and integrating LSTM for parameter optimization into a modified susceptible-infected-quarantined-recovered (SIQR) epidemic model. In \cite{chakraborty2020real}, a regression tree combined with wavelet transform predicts COVID-19 outbreak and assesses its risk on case numbers. In \cite{Bardan20}, a baseline method generates a granular ranking (discrimination) of severe respiratory infection or sepsis on the medical records of the general population, then a decision-tree-based gradient boosting model adjusts the former predicted results in subpopulations by aligning it with the published aggregate fatality rates. 
In addition to the aforementioned methods, other methods and tasks e.g. for innovative pandemic responses are available in the literature. Examples are automated primary care tools to alleviate the shortage of healthcare workers~\cite{tavakoli2020robotics}, expert systems and chatbots for symptom detection and lessening the mental health burden~\cite{miner2020chatbots}, IoT and smart connecting tools to prevent outbreaks, remotely monitoring patients, and prompting enforcement of guidelines and administrative orders to contain future outbreaks~\cite{gupta2021future}.

\textbf{Discussion.} Though various methods of hybridization have been summarized in this section, the relevant research is not systematic, comprehensive, or substantial. This observation applies to hybrid data, hybrid tasks, and hybrid methods. The complex characteristics and challenges of both the virus and disease and of modeling their problems and data, as discussed in Sections \ref{subsec:data}, \ref{subsec:disease} and \ref{subsec:modelcomp} , are substantial. Though overwhelming efforts have been made in modeling COVID-19, the above complexities require significant novel developments through synergizing problems, data, and modeling techniques.

\section{Discussion and Opportunities}
\label{sec:dis-opps}

In the above review of each category of COVID-19 modeling techniques, a brief discussion has been provided on the main limitations, gaps and opportunities in those areas. Here, we expand this specific discussion to broad major gaps in the research on modeling COVID-19. Further, we discuss various open issues and opportunities for future research.

\subsection{Modeling Gap Analyses}
\label{subsec:gaps}

Two major aspects of modeling gaps include: the gaps in understanding the virus and disease nature, and the gaps in modeling their complexities.

\subsubsection{Gaps in understanding the problem nature}
\label{subsubsec:naturegaps}

Since the virus is new and unique, we have limited knowledge on all aspects of the SARS-CoV-2 virus and COVID-19, such as virus characteristics, epidemiological attributes and dynamics, socioeconomic influence, and virus mutations, and so on. Specifically, our poor understanding of the intrinsic and intricate pathological, biomedical and epidemiological attributes of the evolving SARS-CoV-2 and COVID-19 systems limits the modeling attempts and contributions. As a result, our understanding of the virus and disease is still \textit{insufficient} without substantial knowledge and comprehensive evidence on the system complexities; it is \textit{biased} to specific data, conditions or settings; it is \textit{shallow} without deep insights into the virus and disease nature; and it is \textit{partial} without a full picture of the SARS-CoV-2 and COVID-19 complexities, in understanding the COVID complex systems and their data complexities \cite{wang2021complex,aikp_Cao15}. 

To address these issues, the modeling has to start with building a comprehensive understanding of the virus nature and the fundamental complexities of the COVID-19 complex systems. Of the many questions to explore, we highlight the following important unknowns, which require cross-disciplinary scientific explorations by integrating medical science, virology, bio-medicine, and data-driven discovery.
\begin{itemize}
    \item \textit{The hidden nature of SARS-CoV-2 and COVID-19}: How does the virus interact with human and animal hosts? How does the virus genetic system look like? What are the epidemiological attributes of the virus and the disease characteristics after infections under different contexts, e.g., demographics, community (population) scenarios, ethnics, seasonality, and weather conditions? What are the high risk factors or high risk factor combinations of infection and mortality? What causes the different levels of symptom onset and differs asymptomatic and mild symptomatic infections from severe symptomatic infections?
    \item \textit{The mysterious mutation mechanisms of SARS-CoV-2}: What genomic and pathological factors determine the virus transform from one generation to another and over time? What genomic and pathological mechanisms drive the variations? Why the genetic variants differ from regions to regions and between population ethnics? 
    \item \textit{The influence of external factors on virus spread and evolution}: How does the virus evolve under different ethnic, environmental and intervention (including pharmaceutical and non-pharmaceutical) contexts? How do external factors such as demographics, ethnics, environment, and healthcare quality influence the virus transformation? How do personal hygiene, public health systems, public activities, population mobility and daily commuting affect the virus spread and evolution? 
    \item \textit{The virus adaption to vaccine and drug}: What are the relations between key factors such as the various vaccines and drugs available for treating COVID-19, the increasing virus mutants and their more contagious new strains, the widespread delta strain outbreaks, and unpredictable resurgence? How do the virus variants adapt to the vaccines and drugs? How does the vaccination affect the virus evolution? 
    \item \textit{The herd immunity vs. zero tolerance for the virus}: What is the new COVID normalcy, i.e., should a `zero tolerance for the virus' be the target and eventually remove the SARS-COV-2 like we did for SARS or shall `quotidian existence' be the new normal such that humans live with the virus as influenza? For the former, what is the herd immunity threshold for a manageable normal of living with the virus? Where is the manageable risk level of balancing the vaccination rate, public health system capacity, ethnic and community conditions, and acceptable daily numbers of infections and deaths? How does the regional inequality of vaccinations and public health systems affect the global recovery?
\end{itemize}

\subsubsection{Gaps in modeling the system complexities}
\label{subsubsec:modelgaps}

The modeling gaps come from both a poor understanding of the virus and disease nature and the  limitations in modeling their characteristics and complexities. On one hand, even though massive efforts have been made in modeling COVID-19, the existing modeling work is still in its early stage. The weaknesses and limitations of the existing work lie in
\begin{itemize}
    \item an average description of the population-wise coronavirus and the disease's epidemiological characteristics and observations after applying mitigation and control measures, no fine-grained and microlevel analysis and findings are available; 
    \item a direct application of existing (even very simple and classic) modeling methods without COVID-specific and optimal modeling mechanisms, typically by applying overparameterized or independently pretrained deep neural models or complex statistical and compartment models on low-quality and often small COVID data; 
    \item simple data-driven modeling purely motivated by applying advanced models (typically deep models) on COVID-19 data without a deep incorporation of domain and external knowledge and factors; and 
    \item a purposeful design without a comprehensive design or exploration of the multi-faceted COVID-19 characteristics and challenges in one framework or system. 
\end{itemize}

On the other hand, the general applications of existing methods also present unsuitability and incapability in tackling the complexities of the complex virus and disease. Table \ref{tab:models} compares the major modeling methods and their pros and cons in modeling COVID-19.
Consequently, it is common that the existing models and their modeling results 
\begin{itemize}
    \item often only reflect a specific population or cohort-based average estimation or hypothesis of epidemic transmission, losing a personalized applicability to individual cases or scenarios, making it difficult to undertake personalized treatment;
    \item are too specific to expand to other countries and scenarios, hard to reproduce and transfer to other regions without (significant) changes, making it unsuitable for broad applications; 
    \item over- or under-fit the given data and hypothesis settings, they are difficult to validate in a fine-grained way and have weak robustness or generalization for a general but deep understanding of the problems; and 
    \item lack the ability and capacity to disclose the intrinsic nature and general insights about the SARS-CoV-2 virus, COVID-19 disease, and their interventions. 
\end{itemize}

\begin{table*}[htb]
\centering
\caption{Comparison of COVID-19 Modeling Methods.}
\resizebox{1.0\textwidth}{!}{
\begin{tabular}{|  p{0.16\textwidth} | p{0.42\textwidth} | p{0.42\textwidth} |}
    \hline
    \textbf{Methods} & \textbf{Pros} & \textbf{Cons}  \\ \hline
    Time-series analysis & Temporal representations and interaction modelings of periodic and aperiodic components, relations and trends of COVID-19 cases at different states (e.g., new, susceptible, infected, recovered and death) and external temporal factors & Weak modeling power involving other rich factors (e.g., demographics and clinical attributes) and complex data characteristics (e.g., nonstationarity) and discovering the insight of COVID-19 driving forces and interventions \\ \hline
    General machine learning & Multifaceted factor and relation analysis, outlier detection, profiling, classification, prediction and impact analysis for disease diagnosis and case detection on small and poor-quality COVID-19 data, etc. & Poor modeling of weak but complex interactions, couplings, high-dimensional dependencies, heterogeneity, nonstationarity and other data challenges in multisource COVID-19 data \\ \hline
    Statistical modeling & Modeling distributional dynamics, uncertainty and dependency with analytical explanation and parameter settings & Requires informative prior knowledge, high modeling and computation complexity on poor-quality COVID data \\ \hline
    Epidemiological modeling & Built on epidemic knowledge, straightforward but domain-friendly and explainable hypothesis test, strong characterization of infection processes, state transitions, and parameter selection & Captures complex epidemic transmission characteristics, factors, causal relations and processes in COVID-19 developments; hypothesis of homogeneous disease transmissions \\ \hline
    Deep learning & Performs well with large and complex COVID-19 data (e.g., medical imaging)-based case and disease prediction and identification with annotated samples; pretrained model easily adaptable to new tasks & Requires annotated ground-truth of COVID-19 learning targets, easy to overfit small COVID-19 data, vulnerable results, poor interpretability, high computational cost \\ \hline
    Simulation & Imitates and replicates complex COVID-19 mechanisms and processes, cost-effective, reproducible and risk-averse, manually controllable for purposeful test and optimization & Proper knowledge and hypotheses about COVID-19 transmission and factor interactions, high experimental complexity, inactionable for evolving and random real-life scenarios \\ \hline
    Hybrid methods & Flexible and powerful in selecting and combining small COVID-19 multisource data and relevant multi-methods on demand for combined COVID-19 learning tasks and data & Understands constituents for their best ensemble to address specific COVID-19 challenges with appropriate design complexity, less flexible in combination optimization and explanability \\ \hline
    \end{tabular}
    }
\label{tab:models}
\end{table*}

\subsection{Opportunities for AI and Data Science}
\label{subsec:opport}

There are enormous opportunities and future directions in modeling COVID-19, including (1) fundamentally characterizing the system complexities, (2) addressing the aforementioned limitations of existing work, and (3) exploring new directions and alternatives. These are particularly valid for AI, data science and machine learning, which play a dominating role in the data-driven COVID-19 modeling.

\subsubsection{Characterizing the system complexities}
\label{subsubsec:syscomp}

To discover the mysteries of the COVID virus and disease, the most important opportunities come from understanding their nature and system characteristics and complexities, as discussed in Sections \ref{subsec:disease} and \ref{subsubsec:naturegaps}. Combining the domain-driven and data-driven thinking and techniques, there are various directions in characterizing the problem nature and system complexities:
\begin{itemize}
    \item extracting, representing and distinguishing observable and latent factors and metrics to describe the epidemiological, biological (genomic), medical (clinical and pathological) and social attributes, liveliness and dynamic processes of the virus, virus mutations, the disease and its variants from other similar viruses and diseases;
    \item identifying and characterizing external entities and factors (e.g., drugs, vaccines, ethnics, environment) and how they interact with the virus and disease and influence their evolution;
    \item characterizing and simulating the diversified (e.g., explicit vs. implicit, global vs. local, domain-specific vs. general) interactions and relations between the above-extracted explicit and implicit internal and external factors and their dynamics; 
    \item quantifying and simulating the virus and disease's system dynamics and genetic mechanisms (e.g., self-organization, genomic expression, genetic crossover and mutation, interaction and adaptation with external environment) in terms of temporal, dependent variables and major transformations;
    \item simulating and quantifying the virus parasitism, interactions, adaptation and evolution with human, animal and living hosts in a large scale.
\end{itemize}

\subsubsection{Enhancing COVID-19 modeling}
\label{subsubsec:modlimit}

To address the modeling gaps in Section \ref{subsubsec:modelgaps} and those rarely and poorly explored areas and challenges in Section \ref{subsec:data}, we here highlight the following major directions. 

\textit{Rarely to poorly addressed areas.}
First, opportunities to focus on the areas rarely or poorly addressed in the existing COVID-19 modeling include: (1) characterizing the effective NPIS on the variants of the SARS-CoV-2 virus and comparing them with those on the original strains; (2) quantifying the effects of COVID-19 vaccines, pharmaceutical and NPI interventions on the infection control, mobility, mental health, society and the economy, e.g., the efficacy of vaccinated population percentage on herd immunity, and the effect of variable close-contact interactions and individual actions on epidemic de-escalating; (3) balancing the NPI strength and the socioeconomic recovery, e.g., modeling the effect of full vs partial business close-downs and border control on virus confinement at different stages and for different sectors, and characterizing the effect of increasing daily commuting and workforce movement vs working-from-home and telecommuting on the virus confinement; (4) capturing the temporally evolving interplay and interactions between virus propagation and external interventions; and (5) modeling target problems by systemically coupling relevant multisource data and multiple modeling techniques, e.g., by involving pathogen-related, societal, environmental and racist factors and the disparities between developing and developed countries, age groups, and races.

\textit{Hybrid modeling.}
Second, the hybridization of relevant data and techniques offers significant opportunities to improve and expand the existing modeling capacity and results. Examples include integrating (1) coarse-grained and fine-grained modeling, e.g., epidemic modeling by SIR variants to inform further specific NPI's effect analysis; (2) static and dynamic modeling, e.g., from population-based static epidemic modeling to specific NPI-varying and time-varying case forecasting; (3) observable and hidden factors and relations, e.g., multisource-based attributed modeling with deep abstraction and representation of interactions between the multisource factors; (4) local-to-micro-level and global-to-macro-level factors, e.g., involving patient clinical and demographic records with their environmental and socioeconomic context in survival and mortality prediction and medical resource planning; and (5) domain, data and models for domain-specific, interpretable, evidence-based and actionable findings. These typically involve compound modeling objectives, multisource data, and multi-method ensembles. 

\textit{Enhanced COVID-19 modeling.}
Third, another set of new opportunities is to undertake sequential or multi-phase modeling, such as (1) from coarse-grained to fine-grained modeling: e.g., applying epidemic models like SIR and SEIR on COVID-19 in the initial stage and then modeling the impact of NPIs, the mobility and behaviour change of a population on epidemic dynamics; (2) from static to dynamic modeling: e.g., testing constant epidemic parameters and then time-varying settings such as NPI-sensitive varying parameters; and (3) from core to contextual factors: e.g., modeling epidemic processes on case data and then involving pathogen-related, societal and environment (like temperature and humidity) variables to model their influence on epidemic movements.

Lastly, alternative opportunities exist by (1) developing COVID-19-specific modeling methods, benchmarks and evaluation measures to address the virus and disease's challenges and their data challenges for an intrinsic interpretation of the virus and disease nature and dynamics; (2) trans-disciplinarily integrating the relevant domain knowledge and hypotheses from biomedical science, pathology, epidemiology, statistics and computing science to address multifaceted challenges in the virus, disease, data and modeling and to form a comprehensive understanding of the virus and disease; (3) defining multifaceted modeling objectives and tasks to directly address comprehensive epidemiological, clinical, social, economic or political concerns and their challenges in one framework; and (4) ethical and explainable COVID-19 modeling with privacy-preserving and distributed heterogeneous information integration, augmentation, representation and learning by utilizing personal computing devices (e.g., smart phones) and cloud analytics.

\subsubsection{Exploring new opportunities}
\label{subsubsec:newdirt}

In addition to many specific perspectives, such as hybridizing modeling objectives, data and methods in Section \ref{subsubsec:modlimit} and addressing the shortcomings in Section \ref{subsubsec:modelgaps}, we here highlight some other opportunities that may particularly benefit (from) AI, data science and machine learning advances. 

\textit{Quantifying the virus nature and complexities.}
An imperative yet challenging task for the AI, data science and machine learning communities is to `quantify' the nature and complexities of the virus and disease and address the fundamental questions on the virus nature and complexities raised in Section \ref{subsubsec:naturegaps}. Building on multi-disciplinary knowledge such as on epidemiology, genetic computing and theories of complex systems, large-scale agent-based epidemic simulation systems are demanding to test and improve genetic, clinical and epidemiological hypotheses and knowledge about the virus and characterize the virus' genetic evolution mechanisms. 
\begin{itemize}
    \item \textit{Large-scale COVID epidemiological dynamics}: to obtain quantitative results and verification on questions in Section \ref{subsubsec:naturegaps}, such as how does a virus evolve, cross-over and mutate; how do billions of coronaviruses interact, compete, and transform over time; and how do environmental factors affect the virus life and genetic evolution.
    \item \textit{Large-scale human-virus interactions}: to characterize the experiments in relation to questions such as how does the full population of a country interact with the virus under their varied demographic profiles, hygiene protection habits, health conditions, vaccination conditions, mobility settings, etc. by mimicking their physical census data and circumstances in the real world; what is the vaccination threshold to build the herd immunity for a country by considering their specific circumstances; and to compare the simulation results with the reality of various waves of COVID-19 epidemic occurred in the country.   
    \item \textit{Large-scale intervention influence on human-virus interactions}: to quantify and evaluate how does the residents in a country respond to various intervention policies and restrictions on public and household activities over time; how does enforcing or relaxing interventions and restrictions affect the virus spread, infection numbers, and public heath system quality; and what vaccination and intervention preconditions make business reopening possible, etc.
\end{itemize}

\textit{Data-driven discovery of COVID mysteries.}
There is increasing and comprehensive sources of COVID-19 data available publicly and through private providers. Data-driven discovery on this COVID-19 data can substantially leverage other domain-specific research on COVID to disclose the mysteries of COVID. 
\begin{itemize}
    \item \textit{COVID data genomics}: forming the data genomics of COVID for a person, country, community or task by automatically extracting and fusing all possibly relevant data, e.g., contacts, personal health, mobility, clinical reports, exposure to infected people, and household and public activities in a privacy-preserving manner.
    \item \textit{COVID data augmentation}: developing new techniques to address the various data quality issues embedded in the data, as discussed in Section \ref{subsec:data} and novel augmented analytics and learning methods to directly learn from poor quality COVID data.
    \item \textit{All-purpose representation of COVID attributes}: learning the representations on all-relevant COVID data that can be used to describe the full profile of COVID and support diverse learning objectives and tasks in an ethical and privacy-preserving manner.
    \item \textit{Automated COVID screening and diagnosis}: developing techniques and systems to automatically detect, screen, predict and alert potential infection of the virus and disease on the COVID data genomics.
    \item \textit{Virus detection and interaction modeling}: developing personal IoT assistants and sensors to detect the virus, trace its movement and its origin and visualize the 'COVID net' showing its propagation paths, interactions and networking with other viruses and hosts.
    \item \textit{COVID knowledge graph}: generating knowledge graph showing the ontology about the virus; ontological connections between concepts on the virus; relations between knowledge on the virus and its protection, intervention, treatment and influence; and important highlights such as new knowledge discovered and misinformation detected.
    \item \textit{COVID safety and risk management}: developing systems and tools (including mobile apps) for personal and organizational daily management of their COVID safety and risk, e.g., COVID-safe physical and emotional health management, mobility planning, risk estimation and alerting, infection tests, immunity estimation, and compliance management.
    \item \textit{Metasynthetic COVID decision-support systems}: developing evidence-based decision support systems to fuse real-time and relevant big data, simulate and replay the outbreaks, estimate NPI effects, discover evidence from data and modeling, engage domain experts in the modeling and optimization processes, generate recommendations for decision-making, and support the data-driven analytics and management of severe disasters and emergencies.
\end{itemize}

\section{Concluding Remarks}
\label{sec:conclusion}

The COVID-19 pandemic's short-to-long-term influence and impact on public health (both physical and mental health), human daily life, global society, economy and politics is unprecedented, lasting, evolving yet quantified and verified. This review paints a comprehensive picture of the field of COVID-19 modeling. The multidisciplinary methods including mathematical modeling, AI, data science and deep learning on COVID-19 data have deepened our understanding of the SARS-CoV-2 virus and its COVID-19 disease's complexities and nature; contributed to characterizing their propagation, evaluating and assisting in the effect of preventive and control measures, detecting COVID-19 infections, predicting next outbreaks, and estimating the COVID-19 influence and impact on psychological, economic and social aspects.  

The review also highlights the important demands and significant gaps in deeply and systemically characterizing COVID-19-related problems and complexities; and developing effective, interpretable and actionable models to characterize, measure, imitate, evaluate and predict broad-based challenges and problems and to proactively and effectively intervene in them. Such COVID-19 modeling research proposes many significant challenges and opportunities to the multidisciplinary modeling communities in the next decade. These include not only immediately gaining intrinsic knowledge and proactive insight about the evolving coronavirus and its disease outbreak, infection, transmission, influence and intervention; but also preparing to tackle future global health, financial, economic, security-related and other black-swan events and disasters. 

\begin{acks}
This work is partially sponsored by the Australian Research Council Discovery grant DP190101079 and the ARC Future Fellowship grant FT190100734.

We thank Wenfeng Hou, Siyuan Ren, Yawen Zheng, Qinfeng Wang and Yang Yang for their assistance in the literature collection. More information about COVID-19 modeling is in \url{https://datasciences.org/covid19-modeling/}.
\end{acks}

\bibliographystyle{ACM-Reference-Format}
\bibliography{acmart-1}

\end{document}